\documentclass[12pt]{article}
\usepackage{amsmath,amsfonts,epsfig, mathrsfs}
\numberwithin{equation}{section}

\ifx\ltimes\undefined
\newcommand{\ltimes}{{\kern3pt\hbox{\vrule width 0.4pt height 5.30pt
depth .0pt}\kern-1.76pt\times\kern1pt}} \fi

\ifx\lrtimes\undefined
\newcommand{\rtimes}{{\kern1pt\times\kern-4.76pt\kern3pt\hbox{\vrule width 0.4pt height 5.30pt
depth .0pt}}} \fi

\def\Z {\mathbb{Z}}
\def\R {\mathbb{R}}

\def\ti{\tilde}

\def\bid{\hbox{1\hspace{-0.04in}I}} 

\def\a{\alpha}
\def\b{\beta}

\def\g{\gamma}

\def\G{\Gamma}

\def\P{\Pi}

\def\cG{{\cal G}}
\def\cX{{\cal X}}

\begin{document}

\begin{titlepage}
\begin{flushleft}
\hfill  IMPERIAL-TP-2009-CH-01 \\
\hfill  QMUL-PH-09-02
\end{flushleft}

\vspace*{10mm}

\begin{center}

{\Large {\textbf{Non-geometric backgrounds, doubled geometry\\
\vspace*{3mm}
 and generalised T-duality}}} \\

\vspace*{10mm}

{C M Hull$^{1}$ and R A Reid-Edwards$^{2}$} \\
\vspace*{17mm}

{\em $^1$The Blackett Laboratory, Imperial College London} \\
{\em Prince Consort Road, London SW7 2AZ, U.K.} \\

\vspace*{4mm}


{\em $^2$Centre for Research in String Theory}\\
{\em Queen Mary, University of London} \\ {\em Mile End Road, London, E1 4NS, U.K.} \\

\vspace*{12mm}

\end{center}

\begin{abstract}

\noindent
String backgrounds with a local torus fibration such as
T-folds are naturally formulated in a doubled formalism in which the torus fibres are doubled to include dual coordinates conjugate to winding number.
Here we formulate and explore a generalisation of this construction in which all coordinates are doubled, so that the doubled space is a twisted torus, i.e. a compact space constructed from identifying a group manifold under a discrete subgroup. This incorporates reductions with duality twists, T-folds and a class of flux compactifications, together with the non-geometric backgrounds expected to arise from these through T-duality. It also incorporates backgrounds that are not even locally  geometric, and suggests a generalisation of T-duality to a more general context.
We discuss the effective field theory arising from such an internal sector, give a   world-sheet sigma model formulation of string theory on such backgrounds and illustrate our discussion with detailed examples.

\end{abstract}

\vfill

\noindent {$^1$c.hull@imperial.ac.uk\\ $^2$rreidedwards@gmail.com (Visiting Researcher)}

\end{titlepage}

\newpage

\section{Introduction}

String theory can be formulated on certain non-geometrical spaces  \cite{Hellerman:2002ax,Dabholkar ``Duality twists orbifolds and fluxes'',Flournoy:2004vn,Kachru:2002sk,Narain:1990mw,Kumar:1996zx} as well as the familiar geometric spaces that consist of a manifold equipped with a metric and various background fields. An important class of these are the T-folds  \cite{Hull ``A geometry for non-geometric string backgrounds'',Hull ``Doubled geometry and T-folds''}, which are spaces constructed from patches of conventional string backgrounds that have transition functions that include T-dualities. T-folds can arise from taking the T-duals of conventional backgrounds, but there are also some non-trivial examples  that are not related to any conventional background by dualities \cite{Hellerman:2002ax}.

T-duality is a stringy symmetry acting on spaces which have  a torus fibration, so T-folds are constructed from patches that
are a product of a torus with a patch of a base space.
The standard rules for T-duality found by Buscher \cite{Buscher ``A Symmetry of the String Background Field Equations''} require that the $U(1)^d$ torus action on the fibres be isometric and preserve the background fields.
However, there is some evidence that there should be a generalisation of T-duality that applies to the case of a torus fibration where the torus action is not isometric \cite{Dabholkar ``Generalised T-duality and non-geometric backgrounds''}. Such a case arises, for example, on trying to T-dualise a  {three-dimensional} torus with $H$-flux in all three circles \cite{Kachru:2002sk}. More generally, it can take a geometric space or a T-fold to a space with what has been called $R$-flux \cite{Shelton ``Nongeometric flux compactifications'',Shelton:2006fd}.

Non-geometric backgrounds cannot be  {fully understood} using supergravity   or conventional   world-sheet sigma-models, so another approach is needed. One approach has been through the doubled formalism \cite{Hull ``A geometry for non-geometric string backgrounds''}.
Conformal field theory on a $d${-dimensional} torus has a natural formulation on a doubled space, the doubled torus.
States naturally live on the $2d$-dimensional Narain lattice, labeled by integers determining
the momentum and the winding number or string charge.
The T-duality group $O(d,d;\Z)$ acts naturally on this lattice.
Fourier transforming the $d$ quantized momenta and $d$ winding numbers gives $2d$ periodic coordinates of a doubled torus $T^{2d}$ which contains the original torus $T^d$.
Acting with $O(d,d;\Z)$ serves to rotate the physical torus into a different $T^d$ subspace of the doubled torus, which contains all T-duals of the original torus.  In this way, T-duality can be thought of as changing the choice of  $T^d$ subspace of the doubled space that is to be  regarded as \lq physical'. The name polarisation was suggested in \cite{Hull ``A geometry for non-geometric string backgrounds''} for the choice of such a $T^d$ subspace, in analogy with classical mechanics.
The group $O(d,d;\Z)$ acts geometrically through large diffeomorphisms on the doubled torus, allowing a T-duality covariant formulation. This is then broken when a polarisation is chosen.

This is the   basis for the doubled formalism for T-folds  \cite{Hull ``A geometry for non-geometric string backgrounds'',Hull ``Doubled geometry and T-folds''}.
For a T-fold or    geometric   background with a $T^d$ fibration, the $T^d$ fibres are replaced by the doubled tori $T^{2d}$ and, as the   group $O(d,d;\Z)$ acts geometrically through large diffeomorphisms on the doubled fibres, the result is a $T^{2d}$ bundle. If a global polarisation  exists, then it leads to a geometric background by selecting a  submanifold with $T^d$ fibres, while a non-trivial T-fold arises when there is a topological obstruction to choosing a polarisation globally. In such cases, a polarisation can be chosen locally in each patch, but the patches do not fit together to form a submanifold and there is no global spacetime. In \cite{Hull ``A geometry for non-geometric string backgrounds''}, a   world-sheet formulation for strings in such spaces was given, based on a sigma-model whose target is this geometrical space with doubled fibres, together with a self-duality constraint to halve the number of degrees of freedom on the doubled fibres.
This formulation has the virtue of being manifestly duality-covariant, and involves structures that also feature in generalised geometry \cite{Courant,Gualtieri}.

In this doubled picture, the
extra $d$ coordinates that are conjugate to   winding numbers are   auxiliary and no {physical} fields depend on them. However, it was suggested in \cite{Dabholkar ``Generalised T-duality and non-geometric backgrounds''} that   generalising T-duality to the case without isometries on the torus would lead to configurations  in which the background fields have non-trivial dependence on the extra dual coordinates
in this doubled representation.
One of our aims here is to seek a natural doubled geometry for such cases.
There is some evidence that such more general non-geometric backgrounds should arise in string theory.
In \cite{Gregory:1997te} it was argued that T-dualising the NS 5-brane properly leads to
a background with non-trivial dependence on a dual coordinate, and the physical implications were explored. In \cite{Tong:2002rq,Harvey:2005ab,Okuyama:2005gx}, it was argued that this dependence reflected world-sheet instanton effects. In \cite{Dabholkar ``Generalised T-duality and non-geometric backgrounds''} it was suggested that backgrounds depending on both spacetime and dual coordinates would arise natually in string field theory.

Our construction is motivated by   the so-called twisted torus. Consider a
  reduction with duality twist, i.e.  a reduction on a  $d$-torus to give a theory with a duality symmetry, followed by
reduction on a further circle with a duality twist.
This can be thought of as  a stringy version \cite{Dabholkar ``Duality twists orbifolds and fluxes''}  of a Scherk-Schwarz reduction \cite{Scherk ``How To Get Masses From Extra Dimensions''}.
 It was shown in \cite{Dabholkar ``Duality twists orbifolds and fluxes'',Hull ``Massive string theories from M-theory and F-theory''} that if the duality twist is geometric, then
this is equivalent to a compactification on a space which is a $T^d$ bundle over a circle.
Such a torus bundle over a circle is parallelisable and is in fact a $(d+1)$-dimensional group manifold $G$ identified under the action of a discrete group $\Gamma$ \cite{Hull ``Flux compactifications of string theory on twisted tori''}; such a space $G/\Gamma$ is sometimes referred to as a twisted torus in this context. Moreover, the group $G$ is precisely the Kaluza-Klein gauge group that
 arises from compactification of pure gravity on $G/\Gamma$, as we show in section 2.1.

More generally, one can consider a reduction in which the duality twist is in the T-duality group $O(d,d;\Z)$.
Then the doubled formalism is in terms of a $T^{2d}$ bundle over a circle \cite{Hull ``A geometry for non-geometric string backgrounds'',Hull ``Gauge Symmetry T-Duality and Doubled Geometry''}, which is itself a
twisted torus given by the identification of a $(2d+1)$-dimensional group by a discrete subgroup.
We will review this construction in detail and give some illustrative examples.
It is natural to also consider adding a coordinate conjugate to the winding charge
on the base circle, giving a $(2d+2)$-dimensional space. This gives a $(2d+2)$-dimensional twisted torus, but one would expect that the extra doubled coordinate for the base plays a trivial role, in that nothing depends on it.
However, it was argued in \cite{Dabholkar ``Generalised T-duality and non-geometric backgrounds''} that T-duality on the base circle would lead to configurations with a non-trivial dependence on this extra coordinate.
Moreover, we cannot use standard approaches to check this and find the dependence on the dual coordinate. We will here  construct a natural $(2d+2)$-dimensional geometry and attempt to describe different dual formulations in terms of polarisations selecting $d+1$ of the directions locally.
As we will see, this does not always lead to even local patches that are patches of geometric backgrounds, and moreover there can be an unexpected dependence on the extra coordinate doubling the base.

The theory that results from such a reduction with duality twist  has a gauge group ${\cal G}$ which is $(2d+2)$-dimensional, as is familiar from the special cases that have a  field theory truncation in which the dimensional reduction  amounts to a Scherk-Schwarz reduction \cite{Scherk ``How To Get Masses From Extra Dimensions''}.
This suggests considering  the $(2d+2)$-dimensional group manifold ${\cal G}$ identified under a discrete subgroup to give a compact twisted torus $ {\cal G} /\Gamma$.
This was first proposed in \cite{Hull ``Gauge Symmetry T-Duality and Doubled Geometry''} and a related proposal  was considered in \cite{Dall'Agata:2007sr}.
 This gives a natural geometry  which includes a circle that is dual to the base. The idea is that choosing different polarisations of this completely doubled space should give configurations that are dual to one another.
These include the duals of the original configuration obtained by acting  with $O(d,d;\Z)$ T-dualities on the  fibres, but also lead to new configurations with non-trivial dependence on the
dual coordinate of the base by acting with what we refer to as generalised T-dualities.

In this way, a doubled space which is a twisted torus $ {\cal G} /\Gamma$ is a natural generalisation
of the bundles with doubled torus fibres. In this paper we systematically investigate the generalisation of the doubled formalism of \cite{Hull ``A geometry for non-geometric string backgrounds''} to the doubled twisted torus   $ {\cal G} /\Gamma$ for
 general groups $ {\cal G}$ that have a natural metric of signature $(D,D)$.
We discuss   the spacetime picture and give a  world-sheet sigma model with such a doubled target space
together with a constraint that halves the doubled degrees of freedom.
The group structure plays a vital role in the construction.
We discuss the discrete symmetries that replace $O(D,D;\Z)$ (where
$2D$ is the dimension of ${\cal G})$ and the dualisations that arise from different choices of $D$-dimensional polarisation.
The formalism applies readily to  the case  considered above in which $ {\cal G}$ is the gauge group from a reduction with a duality twist. It can also accommodate the non-abelian T-duality of \cite{Giveon   ``On nonAbelian duality''} with $ {\cal G}$ the cotangent bundle of a group $G$,
or the Poisson-Lie duality of \cite{Klimcik  ``Dual Nonabelian Duality And The Drinfel'd Double''}  in the case in which $ {\cal G}$ is a Drinfel'd double.
Such non-abelian dualities are believed not to be symmetries of string theory  \cite{Giveon   ``On nonAbelian duality''} but instead relate distinct string backgrounds, while the generalised  dualities discussed above are expected to be stringy symmetries \cite{Dabholkar ``Generalised T-duality and non-geometric backgrounds''}.
Thus some care is needed in interpreting the formalism and applying it to the general case. However, it is possible that the present formalism may provide new insight into
 non-abelian and Poisson-Lie dualities.

The plan of the paper is as follows. In the following section we   review T-duality twist compactifications. The $(d+1)$-dimensional internal space is described in terms of the doubled torus formalism and the doubled twisted torus formalism. The existence of global polarisations and the role of T-duality in relating different polarisations is discussed. In section three, we apply the formalism of section two to a particular three-dimensional compact manifold - the nilfold - and discuss how this background, and the dual configurations, related to the nilfold by the action of $O(2,2;\Z)$, may be lifted to and recovered from a five-dimensional doubled torus and a six-dimensional doubled twisted torus. Section four reviews the doubled torus formalism from the   world-sheet perspective, as introduced in \cite{Hull ``A geometry for non-geometric string backgrounds''} and a detailed account of how the constraint is imposed in the sigma model theory is given. In section five, this sigma model description is applied to the five-dimensional doubled torus examples considered, from the target space perspective, in section three. Finally, in section six, we introduce a   world-sheet description of the doubled twisted torus formalism. It is shown that the sigma model for the doubled torus introduced in \cite{Hull ``A geometry for non-geometric string backgrounds''} emerges as a particular special case and the   world-sheet description of the $R$-flux background is discussed.

\section{Target Spaces and Doubled Target Spaces }

Consider the theory in $(n+d+1)$-dimensional spacetime with  a metric, two-form gauge field $\widehat{B}_{(2)}$, scalar field $\widehat{\Phi}$ and
the Lagrangian
\begin{equation}
\label{D+d+1 lagrangian} {\cal L}_{n+d+1}=e^{-\widehat{\Phi}}\left( \widehat{R}*1-d\widehat{\Phi} \wedge *d\widehat{\Phi} -
\frac{1}{2}\widehat{G}_{(3)}\wedge *\widehat{G}_{(3)} \right)
\end{equation}
where $\widehat{G}_{(3)}=d\widehat{B}_{(2)}$. The compactification on $T^d$, using the standard Kaluza-Klein ansatz,  gives \cite{Maharana
``Noncompact symmetries in string theory''}  a massless field theory with gauge group $U(1)^{2d}$ and a manifestly $O(d,d)$ invariant Lagrangian
in $(n+1)$ dimensions
\begin{eqnarray}\label{O(d,d) Lagrangian}
{\cal L}_{n+1}&=&e^{-\phi}\left(R*1+*d\phi\wedge d\phi+\frac{1}{2}*G_{(3)}\wedge G_{(3)}+\frac{1}{4}*d{\cal M}^{AB}\wedge d{\cal M}_{AB}\right.
\nonumber\\ &&\left.-\frac{1}{2}{\cal M}_{AB}*{\cal F}^A\wedge{\cal F}^B\right)
\end{eqnarray}
 where ${\cal F}^A=d{\cal A}^A$, and ${\cal A}^A$ are $2d $ abelian gauge fields, with $d$ gauge fields coming from the off-diagonal
parts of the metric and $d$ gauge fields coming from the off-diagonal parts of the 2-form gauge field. The scalar coset space $O(d,d)/O(d)\times
O(d)$ is parameterised by a symmetric $2d\times 2d$ matrix  ${\cal M}_{AB}$, satisfying the constraint
\begin{equation}\label{M=LML}
{\cal M}_{AB}=L_{AC}({\cal M}^{-1})^{CD}L_{BD}
\end{equation}
where $L_{AB}$ is the constant $O(d,d)$-invariant metric, which is used to raise and lower the indices $A,B=1,...,2d$.

The generators $T_A$ of the $U(1)^{2d}$ gauge symmetry, consist of $Z_a$, ($a,b=1,2,...d$)  which generate the   $U(1)^d$ action on
the $T^d$ fibre, and
 $X^a$, which generate antisymmetric tensor transformations for the
$B$-field components with one leg on the $T^d$ and the other in the external spacetime,
 so that
\begin{eqnarray}
T_A=\left(%
\begin{array}{c}
  Z_a \\
  X^a \\
\end{array}%
\right)
\end{eqnarray}
In this basis, the $O(d,d)$ metric is off-diagonal
\begin{equation}
\label{Lis}
L=\left(\begin{array}{cc}0 & 1 \\ 1 & 0
\end{array}\right)
\end{equation}

Next, consider a Scherk-Schwarz reduction on a further circle with periodic coordinate $x \sim x+1$, with an $O(d,d)$ duality twist around the
    circle \cite{Scherk ``How To Get Masses From Extra Dimensions''}. The twist is specified by $N^A{}_B$, a matrix representation of an element of the Lie algebra of $O(d,d)$, and the $x$-dependence is
given in terms of an $O(d,d)$ transformation $\exp (Nx)$, so that the
 $O(d,d)$ monodromy on going around the $x$ circle  is $\exp (N)$.
 In string theory, the monodromy is required to be in the T-duality group $O(d,d;\Z)$ \cite{Hull ``Massive string theories from M-theory and F-theory'',Hull ``Unity of superstring dualities''}.
The reduced theory may be written in a manifestly $O(d+1,d+1)$ covariant way \cite{Kaloper ``The O(dd) story of massive supergravity'',Hull ``Flux compactifications of string theory on twisted tori''}
\begin{eqnarray}\label{O(d+1,d+1) Lagrangian}
{\cal L}_n&=&e^{-\varphi}\left(R*1+*d\varphi\wedge d\varphi+\frac{1}{2}*{\cal H}_{(3)}\wedge {\cal H}_{(3)}+\frac{1}{4}*D{\cal M}_{MN}\wedge
D{\cal M}^{MN}\right. \nonumber\\ &&\left.-\frac{1}{2}{\cal M}_{MN}*{\cal F}^M\wedge{\cal F}^N\right)+V*1
\end{eqnarray}
The theory has a non-abelian gauge symmetry, for which the field strengths for the gauge connections ${\cal A}^M$ are ${\cal F}^M$. The two-form
gauge field $B_{(2)}$ has a three-form field strength ${\cal H}_{(3)}=dB_{(2)}+...$ with Chern-Simons terms. The scalar coset space $O(d+1,d+1)/O(d+1)\times O(d+1)$ is
parameterised by a symmetric  $2(d+1) \times 2(d+1) $ matrix  ${\cal M}_{MN}$, satisfying the constraint
\begin{equation}\label{M=LML2}
{\cal M}_{MN}=L_{MP}({\cal M}^{-1})^{PQ}L_{NQ}
\end{equation}
where $L_{MN}$ is the $O(d+1,d+1)$-invariant metric, which is used to raise and lower the indices $M,N=1,...,2d+2$.
It is a constant $2(d+1)\times 2(d+1)$ matrix given by (\ref{Lis}).
The explicit
relationship between the fields in the $(n+1)$-dimensional abelian theory and the $n$-dimensional non-abelian theory, along with the explicit form
for the scalar potential $V$ may be found in Appendix A of \cite{Hull ``Gauge Symmetry T-Duality and Doubled Geometry''}, or in \cite{Kaloper ``The O(dd) story of massive supergravity''}.

One effect of the duality twist is to give a non-abelian gauge symmetry. With no twist, $N^A{}_B=0$, this would have given a reduction on
$T^{d+1}$ of the same form as (\ref{O(d,d) Lagrangian}), with abelian gauge symmetry $U(1)^{2(d+1)}$ and $2(d+1)$ abelian gauge fields ${\cal
A}^M$. The generators consist of the $2d$ generators $T_A$ together with $Z_x$  {and} $X^x$ from the reduction on the $x$ circle. With a twist,  {$N^A{}_B\neq 0$}, this
algebra is deformed to a non-abelian gauge algebra of the same dimension, $2(d+1)$. The Lagrangian (\ref{O(d+1,d+1) Lagrangian}) has a gauge
symmetry with Lie algebra \cite{Hull ``Gauge Symmetry T-Duality and Doubled Geometry''}
\begin{eqnarray}\label{O(d,d+16) Lie algebra}
\left[Z_x,T_A\right]&=&-N^B{}_AT_B   \qquad\qquad  \left[T_A,T_B\right]=-N_{AB}X^x
\end{eqnarray}
 {where $N_{AB}=L_{AC}N^C{}_B$ is antisymmetric and} all other commutators vanish. Note that the   $T_A$ no longer generate an abelian sub-algebra.

Under the decomposition of $T_A$ into $Z_a$ and $X^a$, the twist matrix may be written as (using $N_{AB}=-N_{BA}$)
\begin{eqnarray}\label{mass}
N^A{}_B=\left(%
\begin{array}{cc}
  f_{xa}{}^b &   Q_x{}^{ab}  \\
 K_{xab}   & -f_{xb}{}^a \\
\end{array}%
\right)
\end{eqnarray}
for some antisymmetric $Q_x{}^{ab}=-Q_x{}^{ba}$  {and} $K_{xab}   =- K_{xba}$. The gauge algebra can then be written as
\begin{eqnarray}
[Z_x,Z_a]=f_{xa}{}^bZ_b+K_{xab}X^b   \qquad  [Z_x,X^a]=-f_{xb}{}^aX^b+Q_x{}^{ab}Z_b\nonumber
\end{eqnarray}
\begin{eqnarray}\label{algebra fHQ}
[Z_a,Z_b ]=K_{xab}X^x  \qquad [X^a,Z_b]=-f_{xb}{}^aX^x \qquad  [X^a,X^b]=Q_x{}^{ab}X^x
\end{eqnarray}
with all other commutators vanishing.  Here $K$ is the $H$-flux, $f$ is sometimes referred to as a geometric flux, and $Q$ is sometimes referred to as a
non-geometric flux.

The matrices of the form (\ref{mass}) are the generators of $O(d,d)$.
We will refer to the  subgroup generated by lower triangular matrices of the form (\ref{mass}) with
 $Q_x{}^{ab}=0$ as the geometric subgroup $\Delta$,
 consisting of $GL(d;\R)$ transformations generated by $f_{xa}{}^b$ and  B-shifts acting on the fibre components of $B$, $B_{ab}\to B_{ab}+ \lambda K_{xab}$.
 This has a discrete subgroup
 $\Delta (\Z)= \Delta \cap  O(d,d;\Z)$.
 If the twist is in  $\Delta (\Z)$,
  then it is geometric, consisting of a $GL(d;\Z)$ twist  acting as a large diffeomorphism of the $T^d$ fibres together with a discrete $B$-shift.
  This is equivalent to the
compactification with $H$-flux on a $T^d$ torus bundle over a circle with monodromy $\exp({f_{xa}{}^b})$ \cite{Hull ``Flux
compactifications of string theory on twisted tori'',Hull ``Flux compactifications of M-theory on twisted tori'',ReidEdwards ``Geometric and non-geometric compactifications of IIB supergravity'',ReidEdwards:2008rd}. For such a geometric twist, this compactification space is a group manifold
$G$, identified under a discrete subgroup $\G\subset G$. The group $G$ is usually non-compact, and $\G$ is chosen, if possible, to be such that
$G/\G$ is compact. A subgroup  $\G$ which satisfies this criterion is said to be cocompact.

\subsection{The Pure Gravity Example}

As an illustrative example, consider the pure gravity theory given by setting the $B$-field and dilaton $\varphi$  to zero. The monodromy of the
reduction is given by (\ref{mass}) with $Q_x{}^{ab}=0$, $K_{xab}=0$ and $f_{xa}{}^b=N^b{}_a$. The $n$-dimensional action (\ref{O(d+1,d+1)
Lagrangian}) reduces to \cite{Scherk ``How To Get Masses From Extra Dimensions''}
\begin{eqnarray} \label{gravact}
{\cal L}_n&=&R*1-\frac{1}{4}g^{mn}g^{pq}Dg_{mp}\wedge *Dg_{nq}-\frac{1}{2}g_{mn}F^m\wedge*F^n+V*1
\end{eqnarray}
The gauge group $G$ has Lie algebra (\ref{O(d,d+16) Lie algebra})
\begin{equation}\label{nilfold algebra}
[Z_x,Z_a]=-N^b{}_aZ_b   \qquad  [Z_a,Z_b]=0
\end{equation}
This can be viewed as compactification on a space ${\cal N}$ that is a $T^d$ bundle over a circle \cite{Hull ``Flux compactifications of string theory on twisted tori''}. This compactifying space
looks locally like the $(d+1)$-dimensional group manifold $G$, and is in fact the group $G$ identified under the action of  a discrete subgroup
$\G\subset G$   which acts from the left \cite{Hull ``Flux compactifications of string theory on twisted tori''}.

Dimensional reduction on a group manifold $G$ would give a theory  with a gauge symmetry $G_L\times G_R$ arising from the isometry group $G_L\times G_R$ of the group manifold, with $G_L$ acting from the left and $G_R$ from the right.
Identifying under the action of a discrete subgroup of  $\Gamma \subset G_L$ acting from the left breaks the $G_L$ symmetry, but the $G_R $ isometry is left intact, so that the theory has at least $G_R$ gauge symmetry.
(Generically, $G_L$ is completely broken, but if there is a subgroup commuting with $\Gamma$, it will   break to that subgroup.) Choosing a vacuum configuration will spontaneously break the gauge symmetry further to the subgroup preserving the vacuum \cite{Hull ``Flux compactifications of string theory on twisted tori''}. The Lagrangian (\ref{gravact}) is a consistent truncation of   that obtained from reduction on a group manifold $G$, in which only the gauge fields $A^m$ for $G_R$ are kept, while the ones for $G_L$ are set to zero. As a result, it is also a consistent truncation  for the reduction on $G/\G$, and contains all the gauge fields for the case in which  identifying under $\G$ breaks all of $G_L$.

 It is useful to consider a matrix representation of the gauge algebra
\begin{equation}\label{duality twist algebra}
[t_x,t_a]=-N^b{}_at_b, \qquad [t_a,t_b]=0
\end{equation}
This algebra can be represented by the $(d+1)\times(d+1)$ matrices
\begin{equation}
t_x=\left(\begin{array}{cc}-N^a{}_b & 0 \\ 0 & 0
\end{array}\right)  \qquad  t_a=\left(\begin{array}{cc}0 & e_a \\ 0 & 0
\end{array}\right)
\end{equation}
where $e_a$ is the $d$-dimensional column vector with a 1 in the a'th  position and zeros everywhere else. Coordinates $x, z^a$ can be
introduced locally for the group manifold $G$, with the group element
 $g=g(x,z^a)\in G$
given by
 \begin{equation}\label{groupG}
g=\left(\begin{array}{cc} \left(e^{-Nx}\right)^a{}_b & z^a \\ 0 & 1
\end{array}\right)
\end{equation}
Then the left-invariant Maurer-Cartan forms, $P=P^mt_m=g^{-1}dg$  are given by
\begin{equation} \label{forms}
P^x=dx \qquad P^a=\left(e^{Nx}\right)^a{}_bdz^b
\end{equation}
The $P^m$ are dual to the left-invariant vector fields
\begin{eqnarray}
Z_x=\frac{\partial}{\partial x}   \qquad Z_a=\left(e^{-Nx}\right)^b{}_a\frac{\partial}{\partial z^b}
\end{eqnarray}
which generate the gauge algebra (\ref{nilfold algebra}) and so the gauge algebra is given a geometric interpretation by the lift of the $n$-dimensional theory to a compactification of an $(n+d+1)$-dimensional theory on a $(d+1)$-dimensional internal space. Note that the left-invariant vector
fields generate the right-acting group $G_R$. We may also define the right-invariant one-forms $\widetilde{P}=\widetilde{P}^mt_m=dgg^{-1}$
\begin{equation} \label{Rforms}
\widetilde{P}^x=dx \qquad \widetilde{P}^a=dz^a+N^a{}_bz^bdx
\end{equation}
which are dual to the right-invariant vector fields
\begin{eqnarray}
\widetilde{Z}_x=\frac{\partial}{\partial x}-N^b{}_az^a\frac{\partial}{\partial z^b}   \qquad \widetilde{Z}_a=\frac{\partial}{\partial z^a}
\end{eqnarray}
which generate the left-acting group $G_L$. The full gauge algebra $G_L\times G_R$ of the group manifold $G$ is generated by the vector fields
$(Z_m,\widetilde{Z}_m)$. The left-invariant $Z_m$ remain globally defined after identifying by the discrete group $\G\subset
G_L$, but the $\widetilde{Z}_m$ generally will not be.

We now turn to the discrete subgroup $\G$. The torus bundle over a circle is obtained from the compactification of this non-compact group
manifold   under the  identification by a discrete subgroup $\G$, acting from the left. The left action of
\begin{equation}
h(\alpha,\beta^a)=\left(\begin{array}{cc} \left(e^{-N\alpha}\right)^a{}_b & \beta^a \\ 0 & 1 \end{array}\right)
\end{equation}
is
\begin{equation}
g(x,z^a)\to h(\alpha,\beta^a)\cdot g(x,z^a)
\end{equation}
and acts on the coordinates through
\begin{equation}
x\to x+\alpha  \qquad  z^a\to (e^{-N\alpha})^a{}_b z^b+\beta^a ~.
\end{equation}
The discrete subgroup is  $\G=\{h(\alpha,\beta^a)\in
G_L\mid \a ,\b^a \in \Z\}$ and we can identify the group manifold $G$ under $\Gamma$. This gives a compact space $G/\Gamma$ \cite{Hull ``Flux
compactifications of string theory on twisted tori''}.

In this example we have seen that the lift of the $n$-dimensional theory to a $(n+d+1)$-dimensional compactification led to a geometric
interpretation of the gauge algebra (\ref{nilfold algebra}). In the following sections we extend this idea and construct backgrounds on which a
part, or all, of the gauge algebra (\ref{O(d,d+16) Lie algebra}) has a natural geometric action. In particular we shall be interested in
generalising the above discussion to compactifications involving a $B$-field, and to non-geometric compactifications.

\subsection{T-Folds and the B-Field}

Reduction on $T^d$ followed by reduction on $S^1$ with a $GL(d,\Z)$ twist is, as we have seen, equivalent to compactification on a $T^d$ bundle
over a circle, which is also a twisted torus. In string theory, however, the twist on the $S^1$ can be by any element of   $O(d,d;\Z)$. For
twists in the  geometric subgroup $\Delta (\Z)$,  this is equivalent to reduction on a  twisted torus with flux. However, for twists involving
T-duality the result is not equivalent to reduction on any geometric space with flux, but can instead be viewed as  reduction on a T-fold, a
non-geometric space with transition functions including T-dualities  \cite{Hull ``A geometry for non-geometric string backgrounds''}. Locally,
these look like $T^d$ bundles, but the transition functions between the fibres on overlaps of patches on the base include $O(d,d;\Z)$
transformations. These twisted reductions over a circle are among the simplest examples of T-folds.

\subsection{The Doubled Torus}


Conventional reduction on $T^d$ with coordinates $z^a$ gives a theory with $O(d,d)$ symmetry, and this symmetry can be made manifest in a doubled formalism in which an
extra $d$ coordinates $\tilde{z}_a$ that are conjugate to the  $d$ winding numbers are introduced, to give a doubled torus $T^{2d}$ \cite{Hull ``A geometry for non-geometric
string backgrounds''} with periodic coordinates $\mathbb{X}^A=(z^a,\tilde{z}_a)$. As  reviewed above, the reduction with a   twist by a $GL(d,\Z)$  torus diffeomorphism is equivalent to compactification
on a space which is a $T^d$ torus bundle over a circle.  More generally, a  non-geometric reduction with twist in $O(d,d;\Z)$ can similarly be
represented as a reduction in the doubled formalism on a $T^{2d}$ bundle over   $S^1$ with monodromy in $O(d,d;\Z)$.  This representation gives the
monodromy a geometric interpretation as an element of the $T^{2d}$ mapping class group, as $O(d,d;\Z)\subset GL(2d;\Z)$. In general, the data
specifying a T-fold over a base $M$ also specifies a doubled torus bundle over $M$ with fibres $T^{2d}$, and the T-fold reduction can be
re-expressed as a compactification in the doubled formalism on the doubled torus bundle over $M$  \cite{Hull ``A geometry for non-geometric
string backgrounds''}.

For the twisted reduction on a circle, the $T^{2d}$ has coordinates $\mathbb{X}^A$ and the base has coordinate $x$, while the set of $2d+1$
natural  one-forms on    the corresponding doubled torus bundle over the circle are
\begin{equation}\label{doubled forms}
{\cal P}^A=\left(e^{Nx}\right)^A{}_Bd\mathbb{X}^B    \qquad\qquad P^x=dx
\end{equation}
These generalise the one-forms (\ref{forms}). This $(2d+1)$-dimensional space   ${\cal T}_{2d+1}$ is a $T^{2d}$ bundle over $S^1$:
\begin{eqnarray}
T^{2d}\hookrightarrow &{\cal T}_{2d+1}&\nonumber\\
&\downarrow&\nonumber\\
&S^1&\nonumber
\end{eqnarray}

The local description of the background in terms of $d+1$ coordinates  is recovered from the duality-covariant doubled torus picture by choosing
a polarisation  \cite{Hull ``A geometry for non-geometric string backgrounds''}, which selects $d$ coordinates $z^a$ from the $2d$ coordinates
$\mathbb{X}^A$ for each point on the base  as coordinates on the physical spacetime.

 More generally, consider   a $T^{2d}$ bundle over a base $M$ (in the examples considered above, $M$  is a circle).
  In a patch
$U_{\alpha}$ of the base $M$ (where $U_{\alpha}$ is open and contractible), the background looks like $U_{\alpha}\times T^{2d}$. To recover the
theory in the physical $(d+1)$-dimensional space, we   choose a projection which determines which $d$ of the $2d$ coordinates $\mathbb{X}^A$
will be treated as spacetime coordinates  and which $d$ coordinates will be treated as conjugate to the winding modes.  In the $T^{2d}$ fibre over the
patch $U_{\alpha}$, a polarisation is specified by a  constant projector $\Pi_{\alpha}$ where $\Pi_{\alpha}:U_{\alpha}\times T^{2d}\rightarrow U_{\alpha}\times
T^{d}$, which selects coordinates $z^a$ on a $T^d$ sub-manifold of $T^{2d}$:
\begin{equation}
z^a=\Pi^a{}_A\mathbb{X}^A
\end{equation}
The physical space with coordinates $z^a$ is required to be maximally isotropic with respect to the $O(d,d)$ metric $L_{AB}$
\begin{equation}
L^{AB}\Pi_A{}^a\Pi_B{}^b=0
\end{equation}
It is useful to define the complement $\widetilde{\Pi}$ which projects onto the auxiliary coordinates
$$\tilde{z}_a=\widetilde{\Pi}_{aA}\mathbb{X}^A$$
It is also useful to introduce the polarisation tensor
$$\Theta^{\hat{A}}{}_A=\left(%
\begin{array}{c}
  \Pi^a{}_A\\  \widetilde{\Pi}_{aA} \\
\end{array}%
\right)$$
 so that
\begin{eqnarray}
\mathbb{X}^{\hat{A}}=\Theta^{\hat{A}}{}_A\mathbb{X}^A:=\left(%
\begin{array}{c}
  z^a \\ \tilde{z}_a \\
\end{array}%
\right)
\end{eqnarray}

For each point on the base, the fibre geometry  is encoded in a \lq generalised metric' ${\cal M}_{AB}$,
which is a symmetric $2d\times 2d$ matrix satisfying the constraint (\ref{M=LML}) so that it
 parameterises the
  coset space $O(d,d)/O(d)\times
O(d)$.
 Given a polarisation, the metric $g_{ab}$ and $B$-field $B_{ab}$ on $T^d$, in each patch
$U_{\alpha}$, are given by  \begin{equation} ({\cal M}^{-1})^{\hat{A}\hat{B}}=\Theta^{\hat{A}}{}_A({\cal M}^{-1})^{AB}\Theta_B{}^{\hat{B}}
\end{equation}
where
\begin{eqnarray}\label{pol}
 ({\cal M}^{-1})^{\hat{A}\hat{B}}= \left(\begin{array}{cc}
g^{ab} & -B_{bc}g^{ac} \\ -B_{ac}g^{bc} & g_{ab}+g^{cd}B_{ac}B_{bd}
\end{array}\right)
\end{eqnarray}

The key point is that backgrounds can be considered in which different polarisations are used in different patches, although they are constant in each patch.
 We then consider a covering by such patches $\{U_{\alpha}\times T^{2d}\}$, each with an associated
projector $\Pi_{\alpha}$. The transition functions on the overlap between patches $U_{\alpha}$ and $U_{\beta}$ are elements
of\footnote{$O(d,d;\Z)$ is the  group of large diffeomorphisms of $T^{2d}$ preserving $L_{AB}$ and $U(1)^{2d}$ is the natural torus action on
$T^{2d}$.} $O(d,d;\Z)\ltimes U(1)^{2d}$. If the $\{U_{\alpha}\times T^d\}$ patch together with transition functions in the geometric subgroup
$\Delta (\Z)\ltimes U(1)^d$ of $O(d,d;\Z)\ltimes U(1)^{2d}$, then the physical space, given this choice of polarisation, is geometric. This is
sufficient for the projector $\Pi$ onto the physical subspace to be globally defined, but this is not sufficient for the complement
$\widetilde{\Pi}$ to be globally defined; this will also be well-defined if  in addition $K=0$ so that $N$ is block diagonal.
If the transition functions are not all  in the geometric subgroup, then the space is a T-fold.

The transition functions can be viewed in two ways \cite{Hull ``A geometry for non-geometric string backgrounds''}. They can be regarded as active,
 with the polarisations defined globally $\Pi_{\alpha}=\Pi_{\beta}$ and on the overlap $U_{\alpha}\cap U_{\beta}$ the coordinates related by
  $\mathbb{X}^I_{\alpha}=(h_{\alpha\beta})^I{}_J\mathbb{X}^J_{\beta}+\alpha^I_{\alpha\beta}$ where $h_{\alpha\beta}\in O(d,d;\Z)$ and
$\alpha_{\alpha\beta}\in U(1)^{2d}$. Alternatively, they can be regarded as passive, with the transition function acting on
the polarisation $\Theta_{\alpha}=h^{-1}_{\alpha\beta}\Theta_{\beta}$ and the coordinates  unchanged,
$\mathbb{X}_{\alpha}=\mathbb{X}_{\beta}$. We will mostly use the passive viewpoint in this paper.

In the examples above with a circle base,   the structure is
encoded in the monodromy of the duality twist reduction. First let us consider the active
perspective. With the identification $x\sim x+1$, the monodromy in the fibre coordinates is given by
\begin{equation}
\mathbb{X}^{\hat{A}}\sim \left(e^{-N}\right)^{\hat{A}}{}_{\hat{B}}\mathbb{X}^{\hat{B}}
\end{equation}
In particular, using the global polarisation, this implies
\begin{equation}
z^a\sim \left(e^{-N}\right)^a{}_bz^b+\left(e^{-N}\right)^{ab}\tilde{z}_b
\end{equation}
so that, if $Q_x{}^{ab}\neq 0$ then $\left(e^N\right)^{ab}\neq 0$ and the monodromy will mix $z^a$   with the $\tilde{z}_a$. Then  the physical space will be non-geometric. From the passive perspective, it is the polarisation which is not globally defined so that if
$Q_x{}^{ab}\neq 0$ no global polarisation will exist.

 Next we consider the issue of the geometrisation of the gauge algebra. The vector fields dual to the one-forms (\ref{doubled forms}) are
\begin{equation}
Z_x=\frac{\partial}{\partial x}      \qquad T_A=\left(e^{-Nx}\right)^B{}_A\frac{\partial}{\partial \mathbb{X}^B}
\end{equation}
These generate, not the gauge algebra (\ref{O(d,d+16) Lie algebra}), but a sub-algebra of a contraction of it, given by
\begin{equation} \label{sdfadfg}
\left[Z_x,T_A\right]=-N^B{}_AT_B    \qquad  \left[T_A,T_B\right]=0 ~ ,
\end{equation}
so that, even though the generators $T_A$ have a geometric action as generators of translations along the $T^{2d}$ fibres,  the gauge algebra
(\ref{O(d,d+16) Lie algebra}) of the $n$-dimensional theory does not have a fully geometric realisation in the doubled torus picture. In
particular, the generator $X^x$ does not have a geometric action on the space. This should not come as a surprise. The symmetries relating to
the components of the $B$-field along the $T^d$ directions have been given a geometric interpretation by doubling the fibres of the torus, but
the $B$-field transformation along the base (generated by $X^x$) does not have a geometric interpretation in the doubled torus formalism, as the $x$ coordinate is not doubled here.

 The doubled space ${\cal T}$ is    in fact a twisted torus of the form ${\cal T}=\cG/\G$ where the $(2d+1)$-dimensional
(non-compact) Lie group $\cG$ is generated by the Lie algebra (\ref{sdfadfg}) and $\G$ is a discrete subgroup of $\cG$ acting from the left.
The elements of this discrete subgroup  are labeled by integers $\a, \b^A$
and
act on the coordinates as
\begin{equation}\label{boundary}
x\to  x+\a    \qquad  \mathbb{X}^A\to  \left(e^{-N\a }\right)^A{}_B\mathbb{X}^B+\b^A
\end{equation}
Taking the left quotient by $\G$ fixes the global structure of ${\cal T}$.

\subsection{The Doubled Twisted Torus}

In the reduction of pure gravity  reviewed in  section 2.1, the reduction with a   twist by a large diffeomorphism of the torus is equivalent to a reduction
on a twisted torus. The gauge group is $(d+1)$-dimensional, and the internal space is the twisted torus given by identifying the group manifold
of the gauge group under a discrete subgroup. The gauge symmetry   then has a manifest geometric origin as the isometry group of the internal
space. For the reduction of string theory with an $O(d,d;\Z)$ twist, it was proposed in
\cite{Hull ``Gauge Symmetry T-Duality and Doubled Geometry''}  that the full $2(d+1)$-dimensional gauge group    be given a geometric
representation as transformations on a $2(d+1)$-dimensional space. This involves doubling the coordinate on the base circle, introducing a
coordinate $\tilde x$ conjugate to the winding number on the $x$-circle, as well as doubling the $d$ fibre coordinates, as in the doubled formalism reviewed above.
 The doubled
space is essentially the group manifold of the gauge group, compactified  by identifying under a discrete subgroup, i.e. it is a
twisted torus $\cX=\cG/\G$ where $\cG$ is the $(2d+2)$-dimensional group manifold for the group generated by the Lie algebra elements satisfying
(\ref{O(d,d+16) Lie algebra}) and $\G$ is a (discrete) cocompact subgroup, acting from the left, which contains information on the global
structure of $\cX$.

The idea \cite{Hull ``Gauge Symmetry T-Duality and Doubled Geometry''} is, then, to seek a doubled space in which all the gauge symmetries are realised as geometric symmetries, and then discuss the way the
{local spacetime} picture emerges from choosing a polarisation. In the doubled torus picture, choosing different
polarisations gives the various T-dual backgrounds. However, in the doubled torus, the only directions which are doubled are torus fibres, while
here the base circle is also doubled.

This gives a general framework in which there is a doubled space that is locally a group manifold. This has been motivated by the case of
reductions with duality twists, in which different T-dual backgrounds arise from different physical slices or polarisations of this doubled
space. This formalism can be applied more generally to theories in $n$ dimensions similar to those discussed above with a $2D$-dimensional gauge
group $\cG$ {with Lie algebra $[T_M,T_N]=t_{MN}{}^PT_P$ where the structure constants $t_{MN}{}^P$ generalise those of (\ref{O(d,d+16)
Lie algebra}).} Then a natural framework \cite{Hull ``Gauge Symmetry T-Duality and Doubled Geometry''} is to consider a doubled internal space given by a group manifold $\cG$ of dimension $2D$, or the
\lq twisted torus' $\cX=\cG/\G$ for some discrete $\G$. When possible, it is natural to choose $\G$ so that $\cX=\cG/\G$ is compact. If the
$n$-dimensional theory arises from a compactification, then the compactifying space will arise as a $D$-dimensional polarisation of
$\cX$. We will in addition require that $\cG$ preserve a metric $L_{MN}$ of signature $(D,D)$  {so that $t_{MNP}=t_{MN}{}^QL_{QP}$ is
totally antisymmetric. Generally ${\cal G}$  will be the semi-direct product of a subgroup of $O(D,D)$ and a group that does not act on the metric $L_{MN}$.

In some cases, different polarisations will give T-dual backgrounds, and these cases will be our main focus here. However, this more general
framework encompasses cases where different polarisations give inequivalent string backgrounds (i.e. the corresponding sigma-models define
distinct conformal field theories). For example, a non-abelian generalisation of T-duality was proposed in \cite{Quevedo
  ``Duality symmetries from nonAbelian isometries in string theory''} and further generalised to
Poisson-Lie duality in \cite{Klimcik  ``Dual Nonabelian Duality And The Drinfel'd Double''}. These give transformations between related backgrounds, but which are not equivalent string backgrounds
\cite{Giveon
  ``On nonAbelian duality''}. The doubled twisted torus $\cX$ in some cases  includes different backgrounds related by Poisson-Lie duality  as different
polarisations of a doubled twisted torus, and the framework also proposes a generalisation of Poisson-Lie duality to include  {$H$- and $R$-fluxes}. We will
discuss briefly  these more general cases here, and further details will be given in \cite{Ron}.

{Let us return to the specific class of examples} arising from  reduction with a duality twist, with gauge algebra (\ref{O(d,d+16) Lie algebra}). The
Lie algebra (\ref{O(d,d+16) Lie algebra}) can be represented in terms of operators acting on the $2(d+1)$ coordinates $(x, \ti x, \mathbb{X}^A)$
of the doubled twisted torus $\cX$, where $\mathbb{X}^A$ are the coordinates on the doubled torus fibre $T^{2d}$, as
\begin{eqnarray}\label{left generators}
Z_x=\frac{\partial}{\partial x}+N^A{}_B\mathbb{X}^B\frac{\partial}{\partial \mathbb{X}^A}   \qquad X^x=\frac{\partial}{\partial {\ti x}} \qquad T_A=\partial_A-\frac{1}{2}N_{AB}\mathbb{X}^B\frac{\partial}{\partial {\ti x}}
\end{eqnarray}
Then $X^x$ generates translations along the new coordinate $\ti x$. The  {left-invariant} one-forms dual to these vector fields   satisfy the Maurer-Cartan
equations
\begin{equation}
d{\cal P}^A-N^A{}_BP^x\wedge {\cal P}^B=0   \qquad \qquad   dQ_x-\frac{1}{2}N_{AB}{\cal P}^A\wedge{\cal P}^B=0  \qquad\qquad    dP^x=0
\end{equation}
which are solved by\footnote{Note that a coordinate redefinition, as described in \cite{Hull ``Gauge Symmetry T-Duality and Doubled Geometry''} has been used to simplify the expressions for the
one-forms.}
\begin{equation}\label{oneformz}
{\cal P}^A=\left(e^{Nx}\right)^A{}_Bd\mathbb{X}^B   \qquad \qquad   Q_x=d\tilde{x}+\frac{1}{2}N_{AB}\mathbb{X}^Ad\mathbb{X}^B  \qquad\qquad
P^x=dx
\end{equation}

It is useful to introduce $\cG$ Lie algebra indices $M,N=1,2,...2d+2$ so that the generators $T_M$
are
\begin{eqnarray}
T_M=\left(%
\begin{array}{c}
  Z_x   \\
  X^x \\
  T_A \\
\end{array}%
\right)
\end{eqnarray}
and the Lie algebra can be written as
\begin{eqnarray}\label{algebra}
[T_M,T_N]=t_{MN}{}^PT_P
\end{eqnarray}
For the $O(d,d;\Z)$ duality-twist reductions, the only non-vanishing components of the structure constants $t_{MN}{}^P$  are
 \begin{equation}
t _{xB}{}^A=-N^A{}_B, \qquad t_{x[AB]}=-N_{AB}
\end{equation}
and those related to these by symmetry.
The dual one-forms (\ref{oneformz}) on $\cX$ can then be written as
 ${\cal P}^M={\cal P}^M{}_Id\mathbb{X}^I$, where  $I,J=1,2,...2d+2$   coordinate indices on the group manifold $\cG$ and twisted torus $\cX$. These one-forms satisfy the  Maurer-Cartan
equations
\begin{eqnarray}
d{\cal P}^M+\frac{1}{2}t_{NP}{}^M{\cal P}^N\wedge{\cal P}^P=0
\end{eqnarray}
so that the space is parallelisable.

\subsubsection{Geometry}

  To formulate dynamics, we introduce a positive definite \lq generalised metric' ${\cal H}_{IJ}$
  and three-form ${\cal K}$ on the doubled twisted torus $\cX=\cG/\G$, in addition to the metric $L{}$ of signature $(D,D)$.
  For cases leading to actions of the form
  (\ref{O(d+1,d+1) Lagrangian}) in which the scalar fields are given by the $2D\times 2D$ matrix ${\cal M}_{MN}$ which is independent of the coordinates on $\cX$, one
   can naturally define the line element  and three-form ${\cal K}$ on  $\cX$ by
$$
ds^2={\cal H}_{IJ}d\mathbb{X}^I\otimes d\mathbb{X}^J	\qquad	{\cal K}=\frac{1}{6}t_{MNP}{\cal P}^M\wedge {\cal P}^N\wedge{\cal P}^N
$$
where the metric ${\cal H}_{IJ}$ is given by
$$
{\cal H}_{IJ}={\cal M}_{MN}{\cal P}^M{}_I{\cal P}^N{}_J
$$
As we shall see in section 4, these can be used to define a sigma-model on $\cX$ with kinetic term determined by
${\cal H}_{IJ}$
and a Wess-Zumino term given by ${\cal K}$, and the normalisation of  ${\cal K}$ is fixed by the
requirement that there be a   self-duality constraint  that can be imposed on the world-sheet fields.
The constant matrix $L_{MN}$ similarly defines a metric ${L}_{IJ}$  of signature $(D,D)$ by
$$
{L}_{IJ}={L}_{MN}{\cal P}^M{}_I{\cal P}^N{}_J
$$
Then
\begin{equation}\label{H=LHL2}
{\cal H}_{IJ}=L_{IK}({\cal H}^{-1})^{KL}L_{LJ}
\end{equation}
Coordinate systems in which
$L_{IJ}$ is a constant matrix given by (\ref{Lis})
are particularly natural, and in such a coordinate system ${\cal P}^M{}_I$ is a matrix in $O(D,D)$.

\subsubsection{Polarisation}

In the doubled torus construction, a polarisation was chosen to specify a physical subspace,  at least locally.  Our aim here is to generalise
this to the curved case but, as we shall see, there are new issues that arise. In this subsection, we will consider polarisations for twisted
tori constructed from general $2D$-dimensional groups $\cG$ preserving the metric $L_{MN}$. We will start by considering polarisations on a
general, {possibly non-compact,}   group manifold $\cG$, then discuss the structures they give rise to on factoring by a discrete subgroup to give a
{(compact)} twisted torus $\cG/\G$.

A natural extension of the polarisation
 $\Pi^a{}_A$ used in the doubled torus formalism
is to introduce a projector $\Pi^m{}_M$ (with $m,n= 1,....,D$) mapping onto a $D$-dimensional subspace of the $2D$
dimensional tangent space,
which is totally null (maximally isotropic) with respect to the metric $L_{MN}$, i.e.
\begin{equation}
L^{MN} \Pi^m{}_M\Pi^n{}_N=0
\end{equation}
Introducing such a projector at the identity element of the group manifold then defines one everywhere; in a natural basis, the projector is constant over the manifold.
As before, the complementary projector is denoted by $ \widetilde{\Pi}_{mM}$.
We say that a subgroup $H$ of $\cG$ is isotropic or null if all of the vector fields on $\cG$
generating $H$ are null with respect to $L_{MN}$, and if the dimension of $H$ is $D$, half that of $\cG$, then we say that it is maximally isotropic.
The polarisation splits the tangent space into two halves, and we
will consider the case in which the frame components $ \Pi^m{}_M$ are locally constant, i.e. there is a constant matrix  $\Pi^m_{(\a)M}$ in each
patch $U_\a$ of $\cG$, but there can be different polarisation matrices in  different patches.

The polarisation projects the left-invariant generators $T_M$
of the right action $\cG_R$ into
\begin{equation}
Z_m = \widetilde{\Pi}_{mM}L^{MN}T_N, \qquad X^m= \Pi^m{}_M L^{MN}T_N
\end{equation}
There is a corresponding split of the dual one-forms into $P^m$ and $Q_m$. If we denote the right-invariant generators of the left acting group
$\cG_L$ by $\widetilde{T}_M$, then the polarisation projects
\begin{equation}
\widetilde{Z}_m = \widetilde{\Pi}_{mM}L^{MN}\widetilde{T}_N, \qquad \widetilde{X}^m= \Pi^m{}_M L^{MN}\widetilde{T}_N
\end{equation}
and   these right-invariant generators play an important role here.
(Recall that on the doubled group manifold $\cG$, both sets
$T_M$ and $\widetilde{T}_M$ are globally defined, but on the doubled twisted torus $\cX=\cG/\G$, where $\G$ acts from the left, generally only
the left-invariant vector fields $T_M$ and one-forms ${\cal P}^M$ will be globally defined.)
The gauge symmetry acts through the right action of $\cG_R$, so we will focus on the $\cG_R$-invariant generators $\widetilde{T}_M$,  which at any given point gives a $\cG_R$-invariant  basis of the tangent space that is split by the polarisation into the vectors
$\widetilde{Z}_m ,\widetilde{X}^m$.
The issue is then whether the split of the tangent vectors defined by the polarisation can be used to define a
 $D$-dimensional submanifold (at least locally) which can be viewed as a patch of spacetime.
 This will be the case provided the distribution defined by the set of $D$ vector fields $\widetilde{X}^m$  is integrable, as we shall now discuss.

An important case is that in which the
$\widetilde{X}^m$   close to form a $D$-dimensional sub-algebra
 \begin{equation}\label{subalg}
[\widetilde{X}^m,\widetilde{X}^n]=t^{mn}{}_p\widetilde{X}^p
\end{equation}
which requires that the structure constants and polarisation tensor satisfy
\begin{equation}\label{no R}
\Pi^p{}_Pt_{MN}{}^P\Pi^{mM}\Pi^{nN}=0
\end{equation}
Then, by Frobenius' theorem, the distribution defined by the $D$ vector fields $\widetilde{X}^m$ is integrable so that the polarisation defines a
submanifold locally. For this group manifold case, it in fact specifies a submanifold globally. The $\widetilde{X}^m$
generate a $D$-dimensional subgroup $\widetilde{G}$ of $\cG$. This acts on $\cG$ through the left action $\widetilde{G}_L\subset \cG _L$ and the
submanifold selected by the polarisation is  the   $D$-dimensional left coset $\cG/\widetilde{G}_L$. There is a natural action of $\cG_R$ on this coset $\cG/\widetilde{G}_L$,
generated by $Z_m$ and $X^m$.
An interesting special case is that in which the $\widetilde{X}$'s generate a subgroup
$\widetilde{G}_L$ and the $\widetilde{Z}$'s also generate a subgroup $G_L$, in which case  the doubled group $\cG$   is a Drinfel'd double with a
Lie algebra { which, as a vector space, is $\mathfrak{g}\oplus\tilde{\mathfrak{g}}$,
where $\mathfrak{g}$ is the Lie algebra of $G_L$ and $\tilde{\mathfrak{g}}$ is that for $\widetilde{G}_L$.
Note that the sub-algebras $\mathfrak{g}$, $\tilde{\mathfrak{g}}$  will not commute in general}  \cite{Klimcik  ``Dual Nonabelian Duality And The Drinfel'd Double''}.

In general the $\widetilde{X}$'s will not generate a sub-algebra, so that the $\widetilde{X}$ distribution will not be integrable and {does} not define a
submanifold. In this case, the polarisation does not pick a subspace even locally, so that it does not
 select a physical subspace in which there is a conventional formulation. If one then tries to lift the polarisation of the Lie algebra to a polarisation of the coordinates and define a geometry on a subspace, then the resulting metric and $B$-field depend explicitly on the auxiliary coordinates and are not ordinary fields on the subspace; this will be seen explicitly in examples in the next section. This is precisely the kind of non-geometric reduction introduced in
\cite{Dabholkar ``Generalised T-duality and non-geometric backgrounds''}. In such cases, if consistent, the theory can only be described in a
doubled formalism, and this will be discussed in later sections.
Similarly, the complementary polarisation will only define a submanifold if the $\widetilde{Z}$'s generate a subgroup $G_L$, and
again this will not be the case in general.

Next we turn to the application of this to {the compact twisted torus} $\cX=\cG/\G$. Consider first the case of a choice of constant $\P$
on $\cG$ in which (\ref{no R}) holds so that the $\widetilde X$ generate a subgroup $\widetilde{G}_L$.
The condition for the action of $\G$ on $\cG$ to induce a well-defined
 action of $\G$ on $\cG/\widetilde{G}_L$ is that $\G$ preserves $\widetilde{G}_L$, i.e.
 for all $\g\in \G$ and $k\in \widetilde{G}_L$
 $$ \g k \g ^{-1} = k'$$
 for some $k' \in \widetilde{G}_L$.
Then taking the quotient of $\cG/\widetilde{G}_L$ by $\G$ is well-defined and defines a subspace of $\cG/\G$.
 The choice  of
polarisation on $\cG$ is then consistent with the action $\G$, so that it is globally well-defined both on $\cG$ and $\cG/\G$ and selects a
geometric subspace of $\cG/\G$.

The discrete group $\G $ acts on the generators $\widetilde{T}_M$ and will map the $\widetilde{X}$'s to linear combinations of
$\widetilde{X}$'s and $\widetilde{Z}$'s. In the geometric case just considered, the   action of $\G$  preserves the subalgebra
$\widetilde{G}_L$ and
 maps the $\widetilde{X}$'s to linear combinations of $\widetilde{X}$'s .
More generally, $\G$ will not preserve the subalgebra  $\widetilde{G}_L$ {and the image of the $\widetilde{X}$'s under the action of $\G$
will include both $\widetilde{X}$'s and $\widetilde{Z}$'s}. In this case the polarisation on $\cG$ is not well-defined on the quotient $\cX$,
and this will give a  non-geometric background.

{ In the non-geometric case in which  the action of $\G$ does not preserve $\widetilde{G}_L$,
taking the quotient by $\G$ is inconsistent with taking the quotient by  $\widetilde{G}_L$.
Then
we cannot expect a global description of the spacetime to
exist, and may only recover a conventional spacetime in local patches.} Suppose then that $\cX=\cG/\G$ is covered by contractible patches
$U_\a$, each of which can be viewed as a contractible patch of the group manifold $\cG$, and with transition functions to be discussed below. In the passive
formulation,
  we choose a different constant polarisation $\P_{(\a)}$ in each patch, related by transitions
consistent with the action of $\G$. Suppose further  that in any given patch, the polarisation selects $\widetilde{X}$'s  that close under
commutation to generate a subalgebra (\ref{subalg}), and so defines an integrable distribution and hence a submanifold of the patch.
 This submanifold has the local structure of a patch of $\cG/\widetilde{G}_L$. The action of the transition functions on the polarisation will mean that in
different patches, different (conjugate) subgroups will be selected, and the submanifolds of each patch will not fit together to form a
submanifold of $\cX$. The result is a non-geometric space, which is constructed from patches each of which is geometric. That is, in each patch,
the polarisation selects a physical spacetime and there is a conventional local formulation, but these do not fit together to give a formulation
on a spacetime manifold. In general, there will be no global choice of polarisation.
 We stress that the condition (\ref{no R}) is a necessary requirement for a conventional
spacetime description to exist locally.

Similarly, the condition for the action of $\widetilde{G}_L$ on $\cG$ to induce a well-defined
 action of $\widetilde{G}_L$ on $\cG/\G$ is that $\widetilde{G}_L$ preserves $\G$, i.e.
 for all $\g\in \G$ and $k\in \widetilde{G}_L$
 $$ k\g k ^ {-1} = \g'$$
 for some $\g ' \in \G$.
Then taking the quotient of $\cG/\G$ by $\widetilde{G}_L$ is well-defined and defines a subspace of $\cG/\G$.

\subsubsection{Physical Interpretation}

We can   think of the doubled formalism as describing a `universal' string background which includes many different string backgrounds, each
given by a different choice of polarisation. In the case in which the different choices of polarisation are related by T-dualities or other
symmetries, they give physically equivalent backgrounds. However, the new formalism on a doubled twisted  torus can also incorporate backgrounds
related by  the non-abelian duality of \cite{Quevedo
  ``Duality symmetries from nonAbelian isometries in string theory''} or the Poisson-Lie duality of
\cite{Klimcik  ``Dual Nonabelian Duality And The Drinfel'd Double''}. These are \lq duality' transformations that relate backgrounds that are not equivalent string
backgrounds \cite{Giveon
  ``On nonAbelian duality''}  so that they are not string symmetries, but instead take one string background to another, physically
inequivalent background \cite{Giveon
  ``On nonAbelian duality''}.

A simple example of a doubled group  is the case in which $\cG=G\times {\widetilde G}$ with $G $ generated by the  $\widetilde{Z}$'s and
${\widetilde G}$ generated by the  $\widetilde{X}$'s. Then one polarisation gives the background given by the group manifold $G$, another gives
the background given by the group manifold ${\widetilde G}$,  but in general giving distinct
string backgrounds. For $\cG/\G$ with $\G= \G_1\times \G_2$ with
 $\cG/\G = ( G/\G_1)\times ({\widetilde G}/\G_2)$, the two polarisations would give
 backgrounds
$G/\G_1$ or ${\widetilde G}/\G_2$.

Some polarisations might lead to   conventional geometric backgrounds, while others might lead to T-folds. Any of these
  can  give consistent string backgrounds, provided other sectors are added to ensure conformal and modular invariance, so that a good string background is rewritten in terms of a polarisation of a doubled twisted torus. However, in general   a doubled twisted torus that
  has some polarisations that give
 good string backgrounds
  may also have other polarisations whose status is less clear.
These will be given by different polarisation projectors in different patches related in overlaps by discrete transformations, and such a
polarisation will lead to a generalisation of a T-fold in which the transition functions involve these discrete transformations.
The key issue is then whether these discrete transformations are symmetries of the string theory: this is essential for these to be candidate backgrounds for string theory.

The natural set of discrete transformations here is the group $Aut(\cG;\G,L)$ of automorphisms
of $\cG$ that preserve $\G$ and the metric $L_{MN}$, and this will then have a natural action on the theory defined on $\cG/\G$.
For example, for $\cG=\R^{2D}$, $\G=\Z^{2D}$ so that $\cG/\G=T^{2D}$, then
$Aut (\cG ;\G,L)=O(D,D;\Z)$, the T-duality group which is a symmetry of the string theory.
We expect that in general different polarisations will be related by the action of this group, and that this group will provide the discrete transition functions relating patches of the twisted torus.
The key issue here is whether $Aut(\cG;\G,L)$ is a symmetry of string theory, or if not, then which subgroup is.
Only transition functions that are symmetries of the physics can lead to
good string backgrounds, and only if polarisations are related by symmetries do they define equivalent backgrounds.
Proper T-dualities acting on torus fibres are symmetries and such transitions give rise to T-folds.
Other discrete transformations arising in this way  include
non-abelian T-dualities or Poisson-Lie dualities.  There is evidence that such `dualities' are  {generally} not symmetries of string theory \cite{Giveon
  ``On nonAbelian duality''},  so   a  background with such transitions would not be a good string background  in general.

The issue is then what subgroup of $Aut (\cG ;\G,L)$ is a symmetry of string theory and can be used
in transition functions. This will clearly contain proper T-dualities, but there is evidence
that certain generalisations of T-duality should also be allowed, although  {generic} Poisson-Lie dualities presumably should not. One of our    motivations is to consider such cases, and to investigate the generalisations of the usual T-dualities that are
suggested by the formalism, such as those proposed in  \cite{Dabholkar ``Generalised T-duality and non-geometric backgrounds''}; these involve
dualising a circle direction which is not isometric, so that conventional T-duality is not possible.

A related issue is that of whether two polarisations are physically equivalent.
In the case of the doubled torus formalism, all polarisations that are related to each other by $O(d,d;\Z)$ T-duality transformations on the $T^{2d}$ doubled torus fibres are physically equivalent.
In the doubled twisted torus formalism, some  polarisations will again be related to others by T-dualities
and will lead to equivalent  representations of the physics. However, others will not be so related, and the question arises as to whether they are  then physically equivalent.
They will typically be related by the action of   $Aut (\cG ;\G,L)$, but only if they are related by proper string symmetries will they be equivalent.

In summary, there are a number of cases. In each $2D$-dimensional patch, there is a polarisation projecting the tangent space at each point onto
a $D$-dimensional subspace. If this distribution is integrable  {(i.e. if (\ref{no R}) is satisfied),} then this selects a $D$-dimensional
submanifold of the patch. { There is then a description of the spacetime in this patch as a patch of $\cG/\widetilde{G}_L$}.
If the polarisation is globally defined on
$\cX$, then it selects a physical subspace which is a submanifold, given by identifying $\cG/\widetilde{G}_L$ under the action of $\G$, and this
gives a geometric background.

If the distribution in each patch is integrable but the polarisation is not globally defined on  $ \cG/\G$ (i.e. not {preserved by} the action of
$\G$), then the result is a  generalisation of a  T-fold, with a good doubled formulation on $ \cG/\G$ but where a $D$-dimensional spacetime can only be
selected locally in each patch, and these spacetime patches do not fit together to give a global spacetime.

Finally, if the distribution selected by the polarisation is not integrable, then although the polarisation splits the tangent space, it does
not define a submanifold even locally, so that no local spacetime
 and no local geometric picture can emerge. More will be said about the interpretation of such cases below.

\subsection{T-Duality and $R$-Flux}

Consider the case of the generic algebra of the form (\ref{O(d,d+16) Lie algebra}) arising from a reduction with a duality twist.
The monodromy  on going around the $x$ circle reflects the fact that translations in the $x$ direction are not an isometry -- the metric and fields depend non-trivially on $x$ -- so that conventional T-duality in the $x$ direction is not possible.
If $x$-translations were an isometry,    a {conventional} T-duality \cite{Buscher ``A Symmetry of the String Background Field Equations''} would have been possible and naturally formulated using the coordinate $\tilde x$  on a dual circle.
 It was  conjectured in \cite{Dabholkar
``Generalised T-duality and non-geometric backgrounds''}  that there should be  a generalised T-duality
in the $x$-dependent case which again involves introducing a dual coordinate
$\tilde x$, and
which exchanges $x$ and $\tilde{x}$.
The result is then a reduction with duality twist monodromy around the $\tilde{x}$ circle, and the duality exchanges
$Z_x$ with $X^x$. This  produces a theory with the gauge algebra
\begin{eqnarray}
\left[X^x,T_A\right]&=&-N^B{}_AT_B   \qquad\qquad  \left[T_A,T_B\right]=-N_{AB}Z_x
\end{eqnarray}
with corresponding one-forms
\begin{equation}\label{P}
{\cal P}^A=\left(e^{N\tilde{x}}\right)^A{}_Bd\mathbb{X}^B   \qquad \qquad   P^x=dx+\frac{1}{2}N_{AB}\mathbb{X}^Ad\mathbb{X}^B \qquad\qquad
Q_x=d\tilde{x}
\end{equation}
Decomposing  $T_A$ into $Z_a$ and
$X^a$, the twist matrix may be written as (using $N_{AB}=-N_{BA}$)
\begin{eqnarray}\label{massa}
N^A{}_B=\left(%
\begin{array}{cc}
  Q_a{}^{bx} &   R^{xab}  \\
 f_{ab}{}^x   & -Q_b{}^{ax} \\
\end{array}%
\right)
\end{eqnarray}
The gauge algebra is then
\begin{eqnarray}\label{algebra 1}
[X^x,Z_a]=-Q_{a}{}^{xb}Z_b+f_{ab}{}^xX^b   \qquad  [X^x,X^a]=Q_{b}{}^{xa}X^b+R^{xab}Z_b\nonumber
\end{eqnarray}
\begin{eqnarray}\label{algebra 2}
[Z_a,Z_b ]=f_{ab}{}^xZ_x  \qquad [X^a,Z_b]=Q_{b}{}^{xa}Z_x \qquad  [X^a,X^b]=R^{xab}Z_x
\end{eqnarray}
with all other commutators vanishing.

From this, we see in particular that the $\tilde x$-twist can incorporate an $R$-flux as well as a $Q$-flux. Consider the case where only $R^{xab}\neq 0$. The left-invariant algebra is then
\begin{eqnarray}\label{algebra 1}
 [X^x,X^a]=R^{xab}Z_b \qquad  [X^a,X^b]=R^{xab}Z_x
\end{eqnarray}
with all other commutators vanishing. The left-invariant one-forms, dual to the vector fields which generate this algebra, are
\begin{eqnarray}
\begin{array}{ll}
 P^x=dx+\frac{1}{2}R^{xab}\tilde{z}_ad\tilde{z}_b  &\qquad  P^a=dz^a+R^{xab}\tilde{x}d\tilde{z}_b\\
  Q_x=d\tilde{x} &\qquad Q_a=d\tilde{z}_a
\end{array}
\end{eqnarray}
The right-invariant generators of the left action $\cG_L$ satisfy the algebra
\begin{eqnarray}\label{algebra 1}
 [\widetilde{X}^x,\widetilde{X}^a]=-R^{xab}\widetilde{Z}_b \qquad  [\widetilde{X}^a,\widetilde{X}^b]=-R^{xab}\widetilde{Z}_x
\end{eqnarray}
with all other commutators vanishing. We see that in this case the generators $\widetilde{X}$ do not close to form a subalgebra and therefore a
conventional target space description cannot be recovered as a coset locally, as described above.
Attempting to choose a polarisation that selects the $Z$'s as the geometric generators does not work, as the corresponding distribution is not integrable.
The only description we have of such backgrounds is
through the doubled formalism.
More generally, the structure constant
$$
\Pi^p{}_Pt_{MN}{}^P\Pi^{mM}\Pi^{nN}=R^{mnp}
$$
is an obstruction to the closure of the algebra generated by $\widetilde{X}^m$ and means that, even locally, this polarisation has no
conventional spacetime description.
R-flux will be discussed further at the end of the next section in the context of particular examples.

\subsection{Drinfel'd doubles and doubled twisted tori}

One case of interest is that in which the generators of $\cG$ consist of $Z_m$ generating a $D$-dimensional subgroup $G$ and $X^m$ generating a $D$-dimensional subgroup $\widetilde{G}$.
The group will not in general be a  product $G\times\widetilde{G}$, but   instead the algebra will have  `cross-terms' $[X^m,Z_n]$ and be of the form
\begin{equation}\label{DD}
[Z_m,Z_n]=f_{mn}{}^pZ_p \qquad  [X^m,Z_n]=f_{np}{}^mX^p-Q_n{}^{mp}Z_p   \qquad  [X^m,X^n]=Q_p{}^{mn}X^p
\end{equation}
Then $G$ and $\widetilde{G}$ are both null with respect to the natural metric $L_{MN}$
(i.e. the generators of $G$ are all null and mutually orthogonal, and similarly for the generators of  $\widetilde{G}$)
and we have
 a triple of  Lie groups $(\cG,G,\widetilde{G})$.  In this context, the triple of Lie groups $(\cG,G,\widetilde{G})$ is often referred to as a Manin triple and the doubled group $\cG$ with metric
 of signature $(D,D)$ is said to be a Drinfel'd double
 \cite{Klimcik:1995ux,Drinfel'd1,Drinfel'd:1986in,SemenovTianShansky:1993ws,Weinstein}.
 The two complementary $D$-dimensional group manifolds $G$ and $\widetilde{G}$ are recovered as the cosets $\cG/\widetilde{G}_L$ and $\cG/G_L$ respectively.

For a given $\cG$, there may be different choices of subgroups
$G,\widetilde{G}\subset\cG$ such that in each case  $(\cG,G,\widetilde{G})$  is a Manin triple,
giving different decompositions of the same Drinfel'd double. In this way different Manin triples may correspond to different choices of polarisation, although not all choices of polarisation will give a Manin triple.

An example of a Drinfel'd double is the cotangent bundle for a $D$-dimensional group $G$, so that $\cG=T^*G$. In this case $\widetilde{G}=\R^D$ and the doubled group is the semi-direct product $\cG=G\ltimes\R^D$ generated by the Lie algebra
\begin{equation}\label{CoTangent}
[Z_m,Z_n]=f_{mn}{}^pZ_p \qquad  [X^m,Z_n]=f_{np}{}^mX^p   \qquad  [X^m,X^n]=0
\end{equation}
This is parameterised by $g\in G$ and coordinates ${\ti x}_m$ on $\widetilde{G}=\R^D$.
A basis of left-invariant one-forms on $\cG$ is
$$
P^m=(g^{-1}dg)^m    \qquad  Q_m=d{\ti x}_m+f_{mn}{}^p{\ti x}_pP^n
$$
where the one-forms $P^m$ and $Q_m$ are dual to the vector fields $Z_m$ and $X^m$ respectively.
The Lie algebra (\ref{CoTangent}) is encoded in the Maurer-Cartan equations for $P^m$ and $Q_m$ as described in previous sections.

 Note that the one-forms $P^m=P^m{}_idx^i$ and $\tilde{\ell}_m=d{\ti x}_m$ are left-invariant one-forms on $G$ and $\widetilde{G}=\R^D$ respectively and therefore give a globally defined basis of left-invariant forms on $G\times\R^D$, but not on $G\ltimes\R^D$. The action of $\widetilde{G}=\R^D$ on $G$ is trivial and so the $P^m$ lift to left-invariant forms on $\cG$; however, the non-trivial action of $G$ on $\widetilde{G}$ means that the $d{\ti x}$ are not globally defined on the double $\cG$. Instead, the globally defined one-forms on $\cG$ are $Q_m$, which are related to the forms $d{\ti x}_m$ on $\widetilde{G}$ by the `twisting' $d{\ti x}_m\rightarrow d{\ti x}_m+\widetilde{b}_{mn}P^n$ where $\widetilde{b}_{mn}=f_{mn}{}^p{\ti x}_p$.
The left-invariant one-forms may be written in a basis independent way as $P=P^mT_m$ and $Q=Q_m\widetilde{T}^m$, where
$$
P=g^{-1}dg  \qquad  Q=d\tilde{x}-[\tilde{x},P]
$$
with $\tilde{x}=\tilde{x}_m\widetilde{T}^m$ and where $T_m$ and $\widetilde{T}^m$ generate a matrix representation of the algebra (\ref{CoTangent}). It is not hard to show that the left action of $\cG$ on the coordinates $x^i$ and $\tilde{x}_m$ is
$$
\delta x^i=(P^{-1})_m{}^i\alpha^m   \qquad  \delta \tilde{x}=g^{-1}\tilde{\alpha}g
$$
with parameters $\a, \ti \a$.
The action of $\widetilde{G}_L$ on the coordinates $x^i$ parameterising $G=\cG/\widetilde{G}_L$ is trivial but   $G$ has a nontrivial   action on the coordinates $\tilde{x}_m$ which parameterise $\widetilde{G}$. It is then easy to see that the natural left-invariant forms $(P,\tilde{\ell})$ on $G\times\widetilde{G}$ are not invariant under $\cG$ but transform as
\begin{equation}\label{ghj}
\delta P=0  \qquad  \delta\tilde{\ell}=-[P,\delta \tilde{x}]
\end{equation}
The $\cG_L$-invariant forms $(P,Q=\tilde{\ell}+[P,\tilde{x}])$ can be thought of as a
  `twist' of $(P,\tilde{\ell})$.

A particular feature of  {the case in which $\cG$ is a Drinfel'd double} is that there are two natural polarisations, one corresponding to the
  coset $\cG/\widetilde{G}_L$ and one leading to the dual one $\cG/G_L$, and the dual one can be treated in the same way as the one corresponding to  $\cG/\widetilde{G}_L$  was in section 2.4.
The doubled twisted torus is given by identifying the Drinfel'd double by a subgroup $\G\subset\cG_L$, so that $\cX=\cG/\G$ is compact.
If the action of $\G$   preserves and is   preserved by $\widetilde{G}$, then the quotient $\cX/\widetilde{G}$
is well-defined and there is a global description of the spacetime resulting from this polarisation, similarly for  $\cX/{G}$. For the example given above in which $\cG=T^*G$, the action of $\G$ preserves and is preserved by $\R^D$ and so the quotient $\cX/\R^D$ is always well-defined and corresponds to the $D$-dimensional twisted torus ${\cal N}=G/\G'$, where $\G'\subset\G$ acts only on the coordinates of $G$ and leaves $\R^D$ invariant. Recovering a conventional spacetime description in this case simply corresponds to the natural bundle projection on $T^*{\cal N}$.

In general, the action of $\G$ need not preserve or be preserved by $\widetilde{G}$ (or $G$),
 in which cases
 the quotients $\cX/\widetilde{G}_L$ (or  $\cX/G_L$)
 will not be well-defined and there will be no global   spacetime from  these choices of polarisation. Conventional spacetime patches can   be recovered locally as   patches of  $\cG/\widetilde{G}_L$ (or
 $\cG/G_L$), as described in sections 2.4.1 and 2.4.2. Only in the cases in which the transitions between patches are through true symmetries of the string theory can such non-geometric backgrounds be  string backgrounds.

The addition of structure constants associated with $H$- and $R$-fluxes deforms the algebra (\ref{DD}) to
$$
[Z_m,Z_n]=f_{mn}{}^pZ_p+K_{mnp}X^p \qquad   [X^m,X^n]=Q_p{}^{mn}X^p+R^{mnp}Z_p
$$
$$
[X^m,Z_n]=f_{np}{}^mX^p-Q_n{}^{mp}Z_p
$$
so that neither $G$ nor $\widetilde{G}$ are subgroups of the doubled group $\cG$. The physics of the doubled geometry corresponding to such $H$- or $R$-twisted Drinfel'd doubles will be explored further in \cite{Ron}. In the absence of $R$-flux, the $X^m$ close to generate $\widetilde {G}$ and this is the case analysed in detail in section 2.4.

\subsubsection{Polarisations and group actions}

For a Drinfel'd double, one can define a basis of right- and left-invariant forms on the groups $G$ and $\widetilde{G}$, which we denote by $(r^m,\ell^m)$ and $(\tilde{r}_m,\tilde{\ell}_m)$ respectively\footnote{$(r^m,\tilde{r}_m)$ are right-invariant and $(\ell^m,\tilde{\ell}_m)$ are left-invariant. For $g\in G$ and ${\ti g}\in\widetilde{G}$, we can write $\ell=g^{-1}dg$, ${\ti \ell}={\ti g}^{-1}d{\ti g}$, $r=dgg^{-1}$ and ${\ti r}=d{\ti g}{\ti g}^{-1}$.}. The     actions of these factor groups on each other   reflect how the two groups are `twisted' together to form $\cG$.

A simple way to characterise the action of the sub-groups on each other is to look at the adjoint action  of $\cG$} on the (matrix) generators $T_m\in\mathfrak{g}$ and $\widetilde{T}^m\in\tilde{\mathfrak{g}}$, selected by   a choice of polarisation, $T_m=\Pi_m{}^MT_M$ and $\widetilde{T}^m=\widetilde{\Pi}^{mM}T_M$. We use $\mathfrak{g}$ and $\tilde{\mathfrak{g}}$ to denote the Lie algebra of $G$ and $\widetilde{G}$ respectively and  the adjoint action of $G$ on the generators of $\cG$  defines matrices $A$, $b$ and $\beta$ by
\begin{eqnarray}\label{adjoint 1}
g^{-1}\left(
        \begin{array}{c}
          T_m \\
          \widetilde{T}_m \\
        \end{array}
      \right)g=\left(
                 \begin{array}{cc}
                   A_m{}^n & b_{mn} \\
                   \beta^{mn} & (A^{-1})^m{}_n \\
                 \end{array}
               \right)\left(
        \begin{array}{c}
          T_n \\
          \widetilde{T}_n \\
        \end{array}
      \right)
 \end{eqnarray}
 Similarly, the adjoint action of $\widetilde{G}$ on the generators of $\cG$   defines matrices   $\widetilde{A}$, $\widetilde{b}$ and $\widetilde{\beta}$ by
\begin{eqnarray}\label{adjoint 2}
       \tilde{g}^{-1}\left(
        \begin{array}{c}
          T_m \\
          \widetilde{T}_m \\
        \end{array}
      \right)\tilde{g}=\left(
                 \begin{array}{cc}
                   (\widetilde{A}^{-1})_m{}^n & \widetilde{b}_{mn} \\
                   \widetilde{\beta}^{mn} & \widetilde{A}^m{}_n \\
                 \end{array}
               \right)\left(
        \begin{array}{c}
          T_n \\
          \widetilde{T}_n \\
        \end{array}
      \right)
\end{eqnarray}
Note that $b,\beta,{\ti b},{\ti \beta}$ are antisymmetric.  The matrices $A$, $b$ and $\beta$ depend on $x^i$ only and encode the adjoint action of $G$ and the matrices $\widetilde{A}$, $\widetilde{b}$ and $\widetilde{\beta}$ depend on ${\ti x}_i$ only and encode the adjoint action of $\widetilde{G}$. $A_m{}^n(g)$ is the adjoint action of $G$ on $\mathfrak{g}$ so that $g^{-1}rg=\ell$ or, in components, $r^nA_n{}^m=\ell^m$. Similarly, $\widetilde{A}^m{}_n(\ti g)$ is the adjoint action of $\widetilde{G}$ on $\mathfrak{\ti g}$ so that ${\ti r}_n\widetilde{A}^n{}_m={\ti \ell}_m$. The adjoint action   preserves the metric  $L_{MN}$ so the
 $2D\times 2D$ matrices whose block form is given in (\ref{adjoint 1}) and (\ref{adjoint 2})
 are      in $O(D,D)$.  The form of the $2D\times 2D$ adjoint matrices is determined by the polarisation chosen and   different choices of polarisation will give different matrices $A$, $b$, $\beta$, $\widetilde{A}$, $\widetilde{b}$ and $\widetilde{\beta}$.

The non-trivial twisting together of $G$ and $\widetilde{G}$ means that the right- and left-invariant one-forms on $\cG$, denoted by $\widetilde{\cal P}^M$ and ${\cal P}^M$ respectively, are not simply
$$
{\cal P}^{\hat{M}}=(\ell^m,\tilde{\ell}_m) \qquad  \widetilde{{\cal P}}^{\hat{M}}=(r^m,\tilde{r}_m)
$$
but are twisted together in a more complicated way.  {In the case of a Drinfel'd double, where the polarisation is such that both $T_m$ and $\widetilde{T}^m$ generate subgroups of $\cG$, the adjoint actions simplify. In particular, the adjoint action of $G$ preserves $\mathfrak{g}$ so that $g^{-1}T_mg=A_m{}^nT_n$ and $b_{mn}=0$. Similarly, the adjoint action of $\widetilde{G}$ preserves $\tilde{\mathfrak{g}}$ so that $\tilde{g}^{-1}\widetilde{T}^m\tilde{g}=\widetilde{A}^m{}_n\widetilde{T}_n$ and $\widetilde{\beta}^{mn}=0$. We shall see that more general groups $\cG$, which are not Drinfel'd doubles, do not simplify in this way.}

 Let us consider group elements $h\in \cG$ that can be written in the form
   $h=g{\ti g}$, where $g\in G$ and ${\ti g}\in\widetilde{G}$. The left-invariant one-form can be written as
$$
{\cal P}=h^{-1}dh=\ell^m({\ti g}^{-1}T_m{\ti g})+{\ti r}_m({\ti g}^{-1}{\widetilde T}^m{\ti g})
$$
or, using the definitions of the adjoint action of $\widetilde{G}$ on the Lie algebra of $\cG$ given above, as
$$
{\cal P}=\left(
           \begin{array}{cc}
             \ell^m & {\ti r}_m \\
           \end{array}
         \right)\left(
                  \begin{array}{cc}
                    (\widetilde{A}^{-1})_m{}^n & \widetilde{b}_{mn} \\
                    0 & \widetilde{A}^m{}_n \\
                  \end{array}
                \right)\left(
                         \begin{array}{c}
                           T_n \\
                           \widetilde{T}^n \\
                         \end{array}
                       \right)
$$
 {It is useful to write this block decomposition of the one-forms   as
$$
{\cal P}=\widetilde{\Phi}^M{\cal W}_M{}^N({\ti x})T_N
$$
where $\widetilde{\Phi}^M=(\ell^m,{\ti r}_m)$.} The information on the twisting together of the two subgroups is contained in ${\cal W}$ which depends only on $\tilde x$.

We now return to the example given above in which $\widetilde{G}$ is abelian (so that ${\ti \ell}_m={\ti r}_m=d{\ti x}_m$) and suppose that $G$ is a  {semi-simple} group with structure constants $f_{mn}{}^p$ so that $\cG=G\ltimes\R^D$. We choose a matrix representation of the generators
\begin{equation}\label{TT}
T_m=\left(
      \begin{array}{cc}
        t_m & 0 \\
        0 & t_m \\
      \end{array}
    \right) \qquad  \widetilde{T}^m=\left(
                                      \begin{array}{cc}
                                        0 & h^{mn}t_n \\
                                        0 & 0 \\
                                      \end{array}
                                    \right)
\end{equation}
where $h_{mn}=\frac{1}{2}f_{mp}{}^qf_{nq}{}^p$ is the non-degenerate Cartan-Killing metric of $G$ which raises and lowers indices on the structure constants and $t_m$ is a $D\times D$ matrix representation of $\mathfrak{g}$ so that $[t_m,t_n]=f_{mn}{}^pt_p$. A general element $h=g{\ti g}$ of $\cG$ may then be written as
$$
h=\left(
    \begin{array}{cc}
      g & 0 \\
      0 & g \\
    \end{array}
  \right)\left(
           \begin{array}{cc}
             1 & \tilde{x} \\
             0 & 1 \\
           \end{array}
         \right)=\left(
                   \begin{array}{cc}
                     g & g\tilde{x} \\
                     0 & g \\
                   \end{array}
                 \right)
$$
and the one-forms (\ref{ghj}) may be read off from
$$
{\cal P}=\left(
           \begin{array}{cc}
             g^{-1}dg & d\tilde{x}+[P,\tilde{x}] \\
             0 & g^{-1}dg \\
           \end{array}
         \right)
$$
We then see explicitly that
$$
(\widetilde{A}^{-1})_m{}^n=\delta_m{}^n  \qquad  \widetilde{b}_{mn}=-f_{mn}{}^p{\ti x}_p
$$
so that the adjoint action of $\widetilde{G}$ on $\mathfrak{\ti g}$ is trivial, as one would expect for an abelian group $\widetilde{G}=\R^D$.

Alternatively, we could consider the parameterisation $h={\ti g}g$ so that the left-invariant one-forms are more naturally written as
$$
{\cal P}=\left(
           \begin{array}{cc}
             r^m & {\ti \ell}_m \\
           \end{array}
         \right)\left(
                  \begin{array}{cc}
                    A_m{}^n & 0 \\
                    \beta^{mn} & (A^{-1})^m{}_n \\
                  \end{array}
                \right)\left(
                         \begin{array}{c}
                           T_n \\
                           {\ti T}^n \\
                         \end{array}
                       \right)
$$
 or schematically
\begin{equation}\label{decomposition}
{\cal P}=\Phi^{\hat{N}}{\cal V}_{\hat{N}}{}^{\hat{M}}(x)T_M
\end{equation}
where $\Phi^{\hat{M}}=(r^m,{\ti \ell}_m)$. The information on the twisting together of the two subgroups is now contained in ${\cal V}_{\hat{N}}{}^{\hat{M}}(x)$ which depends only on the $x^i$. For the polarisation choosing the $x^i$
as physical coordinates, it is this parameterisation leading to a twist ${\cal V}$ depending only on the $x^i$ that is the most useful.

For the $\cG=G\ltimes\R^D$ example, this parametrisation may be written in terms of the basis of generators (\ref{TT}) as
$$
h=\left(
           \begin{array}{cc}
             1 & \tilde{x} \\
             0 & 1 \\
           \end{array}
         \right)\left(
    \begin{array}{cc}
      g & 0 \\
      0 & g \\
    \end{array}
  \right)=\left(
                   \begin{array}{cc}
                     g & \tilde{x}g \\
                     0 & g \\
                   \end{array}
                 \right)
$$
In this parameterisation, the left-invariant one-forms are
$$
{\cal P}=\left(
           \begin{array}{cc}
             g^{-1}dg & g^{-1}d\tilde{x}g \\
             0 & g^{-1}dg \\
           \end{array}
         \right)
$$
so that $P^m=(g^{-1}dg)^m$ and $Q_m=(A^{-1})_m{}^nd{\ti x}_n$. It is not hard to see why this   parameterisation $h={\ti g}g$ is most useful; we shall be interested in recovering a conventional description as the left-acting quotient of the doubled group by $\widetilde{G}$ and the left action of $\widetilde{G}$ on elements of $\cG$ is manifest precisely in the parameterisation $h={\ti g}g$. Similarly, if we were interested in the quotient $\cG/\widetilde{G}_R$, the appropriate parameterisation to consider would be the one with $h=g{\ti g}$  described above.

 \subsubsection{Recovering the physical background fields}

For a Drinfel'd double, using  the parameterisation $h={\ti g}g$ giving
$$
{\cal P}=\Phi^M{\cal V}_M{}^N(x)T_M
$$
 {we can define a $\widetilde{G}_L$-invariant metric which depends on the coordinates $x^i$ only} by
$$
{\cal H}_{MN}(x)={\cal M}_{PQ}{\cal V}^M{}_P{\cal V}^Q{}_N
$$
With a polarisation tensor $\Theta_{\hat{M}}{}^M$,
we can define
$${\cal H}_{\hat{M}\hat{N}}(x)=\Theta_{\hat{M}}{}^M{\cal H}_{MN}(x)\Theta^N{}_{\hat{N}}$$
 whose components define a metric $g_{mn}$ and $B$-field $B_{mn}$ by
\begin{equation}\label{H}
{\cal H}_{\hat{M}\hat{N}}(x)=
\left(\begin{array}{cc}
g_{mn}+B_{mp}g^{pq}B_{qn} & B_{mp}g^{pn} \\
g^{mp}B_{np} & g^{mn}
                          \end{array}\right)
\end{equation}
The metric $g_{mn}(x)$ and $B$-field $B_{mn}(x)$ depend only on the $x^i$ coordinates, are manifestly $\widetilde{G}_L$-invariant and therefore
\begin{equation}\label{physical fields}
g_{ij}=g_{mn}r^m{}_ir^n{}_j  \qquad  B_{ij}=B_{mn} r^m{}_i r^n{}_j
\end{equation}
give a metric and $B$-field which are well-defined on the coset $\cG/\widetilde{G}_L$.  {In general, the} two-form $B$, coming from the $\widetilde{G}_L$-invariant doubled metric (\ref{H}), is not  {necessarily} the only contribution to the physical $H$-field strength. There  {may also be} a contribution  {coming from} the natural $\cG_L\times\cG_R$-invariant three-form on the doubled group
$$
{\cal K}=\frac{1}{6}t_{MNP}{\cal P}^M\wedge{\cal P}^N\wedge{\cal P}^P
$$
It will be shown in section 6 that, when the doubled group is a Drinfel'd double, the physical $H$-field strength on the coset $\cG/\widetilde{G}_L$ is   given by
\begin{equation}\label{H conjecture}
H=dB-\frac{1}{2}d\left(r^m\wedge {\ti \ell}_m\right)+\frac{1}{2}{\cal K}
\end{equation}
This expression may seem surprising but, as we shall see in section 6, its form arises quite naturally from the   world-sheet description of the doubled geometry. Moreover, when the doubled group is a Drinfel'd double, one may use the Maurer-Cartan equations dual to the algebra (\ref{DD}), to show that ${\cal K}=d(r^m\wedge {\ti \ell}_m)$, and so this expression for the $H$-field strength simplifies to
$$
H=dB
$$
and  {so, in the case where the doubled group is given by a Drinfel'd double,} the physical metric and $B$-field may be read off  {directly from} the $\widetilde{G}_L$-invariant metric ${\cal H}_{MN}(x)$.

As an example, we return to the case in which $\widetilde{G}$ is abelian,  {so that}
$\cG = T^*G$.
Then   ${\ti \ell}_m=d{\ti x}_m$ and  let the structure constants of $G$ be $f_{mn}{}^p$ as before.
If we choose the parameterisation $h={\ti g}g\in\cG$, then one can show that $\beta^{mn}=0$ and $A_m{}^n$ is the adjoint action of the group $G$; i.e. $\ell^m=r^nA_n{}^m$. The left-invariant one-form may then be written as
$$
{\cal P}^{\hat{M}}=\left(
           \begin{array}{cc}
             r^m & {\ti \ell}_m \\
           \end{array}
         \right)\left(
                  \begin{array}{cc}
                    A_m{}^n & 0 \\
                    0 & (A^{-1})^m{}_n \\
                  \end{array}
                \right)
$$
so that the metric is given by
$$
ds^2=\delta_{mn}A^m{}_pA^n{}_q r^p\otimes r^q=\delta_{mn}\ell^m\otimes\ell^n
$$
 {and $B=0$. In this case, it is not hard to show that ${\cal K}=d(r^m\wedge {\ti \ell}_m)$ and so the $H$-field vanishes.}

 {Conversely, we can consider the polarisation where we take the physical coordinates $x^m$ to parameterise the abelian group $\R^D$, and $\widetilde{G}$ to have structure constants $Q^{mn}{}_p$ so that $[\widetilde{T}^m,\widetilde{T}^n]=Q^{mn}{}_p\widetilde{T}^p$ and $\cG=T^*\widetilde{G}$. Again, the natural parameterisation to choose is $h={\ti g}g$ so that
$$
{\cal P}=\left(
           \begin{array}{cc}
             r^m & {\ti \ell}_m \\
           \end{array}
         \right)\left(
                  \begin{array}{cc}
                    A_m{}^n & 0 \\
                    \beta^{mn} & (A^{-1})^m{}_n \\
                  \end{array}
                \right)\left(
                         \begin{array}{c}
                           T_n \\
                           \widetilde{T}^n \\
                         \end{array}
                       \right)
$$
where $\beta^{mn}=Q^{mn}{}_px^p$ and $A^m{}_n=\delta^m{}_n$, since $G=\R^D$ and so $\ell^m=r^m=dx^m$. The $\widetilde{G}_L$-invariant metric ${\cal H}_{MN}(x)$ then gives a   metric and $B$-field specified by $(g+B)^{-1}=\bid+\beta$. In the special case where the doubled geometry is six-dimensional and the only non-vanishing structure constant of $\widetilde{G}$ is $Q^{yz}{}_x=m\in\Z$, then $\beta^{yz}=mx$ and the background is a cover of the familiar T-fold. Here too, one may show that ${\cal K}=d(r^m\wedge {\ti \ell}_m)$ and so ${\cal K}$ does not contribute to the physical $H$-field.}

 \subsubsection{General Case}

Consider now the case of a general doubled group $\cG$ which is not a Drinfel'd double, so that there may be H-flux and/or R-flux.
In general  the vielbein ${\cal P}^M{}_I$ will depend on all coordinates, both $x$ and $\tilde x$.
Nonetheless, the $\widetilde{G}_L$-invariant generalised metric
$$
{\cal H}_{\hat{M}\hat{N}}(x)={\cal M}_{\hat{P}\hat{Q}}{\cal V}^{\hat{P}}{}_{\hat{M}}{\cal V}^{\hat{Q}}{}_{\hat{N}}
$$
can still be used to define a metric and $H$-field using (\ref{H}) and (\ref{H conjecture}), but now these fields will depend on both $x$ and $\tilde x$ in general, so their interpretation is unclear.  {In those cases in which ${\cal V}^M{}_N$ and $r^m$  can be chosen to depend only on $x$, the background will be geometric locally. There will be local fields $g_{ij}(x)$, $B_{ij}(x)$, although there may be non-trivial patching as in T-folds.
In other cases where dependence on $\ti x$ cannot be avoided, then the resulting configuration is not even locally geometric, and the $x$-polarisation will involve background fields depending on the dual coordinates $\ti x$.}

For a given polarisation, a natural way of introducing coordinates (in a neighbourhood of the identity) is through the exponential
parameterisation
$$
h= \exp (\tilde x_m \widetilde{T}^m) \exp (x^m T_m)$$
In the case in which the $X^m$ generate a subgroup $\widetilde G$,  then
$\ti g=\exp (\tilde x_m \widetilde{T}^m) \in \widetilde G$. Let $k=\exp (x^m T_m)$ so that $ h = \ti g k$.
 {Then defining $r=dkk^{-1}$, ${\ti \ell}={\ti g}^{-1}d{\ti g}$, we can expand the forms   as
$$
r=r^mT_m+r_m\widetilde{T}^m \qquad  {\ti \ell}={\ti \ell}_m\widetilde{T}^m
$$
where we note that $r$ is in general a linear combination of all generators, since the $k$ are not elements of a subgroup, but are elements of the full doubled group $\cG$. Since the $T_m$ do not generate a subgroup, the adjoint action of $\{k\}$ on $\mathfrak{g}$ does not preserve $\mathfrak{g}$ and we have
$$
k^{-1}T_mk=A_m{}^nT_n+b_{mn}\widetilde{T}^n
$$
so that
$$
{\cal P}^{\hat{M}}=\left(
           \begin{array}{cc}
             r^m & {\ti q}_m \\
           \end{array}
         \right)\left(
                  \begin{array}{cc}
                    A_m{}^n & b_{mn} \\
                    \beta^{mn} & (A^{-1})^m{}_n \\
                  \end{array}
                \right)
$$
where ${\ti q}_m={\ti \ell}_m+r_m$ and we again have an expression of the form
$$
{\cal P}=\Phi^{\hat{M}}{\cal V}_{\hat{M}}{}^{\hat{N}}(x)T_{\hat{N}}
$$
As ${\cal V}$ depends only on $x$,
again a metric and $B$-field depending only on $x$ are obtained using (\ref{H}), but now the $H$-field strength also gets a contribution from the three form ${\cal K}$ and the expression (\ref{H conjecture}) generalises to
\begin{equation}\label{H with K}
H=dB-\frac{1}{2}d\left(r^m\wedge{\ti q}_m\right)+\frac{1}{2}{\cal K}
\end{equation}

As an example,   consider the group generated by the algebra
$$
[Z_m,Z_n]=K_{mnp}X^p    \qquad  [Z_m,X^n]=0 \qquad  [X^m,X^n]=0
$$
The $\cG_L$-invariant one-forms are
$$
P^m=dx^m    \qquad  Q_m=d{\ti x}_m-\frac{1}{2}K_{mnp}x^pdx^n
$$
It is not hard to show that
$$
r^m=dx^m    \qquad  r_m=\frac{1}{2}K_{mnp}x^p r^n    \qquad  {\ti q}_m=Q_m+K_{mnp}x^p r^n
$$
Also, $A_m{}^n=\delta_m{}^n$, $\beta^{mn}=0$ and $b_{mn}=K_{mnp}x^p$. This gives ${\cal P}=\Phi^M{\cal V}_M{}^N(x)T_N$, where
$$
{\cal V}_{\hat{M}}{}^{\hat{N}}(x)=\left(
              \begin{array}{cc}
                \delta_m{}^n & K_{mnp}x^p \\
                0 & \delta^m{}_n \\
              \end{array}
            \right)
$$
and $\Phi^M=(dx^m,{\ti q}_m)$. The metric is given by
$$
ds^2=\delta_{mn}dx^m\otimes dx^n
$$
so that the spacetime is locally $\R^D$. The global structure of the spacetime is determined by $\G$ and in the discussions which follow we shall usually choose $\G$ so that the spacetime is compact which, in this case, gives a $D$-dimensional torus. The physical $H$-field strength on $T^D$ is given by (\ref{H with K}) where
\begin{equation}\label{1}
{\cal K}=-\frac{1}{6}K_{mnp}dx^m\wedge dx^n\wedge dx^p
\end{equation}
We also have the contributions
\begin{equation}\label{2}
db=\frac{1}{2}K_{mnp}dx^m\wedge dx^n\wedge dx^p \qquad  \frac{1}{2}d\left(r^m\wedge{\ti q}_m\right)=\frac{1}{4}K_{mnp}dx^m\wedge dx^n\wedge dx^p
\end{equation}
Substituting (\ref{1}) and (\ref{2}) into (\ref{H with K}), we find that the physical $H$-field strength is
$$
H=\frac{1}{6}K_{mnp}dx^m\wedge dx^n\wedge dx^p
$$

More generally, there may be non-trivial $R$-flux.
Then  the $\widetilde{T}^m$ do not close to give a subalgebra and the expansion of the forms $\tilde{\ell}$ is in general a linear combination of all generators, so that ${\ti \ell}={\ti \ell}_m\widetilde{T}^m+{\ti \ell}^mT_m$. The left-invariant one-forms on $\cG$ can be expanded   to give
$$
{\cal P}^{\hat{M}}=\left(
           \begin{array}{cc}
             p^m & {\ti q}_m \\
           \end{array}
         \right)\left(
                  \begin{array}{cc}
                    A_m{}^n & b_{mn} \\
                    \beta^{mn} & (A^{-1})^m{}_n \\
                  \end{array}
                \right)
$$
where $p^m=r^m+{\ti \ell}^m$ and ${\ti q}_m={\ti \ell}_m+r_m$ and we again have an expression of the form
$$
{\cal P}=\Phi^M{\cal V}_M{}^N(x)T_M
$$
As before, we extract $x$-dependent fields $g_{mn}$ and $B_{mn}$ from ${\cal H}_{MN}(x)$. The main difference in the $R$-flux case, where ${\ti \ell}^m\neq 0$, is that the physical metric
$$
ds^2=g_{mn}(x)p^m\otimes p^n
$$
now depends explicitly on ${\ti x}_i$ through the one-forms $p^m=r^m+{\ti \ell}^m$ and so it is not possible to eliminate the ${\ti x}_i$-dependence completely if the $\widetilde{T}^m$ do not generate a subgroup of $\cG$. Similarly, it is not possible to remove all ${\ti x}_i$-dependence from the $H$-field which is given by the expression
$$
H=dB-\frac{1}{2}d\left(p^m\wedge {\ti q}_m\right)+\frac{1}{2}{\cal K}
$$}
The issues discussed here will be illustrated by further examples in the next section.

\section{Examples}

We shall now apply the formalism developed in the previous section to a specific example. Starting with the three-dimensional
nilfold ${\cal N}$, we explicitly construct the associated five-dimensional doubled torus bundle ${\cal T}$ and the six-dimensional twisted
torus $\cX$. The recovery of a conventional description of the nilfold and its T-duals from these doubled geometries will be explicitly demonstrated in each
case.

The nilfold is a compact three-dimensional manifold. It may be constructed as a $T^2$ bundle over a circle $S^1_x$, where the fibration has
monodromy in the mapping class group, $SL(2;\Z)$, of the torus fibres. Let $z^a=(y,z)$ be the coordinates on the $T^2$ fibre and $x$ the base
circle coordinate with $x\sim x+1$.
The twist of the bundle is given by the $\mathfrak{sl}(2)$ Lie algebra element $f^a{}_b$ which gives a monodromy
$e^{f }\in SL(2;\Z)$, where
\begin{equation}\label{f}
f^a{}_b=\left(%
\begin{array}{cc}
  0 & 0 \\
  -m & 0 \\
\end{array}%
\right)    \qquad  e^{f }=\left(%
\begin{array}{cc}
  1 & 0 \\
  -m & 1 \\
\end{array}%
\right) \qquad m\in\Z
\end{equation}
A globally defined basis of one-forms on the nilfold is
\begin{equation}\label{nilforms}
P^x=dx  \qquad  P^y=dy-mxdz \qquad  P^z=dz
\end{equation}
 The global structure of the nilfold requires the following identification of the local
coordinates
\begin{equation}\label{delta}
(x,y,z)\sim (x+1,y+mz,z)    \qquad  (x,y,z)\sim (x,y+1,z)   \qquad  (x,y,z)\sim (x,y,z+1)
\end{equation}
which leaves the one-forms $P^m=(P^x,P^y,P^z)$ invariant.
A metric  $g =\sum _m P^mP^m$ may be constructed from these one-forms, giving
\begin{equation}\label{nilmetric}
g_{ij}=\left(%
\begin{array}{ccc}
  1 & 0 & 0 \\
  0 & 1 & -mx \\
  0 & -mx & 1+m^2x^2 \\
\end{array}%
\right)
\end{equation}
and it is this metric that is used in the dimensional reduction ansatz discussed in the previous section.

Alternatively, the nilfold may be constructed as a twisted torus  ${\cal
N}=G/\G$ where $G$ is the noncompact Heisenberg group manifold and $\G$ is a discrete subgroup chosen so that $G/\G$ is compact, i.e. $\G$ is cocompact. The generators of the Heisenberg group $G$ satisfy commutation relations
$$
[t_x,t_z]=mt_y  \qquad  [t_y,t_z]=0 \qquad [t_x,t_y]=0
$$
and a useful matrix representation  is
\begin{equation}\label{t's}
t_x=\left(%
\begin{array}{ccc}
  0 & m & 0 \\
  0 & 0 & 0 \\
  0 & 0 & 0 \\
\end{array}%
\right) \qquad  t_y=\left(%
\begin{array}{ccc}
  0 & 0 & 1 \\
  0 & 0 & 0 \\
  0 & 0 & 0 \\
\end{array}%
\right) \qquad  t_z=\left(%
\begin{array}{ccc}
  0 & 0 & 0 \\
  0 & 0 & 1 \\
  0 & 0 & 0 \\
\end{array}%
\right)
\end{equation}
Using the local coordinates $(x,y,z)$ on the group manifold $G$, a general element of the group may be written as
\begin{eqnarray}\label{g}
g=\left(%
\begin{array}{ccc}
  1 & mx & y \\
  0 & 1 & z \\
  0 & 0 & 1 \\
\end{array}%
\right)
\end{eqnarray}
The    one-forms
(\ref{nilforms})  {are given by $P=g^{-1}dg$} and are invariant under the left-action of the group $G$.
The discrete group $\G$ has
 general element $h$  given by
\begin{eqnarray}\label{h}
h=\left(%
\begin{array}{ccc}
  1 & m\alpha & \beta \\
  0 & 1 & \gamma \\
  0 & 0 & 1 \\
\end{array}%
\right)
\end{eqnarray}
where $\alpha$, $\beta$ and $\gamma$ are arbitrary integers. The nilfold is given by the identification of G under the left action of $\G$. Then the identification $g\sim h\cdot g$, with $(\alpha,\beta,\gamma)$ given by
$(1,0,0)$, $(0,1,0)$ or $(0,0,1)$, reproduces the identifications of the coordinates (\ref{delta}).
As the identification is through the left action, the left-invariant one-forms
(\ref{nilforms}) are well-defined on  $G/\G$.

The noncompact group manifold $G$ admits a natural action of the group from the left $G_L$ and from the right $G_R$. The right action, $G_R$, is generated by the left-invariant vector fields
$$
Z_x=\frac{\partial}{\partial x} \qquad  Z_y=\frac{\partial}{\partial y} \qquad  Z_z=\frac{\partial}{\partial z}+mx\frac{\partial}{\partial y}
$$
and   $Z_y$ and $Z_z$ are Killing vectors of the metric (\ref{nilmetric}), whilst $Z_x$ is not. Note that the Cartan-Killing metric for the Heisenberg group, which would automatically be invariant under $G_L\times G_R$, is identically zero, and we are using a non-degenerate  metric  (\ref{nilmetric}) which is only invariant under  a subgroup of $G_L\times G_R$.
Furthermore, the vector fields
$Z_m=(Z_x,Z_y,Z_z)$ are invariant under the action of $\G \subset G_L$ and so are well-defined on the quotient ${\cal N}=G/\G$.
The one-forms (\ref{nilforms}), dual to these vectors, are also well defined on ${\cal N}$.

The right-invariant vector fields $\widetilde{Z}_m=(\widetilde{Z}_x,\widetilde{Z}_y,\widetilde{Z}_z)$ on the group manifold $G$ generate the left action $G_L$
and are given by
\begin{eqnarray}\label{nilsymL}
 \widetilde{Z}_x=\frac{\partial}{\partial x}+mz\frac{\partial}{\partial y}  \qquad  \widetilde{Z}_y=\frac{\partial}{\partial y} \qquad \widetilde{Z}_z=\frac{\partial}{\partial z}
\end{eqnarray}
Note that  $\widetilde{Z}_y$ and $\widetilde{Z}_z$ are Killing vectors of the metric (\ref{nilmetric}), whilst $\widetilde{Z}_x$ is not.
These vector fields are not invariant under the action of $\G{}$ and transform as
\begin{eqnarray}
\begin{array}{lll}
 \widetilde{Z}_x\rightarrow \widetilde{Z}_x+m \gamma\widetilde{Z}_y  &\qquad  \widetilde{Z}_y\rightarrow \widetilde{Z}_y &\qquad \widetilde{Z}_z\rightarrow \widetilde{Z}_z-m\alpha \widetilde{Z}_y \\
\end{array}
\end{eqnarray}
Then
although the three right-invariant vector
fields $\widetilde{Z}_m$ are globally defined on $G$, only $\widetilde{Z}_y $  is well-defined on the quotient ${\cal N}=G/\G{
}$. Of particular
importance is the fact that the generator $\widetilde{Z}_z$ is not preserved by $\G_{}$. The fact that $\widetilde{Z}_z$
is locally defined (on each $T^2$ fibre) but not globally defined on ${\cal N}$ leads to a T-dual description of the background, given by
dualising
along the $z$ direction (i.e. with respect to the generator $\widetilde{Z}_z$), which is a T-fold.

Of particular interest is the use of twisted tori, such as the nilfold, as internal manifolds in  conventional compactifications of string
theory and supergravity. Compactification of a supergravity theory with metric, $B$-field and dilaton,  with action  of the form (\ref{D+d+1 lagrangian}) (plus terms involving   other fields), on the nilfold gives a massive supergravity of the
form (\ref{O(d+1,d+1) Lagrangian}) with a  non-abelian gauge algebra  given by
\begin{equation}\label{galgebra}
[Z_x,Z_z]=mZ_y  \qquad  [Z_x,X^y]=mX^z  \qquad  [Z_z,X^y]=-mX^x
\end{equation}
where all other commutators vanish. The symmetries  generated by $Z_m=(Z_x,Z_y,Z_z)$  arise from the action of the  left-invariant  vector fields $Z_m$ on the nilfold given above, while
the symmetries generated by $X^m$ arise from $B$-field antisymmetric tensor transformations. The gauge algebra is that of the six-dimensional group
$G\ltimes \R^3$ where $G$ is the Heisenberg group. This compactification can be equivalently constructed as a duality twist reduction of the
supergravity, as described in the previous section, where the twist matrix (\ref{mass}) is given by
\begin{equation}\label{nilodromy}
N^A{}_B=\left(%
\begin{array}{cc}
  f^a{}_b & 0 \\
  0 & -f_a{}^b \\
\end{array}%
\right)
\end{equation}
with $f^a{}_b$ given in (\ref{f}).

\subsection{ T-duality}

\noindent With the choice of metric (\ref{nilmetric}), the Heisenberg group manifold has
the geometry
\begin{equation}\label{nil}
ds^2_{\cal N}=dx^2+(dy-mxdz)^2+dz^2   \qquad  B=0
\end{equation}
and has   Killing vectors $Z_y=\widetilde Z_y, Z_z, \widetilde Z_z $. On taking the quotient  by $\G$ to obtain the  nilfold background,  $Z_y=\widetilde{Z}_y$ and $Z_z $
are left-invariant and so are Killing vector fields of the nilfold, while $\widetilde Z_z $
  remains as  a local solution to Killing's equation, but does not
extend to a globally-defined vector field on the nilfold.

In  Buscher's formulation of T-duality \cite{Buscher ``A Symmetry of the String Background Field Equations''}, the starting point is a sigma-model whose target is a torus bundle with  a compact
 abelian isometry group, preserving the $H$-field and dilaton as well as the metric.
The isometry is then gauged, and the gauge connection constrained to be trivial. Eliminating the gauge field recovers the original theory, while integrating out the torus fibres gives the T-dual target.
Buscher T-duality then requires a compact abelian isometry  {which leaves the background invariant}.

 {The sigma model with the nilfold as  target space is constructed from the pull-back of the left-invariant one-forms (\ref{nilforms}) to the   world-sheet and as such there is a manifest rigid $G_L$ symmetry in the   world-sheet theory. The application of Buscher's construction then requires that there is an abelian subgroup of this rigid $G_L$ symmetry which generates an invariance of the full background.} There is such an  invariance of the nilfold background given by the $U(1)$ isometry $ y\rightarrow y+\epsilon$ generated by
$\widetilde{Z}_y=\partial_y$. The vector field $\widetilde{Z}_y$ is preserved by $\G{}$ and therefore is well-defined on the nilfold.
Applying the Buscher construction it was shown in \cite{Kachru:2002sk,Hull ``Global Aspects of T-Duality Gauged Sigma Models and T-Folds''} that the T-dual   of the nilfold background (\ref{nil}) is given by a
three-dimensional torus with non-trivial $B$-field
\begin{equation}\label{gH}
ds^2_{T^3}=dx^2+dy^2+dz^2   \qquad  B=mxdy\wedge dz
\end{equation}
The $B$-field gives a constant $H$-flux, with  $H=mdx\wedge dy\wedge dz$. The global structure of the torus is read off from the identifications of the coordinates
$$
(x,y,z)\sim (x+1,y,z)  \qquad  (x,y,z)\sim (x,y+1,z) \qquad  (x,y,z)\sim (x,y,z+1)
$$

A second invariance of the Heisenberg group manifold $G$
 is   the abelian isometry $z\rightarrow z+\epsilon$. The generator
$\widetilde{Z}_z=\partial_z$ is globally defined on the Heisenberg group manifold $G$ but is not globally defined on the nilfold ${\cal N}$.
Under the shift of the coordinate $x\rightarrow x+1$, the vector field is not invariant but transforms as
$$
\widetilde{Z}_z\rightarrow \widetilde{Z}_z-m\widetilde{Z}_y
$$
and so $\widetilde{Z}_z$ is not periodic on ${\cal N}$. Strictly speaking, the Buscher rules cannot be applied to this case, as the Killing
vector is not globally well-defined on the nilfold and is, at best, multi-valued. This problem can be avoided   by going to a covering space in
which the periodicity of $x$ is dropped. This covering space is $C_{\cal N}=G/\G '$ where $\G'$ is the  subgroup of
$\G$ given by elements of the form (\ref{h}) with $\a=0$. This gives the periodic identifications $y\sim y+1$ and $z\sim z+1$ while leaving $x$  non-compact, so that $C_{\cal N}$ has topology $\R\times T^2$.

 On the covering space $C_{\cal N}$, $\widetilde{Z}_z$ is globally defined and we can consider T-duality
along the $z$ direction using the Buscher rules. Performing the T-duality gives a smooth manifold $C_T$ which  again has topology $T^2\times \R$ with
metric and $B$-field given by
\begin{equation}\label{gb}
ds^2_{T-Fold}=dx^2+\frac{1}{1+(mx)^2}(dy^2+dz^2)   \qquad  B=\frac{mx}{1+(mx)^2}dy\wedge dz
\end{equation}
This background is a conventional geometry, with a non-trivial $B$-field. However, we are interested in the background T-dual to the nilfold,
with periodic $x$, suggesting that we now try to make $x$ periodic.
The metric and $B$ field (\ref{gb}) are clearly not periodic in $x$, so this could not  lead to a smooth geometry.

To better understand this background, consider first the T-dual of $C_{\cal N}$ given by the covering space $C_{T^3}$ of the $T^3$ with $H$-flux $m$, which is $\R\times T^2$ with $x$ the non-compact coordinate with metric and $B$-field (\ref{gH}).
Consider a particular $T^2$ fibre at some fixed $x$, with metric $\tilde{g}$ and $B$-field $\tilde{B}$, so that $\tilde{E}=\tilde{g}+\tilde{B}$
is a $2\times 2$ matrix  given  by
$$
  \tilde{E}=\left(%
\begin{array}{cc}
  1 & mx \\
  -mx & 1 \\
\end{array}%
\right)=1+x \Omega
$$
where
$$\Omega
=\left(%
\begin{array}{cc}
  0 & m \\
  -m & 0 \\
\end{array}%
\right)$$
T-dualising along the $y$ and $z$ directions of the $T^2$ leads to a dual torus background with metric $g$ and  $B$-field ${B}$ with $E=g+B$ given by ${E} = \tilde E^{-1}$, so that
$$
E=\frac{1}{1+(mx)^2}\left(%
\begin{array}{cc}
  1 & mx \\
  -mx & 1 \\
\end{array}%
\right)
$$
This is the same result  as is  obtained by T-dualising $C_{\cal N}$ in the $z$ direction.

Under the shift $x\rightarrow x+1$, the $B$-field of the dual background is shifted
$\tilde{B}_{yz}\rightarrow\tilde{B}_{yz}+m$ and so we see that periodically identifying the $x$ coordinate of $C_{T^3}$ gives
a space with an $x$-monodromy that is a shift of the $B$-field, $\tilde{B}_{yz}\rightarrow\tilde{B}_{yz}+m$. This of course gives a $T^3$ with $H$-flux $m$. Now for the dual space (\ref{gb}),
under  the shift $x\rightarrow x+1$,
$$E= \tilde E
 ^{-1}  \rightarrow  ( \tilde E+ \Omega
)^{-1} $$
which is a T-duality transformation of $E$, in $O(2,2;\Z)$. Then the monodromy is a non-geometric  T-duality transformation, resulting in a T-fold.
This amounts to what is sometimes described as applying the
duality fibrewise.

Locally, the T-fold is a conventional geometry, but the global structure cannot be understood as a manifold since the monodromy is not in  the $SL(2;\Z)$ mapping class group of the $T^2$ fibres. The non-geometric monodromy of the $T^2$ fibres of the T-fold
background can be recast as a geometric monodromy of the $T^4$ fibres of a doubled torus bundle ${\cal T}$, in which auxiliary coordinates are
introduced as described in the previous section. We now turn to this doubled formulation of our example.

\subsection{The Doubled Torus Fibration ${\cal T}$}

\noindent As discussed in section 2, a string background which is  a torus bundle or T-fold also admits a description as a doubled
torus bundle ${\cal T}$. For the current example with
$T^2$ fibres,
this doubled torus bundle with $T^4$ fibres  is constructed by introducing auxiliary coordinates
$\tilde{z}_a=(\tilde{y},\tilde{z})$ for the torus T-dual to the physical torus, so that
$$
z^a=(y,z)\rightarrow \mathbb{X}^A=(y,z,\tilde{y},\tilde{z})
$$
Then $z^a=(y,z)$ are coordinates for the physical fibre $T^2\subset T^4$ and
$\tilde{z}_a=(\tilde{y},\tilde{z})$ are coordinates on the T-dual torus $\widetilde{T}^2\subset T^4$.

In this description, all monodromies have a geometric action on the doubled fibres as a large diffeomorphism since $O(2,2;\Z)\subset GL(4;\Z)$.
The monodromy of the doubled torus fibres is
$$
x\rightarrow x+1    \qquad  \mathbb{X}^A\rightarrow \left(e^{-N}\right)^A{}_B\mathbb{X}^B
$$
which in the case of the nilfold is given by (\ref{massa}) where $K_{xab}$ and $Q_x{}^{ab}$ are both zero and $f_{xz}{}^y=m\in\Z$, so that on taking
$x\rightarrow x+1$,
\begin{eqnarray}
\left(%
\begin{array}{c}
  y \\
 z \\
  \tilde{y}\\
  \tilde{z} \\
\end{array}%
\right)\rightarrow\left(%
\begin{array}{cccc}
  1 & m & 0 & 0 \\
  0 & 1 & 0 & 0 \\
  0 & 0 & 1 & 0 \\
  0 & 0 & -m & 1 \\
\end{array}%
\right)\left(%
\begin{array}{c}
  y \\
 z \\
  \tilde{y}\\
  \tilde{z} \\
\end{array}%
\right)
\end{eqnarray}

The metric $g$ and $B$-field of the $T^2$ fibres specify a generalised  metric ${\cal H}$ on the doubled  $T^4$ fibres of ${\cal T}$,  a $4\times 4 $ matrix  with components in the $y,z,\tilde{y},\tilde{z}$ basis given by
\begin{equation}\label{genmet}
{\cal H} =\left(%
\begin{array}{cc}
  g-Bg^{-1}B & Bg^{-1} \\
  -g^{-1}B & g^{-1} \\
\end{array}%
\right)
\end{equation}
For the nilfold with metric (\ref{nilmetric}) and $B=0$, the generalised metric on the $T^4$ fibres of ${\cal T}$ is
$$
{\cal H}_{\cal N}=\left(%
\begin{array}{cccc}
 1 & -mx & 0 & 0 \\
 -mx & 1+m^2x^2 & 0 & 0 \\
 0 & 0 & 1+m^2x^2 & mx \\
 0 & 0 & mx & 1 \\
\end{array}%
\right)
$$

This $x$-dependent metric ${\cal H}$ is related to the $x$-independent metric ${\cal M}$ appearing  in (\ref{O(d,d) Lagrangian}) by ${\cal H}_{AB}(x)=(e^{Nx})_A{}^C{\cal M}_{CD}(e^{Nx})^D{}_B$ where, in this case, ${\cal M}_{AB}=\delta_{AB}$. As for  the nilfold, the doubled torus bundle ${\cal T}$ can be thought of either as a $T^4$ bundle over $S^1_x$ or  as a twisted torus
${\cal T}=\cG/\G_{}$, given by identifying a  certain group manifold $\cG$ under the action of a discrete subgroup $\G_{}$. The
five-dimensional group $\cG$ is that generated by (\ref{sdfadfg}).
Using the same coordinates $(x,y,z,\tilde{y},\tilde{z})$ as above,
the general element $g\in\cG$ is
 \begin{eqnarray}\label{g2}
g(x,y,z,\tilde{y},\tilde{z})=\left(%
\begin{array}{ccccc}
  1 & mx & 0 & 0 & y \\
  0 & 1 & 0 & 0 & z \\
  0 & 0 & 1 & 0 & \tilde{y} \\
  0 & 0 & -mx & 1 & \tilde{z} \\
  0 & 0 & 0 & 0 & 1 \\
\end{array}%
\right)
 \end{eqnarray}
The global structure of ${\cal T}$ is given by taking the quotient by a discrete subgroup.  The relevant discrete subgroup
$\G_{}$ consists of elements of the form
\begin{eqnarray}\label{h2}
h=\left(%
\begin{array}{ccccc}
  1 & m\alpha & 0 & 0 & \beta \\
  0 & 1 & 0 & 0 & \gamma \\
  0 & 0 & 1 & 0 & \tilde{\beta} \\
  0 & 0 & -m\alpha & 1 & \tilde{\gamma} \\
  0 & 0 & 0 & 0 & 1 \\
\end{array}%
\right)
\end{eqnarray}
where $\alpha$, $\beta,\g$, $\tilde{\beta}$ and $\tilde{\gamma}$ are arbitrary integers. The left action of $h$ is $g\rightarrow h\cdot g$ and acts on
the coordinates through
\begin{eqnarray}
\begin{array}{lll}
 x\rightarrow x+\alpha  &\qquad  y\rightarrow y+m\alpha z+\beta   &\qquad  z\rightarrow z+\gamma \\
 &\qquad  \tilde{y}\rightarrow \tilde{y}+\tilde{\beta}  &\qquad\tilde{z}\rightarrow\tilde{z}-m\alpha {\ti y}+\tilde{\gamma}
\end{array}
\end{eqnarray}
We identify $\cG$ under the left action of $\G_{}$ so that the coordinates are subject to the identifications
\begin{eqnarray}
(x,y,z,\tilde{y},\tilde{z})&\sim& (x+1,y+mz,z,\tilde{y},\tilde{z}-m\tilde{y}) \nonumber\\
  (x,y,z,\tilde{y},\tilde{z})&\sim&
(x,y+1,z,\tilde{y},\tilde{z})\nonumber\\
(x,y,z,\tilde{y},\tilde{z})&\sim& (x,y,z+1,\tilde{y},\tilde{z}) \nonumber\\
 (x,y,z,\tilde{y},\tilde{z})&\sim&
(x,y,z,\tilde{y}+1,\tilde{z})\nonumber\\
(x,y,z,\tilde{y},\tilde{z})&\sim& (x,y,z,\tilde{y},\tilde{z}+1)
\end{eqnarray}

There is a natural action of $\cG_L\times\cG_R$ on the group manifold $\cG$, generated by associated right- and left-invariant vector fields. The right action
$\cG_R$ is generated by the left-invariant vector fields (i.e. invariant under  $\cG_L$)
\begin{eqnarray}\label{doublevecL}
 Z_x=\frac{\partial}{\partial x}  \qquad  Z_y&=&\frac{\partial}{\partial y} \qquad\qquad\qquad Z_z=\frac{\partial}{\partial z}+mx\frac{\partial}{\partial y}
\nonumber\\
 X^y&=&\frac{\partial}{\partial \tilde{y}}-mx\frac{\partial}{\partial \tilde{z}} \qquad X^z=\frac{\partial}{\partial \tilde{z}}
\end{eqnarray}
which satisfy the commutation relations
\begin{equation}
[Z_x,Z_z]=mZ_y  \qquad  [Z_x,X^y]=mX^z
\end{equation}
where all other commutators vanish. Note that this is not the gauge algebra of the  {field theory} (\ref{O(d,d+16) Lie algebra}) obtained by compactification on the nilfold, but is a subgroup of a contraction of
the  {gauge algebra} (\ref{sdfadfg}). The fact that the vector field is left-invariant means that it is invariant under the action of the
discrete group $\G$ and so well-defined on the quotient ${\cal T}=\cG/\G$. Indeed, the left-invariant vector fields are dual
to the left-invariant one-forms
\begin{eqnarray}\label{nilforms 2}
\begin{array}{lll}
 P^x=dx  &\qquad  P^y=dy-mxdz &\qquad P^z=dz \\
 &\qquad  Q_y=d\tilde{y} &\qquad Q_z=d\tilde{z}+mxd\tilde{y}
\end{array}
\end{eqnarray}
which are also well-defined on the quotient $\cG/\G$.

By contrast, the generators of the left action $\cG_L$
\begin{eqnarray}
 \widetilde{Z}_x=\frac{\partial}{\partial x}+mz\frac{\partial}{\partial y}-m\tilde{y}\frac{\partial}{\partial \tilde{z}}  \qquad  \widetilde{Z}_y&=&\frac{\partial}{\partial y} \qquad \widetilde{Z}_z=\frac{\partial}{\partial z}
\nonumber\\
\widetilde{X}^y&=&\frac{\partial}{\partial \tilde{y}} \qquad \widetilde{X}^z=\frac{\partial}{\partial \tilde{z}}
\end{eqnarray}
are   globally defined on the group $\cG$, but are not invariant under the action of $\G$, which acts as
\begin{eqnarray}
\begin{array}{lll}
 \widetilde{Z}_x\rightarrow \widetilde{Z}_x+m\gamma \widetilde{Z}_y-m \tilde{\beta}\widetilde{X}^z  &\qquad  \widetilde{Z}_y\rightarrow \widetilde{Z}_y &\qquad \widetilde{Z}_z\rightarrow \widetilde{Z}_z-m \alpha\widetilde{Z}_y \\
 &\qquad  \widetilde{X}^y\rightarrow \widetilde{X}^y+m \alpha\widetilde{X}^z &\qquad \widetilde{X}^z\rightarrow
\widetilde{X}^z
\end{array}
\end{eqnarray}
These vector fields are therefore not globally defined on the twisted torus   ${\cal T}\simeq{\cal G}/\G$. The discrete group
$\G_{}$ does however preserve the subgroup $\widetilde{G}_L\simeq\R^2\subset\cG_L$ generated by
$\widetilde{X}^a=(\widetilde{X}^y,\widetilde{X}^z)$
$$
\G_{ }:\widetilde{G}_L\rightarrow\widetilde{G}_L
$$
(It also preserves $\tilde Z_y$.)
The subgroup $\widetilde{G}_L$
consists of matrices $f$  of the form (\ref{g2}) with $x=y=z=0$,
 $$
f(\tilde{y},\tilde{z})=\left(%
\begin{array}{ccccc}
  1 & 0 & 0 & 0 & 0 \\
  0 & 1 & 0 & 0 & 0 \\
  0 & 0 & 1 & 0 & \tilde{y} \\
  0 & 0 & 0 & 1 & \tilde{z} \\
  0 & 0 & 0 & 0 & 1 \\
\end{array}%
\right)
 $$
Taking the quotient of the group $\cG$ by the left action of $\widetilde{G}_L$ gives the coset
$\cG /\widetilde{G}_L$ which is just the Heisenberg group $G$, represented by group elements
of the form (\ref{g2}) with $ \ti y=\ti z =0$.
The subgroup  $\widetilde{G}_L$ has the property that, for all $h\in \G$ and all $f\in \widetilde{G}_L $,
$$
fhf^{-1}\in \G
$$
i.e. there is an $h'\in \G$ so that
$$
hf=fh'$$
This implies that  $\widetilde{G}_L$ has a well-defined action on
the coset ${\cal T}=\cG/\G$
so that identifying ${\cal T}$ under the action of $\widetilde{G}_L$ is well-defined. This quotient gives the nilfold, ${\cal N}={\cal T}/\widetilde{G}_L$.
It can also be viewed as the quotient of $\cG$ by the left action of the subgroup of elements of the form (\ref{h2}) with
$\alpha$, $\beta,\g$   arbitrary integers   and $\tilde{\beta}$,$\tilde{\gamma}$ arbitrary real numbers.

\subsubsection{Polarisations}

We have seen that the data given by the nilfold background specifies a doubled torus bundle ${\cal T}$ and that the nilfold
geometry can be recovered as the quotient ${\cal N}={\cal T}/\widetilde{G}_L$.
The T-duals of the nilfold are  the $T^3$ with $H$-flux and the T-fold, and these can also be recovered from the same
doubled geometry ${\cal T}$ through different choices of polarisation. Then ${\cal T}$ is a universal geometry containing  the original space and   its T-duals, as discussed in Section 2.
Using the notation  $\mathbb{X}^A$ ($A=1,2,3,4$)
 for  the  four coordinates  of  ${\cal T}$,  each
polarisation selects two of the four coordinates $\mathbb{X}^A$ to be the \lq physical' coordinates $z^a=(y,z)$ and the other two to be the \lq
auxiliary' $\tilde{z}_a=(\tilde{y},\tilde{z})$.
In Section 2,  polarisations were defined in open, contractible patches of the base, which here is the circle $S^1_x$.  We will consider polarisations defined on  the interval $I$ with $0< x < 1$, so that
the polarisation defines a 3-dimensional subspace, topologically $I\times T^2$,
of the doubled space that has  topology $I\times T^4$.

A polarisation selects a maximally isotropic choice of $T^2\subset T^4$ as the physical space, defined by a constant projector $\Pi^a{}_A$ so that  $z^a=(y,z)=\Pi^a{}_A\mathbb{X}^A$ are the coordinates of the physical $T^2$.
The complementary projector $\widetilde{\Pi}_{aA}$ defines the auxiliary $T^2$ with
coordinates  $\tilde{z}_a=(\tilde{y},\tilde{z})=\widetilde{\Pi}_{aA}\mathbb{X}^A$.
 It is useful
to define the polarisation tensor $\Theta^{\hat{A}}{}_A$ so that
$$
\Theta^{\hat{A}}{}_A=\left(%
\begin{array}{cc}
  \Pi^a{}_A & \widetilde{\Pi}_{aA} \\
\end{array}%
\right) \qquad
\mathbb{X}^{\hat{A}}=\Theta^{\hat{A}}{}_A\mathbb{X}^A=\left(%
\begin{array}{c}
  y \\ z \\ \tilde{y} \\ \tilde{z} \\
\end{array}%
\right)
$$

The polarisation is constant over $I$, so that it   selects a subspace  $I\times T^2$ of $I\times T^4$. This then   can be continued in $x$
so that it selects a subspace  $\R\times T^2$ of $\R\times T^4$.
We will see that the various choices of subspace $\R\times T^2$ will give the covering spaces $C_{\cal N}$, $C_T$, $C_{T^3}$.
 The $O(2,2;\Z)$ transition functions of section 2 are now seen,   after the identification $x\sim x+1$, as an $O(2,2;\Z)$ monodromy round the $x$ circle.

The effect of a T-duality was analysed in \cite{Hull ``A geometry for non-geometric string backgrounds''}.
Acting with the $O(2,2;\Z)$ element
$
{\cal O}^A{}_B$ changes the polarisation
$$ \Theta^{\hat{A}}{}_A \to \Theta ' {}^{\hat{A}}{}_A =\Theta^{\hat{A}}{}_B {\cal O}^B{}_A
$$
and the new physical coordinates are $y',z'$, where
$$
\mathbb{X}'{}^{\hat{A}}=\Theta'{}^{\hat{A}}{}_A\mathbb{X}^A=\left(%
\begin{array}{c}
  y' \\ z' \\ \tilde{y}' \\ \tilde{z} '\\
\end{array}%
\right)
$$
The generalised metric transforms as
$$ {\cal H} _{AB} \to {\cal H} '{}_{AB}= ({\cal O}^t)_A{}^C{\cal H} _{CD} {\cal O}^D{}_B
$$
and the new
 metric $g'$ and $B$-field $B'$ of the $T^2$ fibres
 can be read off from
 \begin{equation}\label{genmetpr}
{\cal H} '=\left(%
\begin{array}{cc}
  g'-B'g'{}^{-1}B' & B'g'{}^{-1} \\
  -g'{}^{-1}B' & g'{}^{-1} \\
\end{array}%
\right)
\end{equation}

We shall now consider how this works in the example of the doubled space ${\cal T}$ constructed above.
For the nilfold, the polarisation is
\begin{eqnarray}\label{nilpol}
\begin{array}{ll}
  y=\Pi^y{}_A\mathbb{X}^A=\mathbb{X}^1 &\qquad \tilde{y}=\widetilde{\Pi}_{yA}\mathbb{X}^A=\mathbb{X}^3 \\
  z=\Pi^z{}_A\mathbb{X}^A=\mathbb{X}^2 &\qquad \tilde{z}=\widetilde{\Pi}_{zA}\mathbb{X}^A=\mathbb{X}^4
\end{array}
\end{eqnarray}
and the polarisation tensor is just the identity matrix
$$
\Theta^{\hat{A}}{}_A=\left(%
\begin{array}{cccc}
  1 & 0 & 0 & 0 \\
  0 & 1 & 0 & 0 \\
  0 & 0 & 1 & 0 \\
  0 & 0 & 0 & 1 \\
\end{array}%
\right)
$$
The polarisation selects the subspace with coordinates $x,y,z$, and this gives the nilfold on identifying the $x$ coordinate.

\noindent\textbf{$T^3$ with $H$-Flux}

Acting with the $O(2,2;\Z)$ element
$$
{\cal O}^A{}_B=\left(%
\begin{array}{cccc}
  0 & 0 & 1 & 0 \\
  0 & 1 & 0 & 0 \\
  1 & 0 & 0 & 0 \\
  0 & 0 & 0 & 1 \\
\end{array}%
\right)
$$
which corresponds to a T-duality in the $y$ direction,   the polarisation becomes (dropping primes)
$$
\Theta^{\hat{A}}{}_A=\left(%
\begin{array}{cccc}
  0 & 0 & 1 & 0 \\
  0 & 1 & 0 & 0 \\
  1 & 0 & 0 & 0 \\
  0 & 0 & 0 & 1 \\
\end{array}%
\right)
$$
The polarisation in the $T^4$ fibres of ${\cal T}$ is then
\begin{eqnarray}
\begin{array}{ll}
  y=\Pi^y{}_A\mathbb{X}^A=\mathbb{X}^3 &\qquad \tilde{y}=\widetilde{\Pi}_{yA}\mathbb{X}^A=\mathbb{X}^1 \\
  z=\Pi^z{}_A\mathbb{X}^A=\mathbb{X}^2 &\qquad \tilde{z}=\widetilde{\Pi}_{zA}\mathbb{X}^A=\mathbb{X}^4
\end{array}
\end{eqnarray}
Note that, compared with (\ref{nilpol}), the duality interchanges $\Pi^y{}_A$ and $\widetilde{\Pi}_{yA}$ in the passive perspective or
equivalently, $\mathbb{X}^1$ and $\mathbb{X}^3$ in the active perspective. The generalised metric, in this polarisation, may be written as
$$
{\cal H}_{T^3}=\left(%
\begin{array}{cccc}
  1+m^2x^2 & 0 & 0 & mx \\
  0 & 1+m^2x^2 & -mx & 0 \\
  0 & -mx & 1 & 0 \\
  mx & 0 & 0 & 1 \\
\end{array}%
\right)
$$
The metric and $B$-field in the $T^2\subset T^4$ fibre can then be read off by comparison with (\ref{genmetpr}) and we recover the expected
background
$$
g_{ab}=\left(%
\begin{array}{cc}
  1 & 0 \\
  0 & 1 \\
\end{array}%
\right) \qquad B_{ab}=\left(%
\begin{array}{cc}
  0 & mx \\
  -mx & 0 \\
\end{array}%
\right)
$$

The global structure is given by the identifications of the coordinates
\begin{eqnarray}
(x,y,z,\tilde{y},\tilde{z})&\sim& (x+1,y,z,\tilde{y}+mz,\tilde{z}-my) \nonumber\\
  (x,y,z,\tilde{y},\tilde{z})&\sim&
(x,y+1,z,\tilde{y},\tilde{z})\nonumber\\
(x,y,z,\tilde{y},\tilde{z})&\sim& (x,y,z+1,\tilde{y},\tilde{z}) \nonumber\\
 (x,y,z,\tilde{y},\tilde{z})&\sim&
(x,y,z,\tilde{y}+1,\tilde{z})\nonumber\\
(x,y,z,\tilde{y},\tilde{z})&\sim& (x,y,z,\tilde{y},\tilde{z}+1)
\end{eqnarray}
so that the physical coordinates $(x,y,z)$ are periodic and parameterise a $T^3$, as expected.
The structure
 is encoded in the monodromy matrix $e^N$ which in this polarisation is given by (\ref{mass}) where $f_{xa}{}^b=Q_x{}^{ab}=0$ and
\begin{equation}
K_{xab}=\left(%
\begin{array}{cc}
  0 & m \\
  -m & 0 \\
\end{array}%
\right)
\end{equation}
The twist matrix $N$ is upper triangular and the monodromy is just a shift of the $B$-field, corresponding to non-trivial $H$-flux.
It is a geometric transformation in  $\Delta(\Z)$ and the discrete
subgroup preserves the polarisation.

In this polarisation, the generators of the left action, $\cG_L$, are
\begin{eqnarray}
 \widetilde{Z}_x=\frac{\partial}{\partial x}+mz\frac{\partial}{\partial \tilde{y}}-my\frac{\partial}{\partial \tilde{z}}  \qquad  \widetilde{Z}_y&=&\frac{\partial}{\partial y} \qquad \widetilde{Z}_z=\frac{\partial}{\partial z}
\nonumber\\
\widetilde{X}^y&=&\frac{\partial}{\partial \tilde{y}} \qquad \widetilde{X}^z=\frac{\partial}{\partial \tilde{z}}
\end{eqnarray}
These are not preserved by the action of $\G$ and transform as
\begin{eqnarray}
\begin{array}{lll}
 \widetilde{Z}_x\rightarrow \widetilde{Z}_x+m\gamma \widetilde{X}^y-m\beta \widetilde{X}^z  &\qquad  \widetilde{Z}_y\rightarrow \widetilde{Z}_y+m\alpha \widetilde{X}^z &\qquad \widetilde{Z}_z\rightarrow \widetilde{Z}_z-m\alpha \widetilde{X}^y \\
&\qquad  \widetilde{X}^y\rightarrow \widetilde{X}^y &\qquad \widetilde{X}^z\rightarrow \widetilde{X}^z
\end{array}
\end{eqnarray}
We see that $\G$ preserves the subgroup $\widetilde{G}_L\simeq \R^2\subset\cG_L$ generated by $(\widetilde{X}^y,\widetilde{X}^z)$ and
the physical space is therefore be recovered as the quotient $T^3={\cal T}/\widetilde{G}_L$.

\noindent\textbf{T-Fold}

We have seen that a T-duality along the $y$-direction relates the nilfold and the $T^3$ with $H$-flux. If instead we act on the nilfold
polarisation with the element of $O(2,2;\Z)$
$$
{\cal O}^A{}_B=\left(%
\begin{array}{cccc}
  1 & 0 & 0 & 0 \\
  0 & 0 & 0 & 1 \\
  0 & 0 & 1 & 0 \\
  0 & 1 & 0 & 0 \\
\end{array}%
\right)
$$
which corresponds to T-duality along the $z$ direction, we find the polarisation tensor is now
$$
\Theta^{\hat{A}}{}_A=\left(%
\begin{array}{cccc}
  1 & 0 & 0 & 0 \\
  0 & 0 & 0 & 1 \\
  0 & 0 & 1 & 0 \\
  0 & 1 & 0 & 0 \\
\end{array}%
\right)
$$
In this polarisation, the coordinates are
\begin{eqnarray}
\begin{array}{ll}
  y=\Pi^y{}_A\mathbb{X}^A=\mathbb{X}^1 &\qquad \tilde{y}=\widetilde{\Pi}_{yA}\mathbb{X}^A=\mathbb{X}^3 \\
  z=\Pi^z{}_A\mathbb{X}^A=\mathbb{X}^4 &\qquad \tilde{z}=\widetilde{\Pi}_{zA}\mathbb{X}^A=\mathbb{X}^2
\end{array}
\end{eqnarray}
and the generalised metric on the $T^4$ fibres is
\begin{eqnarray}\label{HT}
{\cal H}_{T-Fold}=\left(%
\begin{array}{cccc}
  1 & 0 & 0 & -mx \\
  0 & 1 & mx & 0 \\
  0 & mx & 1+m^2x^2 & 0 \\
  -mx & 0 & 0 & 1+m^2x^2 \\
\end{array}%
\right)
\end{eqnarray}
from which the metric and B-field on the physical $T^2$ fibres may be read off
$$
g_{ab}=\frac{1}{1+m^2x^2}\left(%
\begin{array}{cc}
  1 & 0 \\
  0 & 1 \\
\end{array}%
\right) \qquad  B_{ab}=\frac{mx}{1+m^2x^2}\left(%
\begin{array}{cc}
  0 & 1 \\
  -1 & 0 \\
\end{array}%
\right)
$$

The monodromy matrix $e^N$ in this polarisation is given by (\ref{mass}) where $f_{xa}{}^b=K_{xab}=0$ and
\begin{equation}
Q_x{}^{ab}=\left(%
\begin{array}{cc}
  0 & m \\
  -m & 0 \\
\end{array}%
\right)
\end{equation}
which is not in the geometric group $\Delta(\Z)$. The monodromy then includes a T-duality acting on the physical
$T^2$ fibres and so the global structure, which is determined by $\G$, requires the following identifications of the coordinates
\begin{eqnarray}\label{doubledtorusgamma}
(x,y,z,\tilde{y},\tilde{z})&\sim& (x+1,y+m\tilde{z},z-m\tilde{y},\tilde{y},\tilde{z}) \nonumber\\
  (x,y,z,\tilde{y},\tilde{z})&\sim&
(x,y+1,z,\tilde{y},\tilde{z})\nonumber\\
(x,y,z,\tilde{y},\tilde{z})&\sim& (x,y,z+1,\tilde{y},\tilde{z}) \nonumber\\
 (x,y,z,\tilde{y},\tilde{z})&\sim&
(x,y,z,\tilde{y}+1,\tilde{z})\nonumber\\
(x,y,z,\tilde{y},\tilde{z})&\sim& (x,y,z,\tilde{y},\tilde{z}+1)
\end{eqnarray}
Here it is clear from the identifications
$$
x\sim x+1   \qquad  y\sim y+m\tilde{z}\qquad    z\sim z-m\tilde{y}
$$
that one cannot distinguish globally between the coordinates $(y,z)$ on $T^2$ and the coordinates $(\tilde{y},\tilde{z})$ on the dual torus
$\widetilde{T}^2$, as they get mixed by the monodromy.

 The
generators of the left action, $\cG_L$, are
\begin{eqnarray}
 \widetilde{Z}_x=\frac{\partial}{\partial x}+m\tilde{z}\frac{\partial}{\partial y}-m\tilde{y}\frac{\partial}{\partial z}  \qquad  \widetilde{Z}_y&=&\frac{\partial}{\partial y} \qquad \widetilde{Z}_z=\frac{\partial}{\partial z} \nonumber\\
 \widetilde{X}^y&=&\frac{\partial}{\partial \tilde{y}} \qquad
\widetilde{X}^z=\frac{\partial}{\partial \tilde{z}}
\end{eqnarray}
These are not invariant under $\G$, but transform as
\begin{eqnarray}
\begin{array}{lll}
 \widetilde{Z}_x\rightarrow \widetilde{Z}_x+m\tilde{\gamma} \widetilde{Z}_y-m\tilde{\beta} \widetilde{Z}_z  &\qquad  \widetilde{Z}_y\rightarrow \widetilde{Z}_y &\qquad \widetilde{Z}_z\rightarrow\widetilde{Z}_z \\
 &\qquad  \widetilde{X}^y\rightarrow \widetilde{X}^y+m \alpha\widetilde{Z}_z &\qquad \widetilde{X}^z\rightarrow
\widetilde{X}^z-m\alpha \widetilde{Z}_y
\end{array}\nonumber
\end{eqnarray}
We see that $\G$ does not preserve the subgroup $\widetilde{G}_L\simeq\R^2\subset\cG_L$ generated by $(\widetilde{X}^y,\widetilde{X}^z)$.

The metric and $B$-field (\ref{gb}) on $I\times T^4$ can be extended  to $\R\times T^4$ by continuing in $x$.
This gives a covering space ${\cal C}$ of $\cal T$ in which the first identification in (\ref{doubledtorusgamma}) is dropped. It is obtained by
identifying $\cG$ under $\G_{\cal C}$ where $\G_{\cal C}$ is the subgroup of $\G$ with $\a=0$.
The subgroup  $\widetilde{G}_L\simeq\R^2\subset\cG_L$ generated by $(\widetilde{X}^y,\widetilde{X}^z)$
is   preserved by $\G_{\cal C}$,
so that the coset ${\cal C}/ \widetilde{G}_L$ is well-defined, and gives the covering space $C_T$ of the T-fold
considered previously.

Finally, consider the identification $x\sim x+1$, so that the   fibres at $x=0$ and $x=1$ are glued together with an
$O(2,2;\Z)$ transformation.
For the doubled space $I\times T^4$, the
$O(2,2;\Z)$ gluing is a diffeomorphism of the $T^4$, giving a $T^4$ bundle over $S^1$, which is precisely $\cal T$ with
 the coordinate identification given  in
(\ref{doubledtorusgamma}).
 For  $I\times T^2$, the
$O(2,2;\Z)$ gluing is a T-duality giving a T-fold. The local structure of the T-fold is that of    the coset ${\cal C}/ \widetilde{G}_L$.

\subsection{The Doubled Twisted Torus $\cX$}

\noindent The doubled torus geometry ${\cal T}$ gives a geometric interpretation to the action of the twist matrix $N^A{}_B$ but does not give a
geometric interpretation for the full gauge algebra (\ref{galgebra}). In the nilfold polarisation, the natural left-invariant vector fields
(\ref{doublevecL}) on ${\cal T}$ satisfy the algebra
$$
[Z_x,Z_z]=mZ_y  \qquad  [Z_x,X^y]=mX^z
$$
where all other commutators vanish. This algebra is a  subgroup  of a contraction of the full gauge algebra of the  {theory} (\ref{O(d+1,d+1) Lagrangian}), which is
\begin{equation}\label{nalgebra}
[Z_x,Z_z]=mZ_y  \qquad  [Z_x,X^y]=mX^z  \qquad  [Z_z,X^y]=-mX^x
\end{equation}
where all other commutators vanish. This highlights the fact that the doubled torus formulation does not    encode all of the information of the
 {field theory} (\ref{O(d+1,d+1) Lagrangian}) in its geometry.

This is not surprising since, as discussed in \cite{Hull ``Gauge Symmetry T-Duality and Doubled Geometry''} and reviewed in Section 2,  the generator $X^x$ of
$B$-shifts with one leg along $S^1_x$ does not have a geometric interpretation in the ${\cal T}$ construction. It can be geometrised by introducing
an auxiliary coordinate for the base coordinate $x$ so that
 $(x,\mathbb{X}^A)\rightarrow (x,\tilde{x},\mathbb{X}^A)$.
 Indeed, it is natural to introduce a variable $\tilde{x}$ conjugate to  winding modes on the $x$ circle.
 The natural doubled geometry, encoding the full gauge group,
 is  {given by}
 the six-dimensional noncompact group manifold $\cG$ with  Lie algebra (\ref{nalgebra})
 and  {then taking the quotient} by some
 discrete subgroup $\G$ to obtain a compact
   six-dimensional doubled twisted torus
 $\cX=\cG/\G$.

The   six-dimensional doubled group   is $\cG=G\ltimes \R^3$ where $G$ is the
three-dimensional Heisenberg group, so that  $\cG$
is the cotangent bundle $\cG=T^*G $ of the Heisenberg group.
A matrix representation for the Lie algebra (\ref{nalgebra}) can be given in terms of the
 $t_m$ (\ref{t's}) as
$$
T_x=\left(%
\begin{array}{cc}
  t_x & 0 \\
  0 & t_x \\
\end{array}%
\right) \qquad  T_y=\left(%
\begin{array}{cc}
  t_y & 0 \\
  0 & t_y \\
\end{array}%
\right) \qquad  T_z=\left(%
\begin{array}{cc}
  t_z & 0 \\
  0 & t_z \\
\end{array}%
\right)
$$
$$
T^x=\left(%
\begin{array}{cc}
  0 & 0 \\
  t_y & 0 \\
\end{array}%
\right) \qquad  T^y=\left(%
\begin{array}{cc}
  0 & -t_z \\
  -t_x & 0 \\
\end{array}%
\right) \qquad  T^z=\left(%
\begin{array}{cc}
  0 & t_y \\
  0 & 0 \\
\end{array}%
\right)
$$
and so a general element of the doubled group may be written as
\begin{equation}\label{twistedtorusg}
g=\left(%
\begin{array}{cccccc}
  1 & mx & y & 0 & 0 & \tilde{z} \\
  0 & 1 & z & 0 & 0 & -\tilde{y} \\
  0 & 0 & 1 & 0 & 0 & 0 \\
  0 & -m\tilde{y} & \tilde{x}-mz\tilde{y} & 1 & mx & y+\frac{1}{2}m\tilde{y}^2 \\
  0 & 0 & 0 & 0 & 1 & z \\
  0 & 0 & 0 & 0 & 0 & 1 \\
\end{array}%
\right)
\end{equation}
showing the dependence on the coordinates $(x,y,z,\tilde{x},\tilde{y},\tilde{z})$.

Taking the quotient of $\cG$ by the action of $X^x$ eliminates $\tilde x$ and gives the five-dimensional group manifold
used in the doubled torus construction considered in the previous subsection.
The choice of discrete cocompact subgroup $\G$ of the six-dimensional group $\cG$ is largely dictated by requiring that it be compatible with the five-dimensional quotient  {used in the doubled torus construction}.
The global structure of the twisted torus $\cG/\G$ is then given by the identification $g\sim h\cdot g$, where a generic element $h\in \G$ is given by
$$
h=\left(%
\begin{array}{cccccc}
  1 & m\alpha & \beta & 0 & 0 & \tilde{\beta} \\
  0 & 1 & \gamma & 0 & 0 & -\tilde{\gamma} \\
  0 & 0 & 1 & 0 & 0 & 0 \\
  0 & -m\tilde{\beta} & \tilde{\alpha}-m\beta\tilde{\gamma} & 1 & m\alpha & \beta+\frac{1}{2}m\tilde{\beta}^2 \\
  0 & 0 & 0 & 0 & 1 & \gamma \\
  0 & 0 & 0 & 0 & 0 & 1 \\
\end{array}%
\right)
$$
where $(\alpha,\beta,\gamma,\tilde{\alpha},\tilde{\beta},\tilde{\gamma})$ are arbitrary integers. The left action of $h$ is $g\rightarrow h\cdot
g$ and acts on the coordinates through
\begin{eqnarray}
\begin{array}{lll}
 x\rightarrow x+\alpha  &\qquad  y\rightarrow y+m\alpha z+\beta   &\qquad  z\rightarrow z+\gamma \\
\tilde{x}\rightarrow \tilde{x}+m\gamma\tilde{y}+\tilde{\alpha} &\qquad  \tilde{y}\rightarrow \tilde{y}+\tilde{\beta}
&\qquad\tilde{z}\rightarrow\tilde{z}-m\alpha {\ti y}+\tilde{\gamma}
\end{array}
\end{eqnarray}
Identifying $\cG$ under the action of $\G$ implies that the coordinates are subject to the identifications
\begin{eqnarray}
(x,y,z,\tilde{x},\tilde{y},\tilde{z})&\sim& (x+1,y+mz,z,\tilde{x},\tilde{y},\tilde{z}-m\tilde{y}) \nonumber\\
  (x,y,z,\tilde{x},\tilde{y},\tilde{z})&\sim&
(x,y+1,z,\tilde{x},\tilde{y},\tilde{z})\nonumber\\
(x,y,z,\tilde{x},\tilde{y},\tilde{z})&\sim& (x,y,z+1,\tilde{x}+m\tilde{y},\tilde{y},\tilde{z}) \nonumber\\
 (x,y,z,\tilde{x},\tilde{y},\tilde{z})&\sim&
(x,y,z,\tilde{x}+1,\tilde{y},\tilde{z})\nonumber\\
 (x,y,z,\tilde{x},\tilde{y},\tilde{z})&\sim&
(x,y,z,\tilde{x},\tilde{y}+1,\tilde{z})\nonumber\\
(x,y,z,\tilde{x},\tilde{y},\tilde{z})&\sim& (x,y,z,\tilde{x},\tilde{y},\tilde{z}+1)
\end{eqnarray}

 The doubled group $\cG$ has a natural action of $\cG_L\times \cG_R$ which is generated by right- and left-invariant
vector fields. The right action $\cG_R$ is generated by the left-invariant vector fields
\begin{eqnarray}
 Z_x=\frac{\partial}{\partial x}  \qquad  Z_y&=&\frac{\partial}{\partial y} \qquad\qquad\qquad\qquad\qquad Z_z=\frac{\partial}{\partial z}+mx\frac{\partial}{\partial y}\nonumber\\
X^x=\frac{\partial}{\partial \tilde{x}} \qquad  X^y&=&\frac{\partial}{\partial \tilde{y}}+mz\frac{\partial}{\partial
\tilde{x}}-mx\frac{\partial}{\partial \tilde{z}} \qquad X^z=\frac{\partial}{\partial \tilde{z}}
\end{eqnarray}
which satisfy the commutation relations of the full gauge algebra of the supergravity (\ref{nalgebra}). Since the cocompact subgroup
$\G$ acts from the left, the left-invariant $(Z_m,X^m)$ are globally defined on $\cX$. The left-invariant one-forms $(P^m,Q_m)$, dual
to the vector fields $(Z_m,X^m)$, are
\begin{eqnarray}
\begin{array}{lll}
 P^x=dx   &\qquad P^y=dy-mxdz   &\qquad P^z=dz \\
Q_x=d\tilde{x}-mzd\tilde{y} &\qquad  Q_y=d\tilde{y} &\qquad Q_z=d\tilde{z}+mxd\tilde{y}
\end{array}
\end{eqnarray}
These one-forms generalise those of the nilfold (\ref{nilforms}) and the doubled torus (\ref{nilforms 2}). The generators of the left action,
$\cG_L$, are
\begin{eqnarray}
 \widetilde{Z}_x=\frac{\partial}{\partial x}+mz\frac{\partial}{\partial y}-m\tilde{y}\frac{\partial}{\partial \tilde{z}}  \qquad  \widetilde{Z}_y&=&\frac{\partial}{\partial y} \qquad \widetilde{Z}_z=\frac{\partial}{\partial z}+m\tilde{y}\frac{\partial}{\partial \tilde{x}}
\nonumber\\
\widetilde{X}^x=\frac{\partial}{\partial \tilde{x}} \qquad  \widetilde{X}^y&=&\frac{\partial}{\partial \tilde{y}} \qquad
\widetilde{X}^z=\frac{\partial}{\partial \tilde{z}}
\end{eqnarray}
These are not invariant under $\G$ and transform as
\begin{eqnarray}
\begin{array}{lll}
 \widetilde{Z}_x\rightarrow \widetilde{Z}_x+m\gamma \widetilde{Z}_y-m \tilde{\beta}\widetilde{X}^z  &\qquad  \widetilde{Z}_y\rightarrow \widetilde{Z}_y &\qquad \widetilde{Z}_z\rightarrow \widetilde{Z}_z-m\alpha \widetilde{Z}_y+m\tilde{\beta} \widetilde{X}^x \\
\widetilde{X}^x\rightarrow \widetilde{X}^x &\qquad  \widetilde{X}^y\rightarrow \widetilde{X}^y+m\alpha \widetilde{X}^z-m\gamma \widetilde{X}^x
&\qquad \widetilde{X}^z\rightarrow \widetilde{X}^z\nonumber
\end{array}
\end{eqnarray}
We see that the $\widetilde{X}^m$ close to generate an abelian sub-group $\widetilde{G}_L\simeq \R^3\subset\cG_L$. Since $\G$ preserves
$\widetilde{G}_L$, the nilfold geometry is in fact recovered globally as the quotient
$$
{\cal N}\simeq\cX/\widetilde{G}_L
$$
To see this, the group element (\ref{twistedtorusg}) can be decomposed as
$$
g=\widetilde{g}\cdot k
$$
where $\widetilde{g}\in \widetilde{G}_L$ and the coset representative $k$
$$
k=\left(%
\begin{array}{cc}
  g' & 0 \\
  0 & g' \\
\end{array}%
\right)\in \cG/\widetilde{G}_f
$$
where $g'$ is an element of the Heisenberg group $G$ as given by (\ref{g}). Since $\G$ preserves $\widetilde{G}_L$ we can recover the
global structure of the nilfold by the action of $\G$ on $(x,y,z)$.

\subsubsection{Polarisations}

Following Section 2,
it is useful to
 introduce a polarisation  of the Lie algebra, with  a projector $\widetilde{\Pi}_{mM}$  projecting  the generators $T^M$ onto $Z_m$, associated with an action
 on the coordinates $x,y,z$, and the complementary projector $ \Pi^m{}_M$ onto the $X^m$.
 These combine into a polarisation tensor
 $$
\Theta^{\hat{M}}{}_M=\left(%
\begin{array}{c}
  \Pi^m{}_M \\ \widetilde{\Pi}_{mM} \\
\end{array}%
\right)   $$
so that
$$T^{\hat{M}}=\Theta^{\hat{M}}{}_M T^M=
\left(%
\begin{array}{c}
X^x \\ X^y \\ X^z \\ Z_x \\ Z_y \\ Z_z
 \\
\end{array}%
\right)
$$
The left-invariant forms $\mathcal{P}^M$ on $\cG$ are similarly projected onto $P^m$ and $Q_m$,
$$
  {\cal P}^{\hat{M}}=\Theta^{\hat{M}}{}_M\mathcal{P}^M=
\left(
  \begin{array}{c}
P^m \\ Q_m
 \\
\end{array}%
\right)
$$
Similarly, the same projectors split the right-invariant  $\widetilde T^{M}$ into $\widetilde Z_m$ and $\widetilde X^m$.
As in section 2, if the $ \widetilde X^m$ generate a closed algebra then   they constitute an integrable distribution and define a submanifold, at least locally.

We can also  define a polarisation of the coordinates.
Let the coordinates of $\cX$ be  $\mathbb{X}^I$, $I=1,2,..,6$.
A polarisation $x^i=(x,y,z)=\Pi^i{}_I\mathbb{X}^I$ then locally selects which three of the six $\mathbb{X}^I$ are to be treated as the physical
coordinates $(x,y,z)$, and which  three are to be treated  as auxiliary,  $\tilde{x}_i=(\tilde{x},\tilde{y},\tilde{z})=\widetilde{\Pi}_{iI}\mathbb{X}^I$.
As we shall see, for a geometric background or for one that is locally geometric (e.g. a T-fold) the background configuration is given in terms of
fields that depend on  $(x,y,z)$ but not  $(\tilde x, \tilde y, \tilde z)$.
 It is useful to  introduce a polarisation tensor for the coordinates
$$
\Theta^{\hat{I}}{}_I=\left(%
\begin{array}{c}
  \Pi^i{}_I \\ \widetilde{\Pi}_{iI} \\
\end{array}%
\right)    \qquad  \mathbb{X}^{\hat{I}}=\Theta^{\hat{I}}{}_I\mathbb{X}^I=
\left(%
\begin{array}{c}
x\\  y \\ z \\ \ti x \\ \tilde{y} \\ \tilde{z} \\
\end{array}%
\right)
$$
The different polarisations we shall consider will just correspond to a relabeling of the coordinates, choosing different subsets of three of the six coordinates to be physical.

\noindent\textbf{Nilfold}

The nilfold is recovered   by the polarisation choice
\begin{eqnarray}
\begin{array}{lll}
 x=\Pi^x{}_I\mathbb{X}^I=\mathbb{X}^1  &\qquad  y=\Pi^y{}_I\mathbb{X}^I=\mathbb{X}^2 &\qquad z=\Pi^z{}_I\mathbb{X}^I=\mathbb{X}^3 \\
\tilde{x}=\widetilde{\Pi}_{xI}\mathbb{X}^I=\mathbb{X}^4 &\qquad  \tilde{y}=\widetilde{\Pi}_{yI}\mathbb{X}^I=\mathbb{X}^5 &\qquad
\tilde{z}=\widetilde{\Pi}_{zI}\mathbb{X}^I=\mathbb{X}^6
\end{array}
\end{eqnarray}
which corresponds to the polarisation tensor
$$
\Theta=\left(%
\begin{array}{cccccc}
  1 & 0 & 0 & 0 & 0 & 0 \\
  0 & 1 & 0 & 0 & 0 & 0 \\
  0 & 0 & 1 & 0 & 0 & 0 \\
  0 & 0 & 0 & 1 & 0 & 0 \\
  0 & 0 & 0 & 0 & 1 & 0 \\
  0 & 0 & 0 & 0 & 0 & 1 \\
\end{array}%
\right)
$$
This selects the submanifold with coordinates $x,y,z$, which is of course the nilfold.
The doubled twisted torus is the cotangent bundle for the Heisenberg group, $\cG=T^*G=G\ltimes
\R^3$, modded out by $\G$. The polarisation projection $\Pi$ can then be identified as the natural bundle projection of the cotangent
bundle to the base $\Pi:T^*G\rightarrow G$. As in the doubled torus construction, we can recover the dual formulation as a $T^3$ with $H$-flux or as a  T-fold
as different polarisations of the same doubled background.

As before, acting with a linear transformation ${\cal O}$ changes the polarisation
$$ \Theta \to  \Theta  '= \Theta {\cal O}
$$
and any two polarisations are  related in this way.
This changes the polarisation of the Lie algebra in the way described above, and we shall accompany this with the relabeling of the coordinates associated with the change of coordinate polarisation, using the same linear transformation ${\cal O}$.
If ${\cal O}\in O(2,2;\Z)$, the T-duality group acting on the coordinates $y,z,\ti y , \ti z$, then the two polarisations give two backgrounds related by  a   T-duality, as we shall review below, and these give two physically equivalent string backgrounds.
However, other polarisations seem possible, with  ${\cal O}$ not in $O(2,2;\Z)$, and the question arises as to whether they give physically equivalent backgrounds. In \cite{Dabholkar ``Generalised T-duality and non-geometric backgrounds''}, it was conjectured that there are generalised T-dualities acting in precisely this way, and we will see below that the doubled twisted torus formalism suggests a natural form for these.

The one-forms in this polarisation are
$$
\begin{array}{lll}
P^x=dx &\qquad P^y=dy-mxdz &\qquad P^z=dz \\
Q_x=d\tilde{x}-mzd\tilde{y} &\qquad Q_y=d\tilde{y} &\qquad Q_z=d\tilde{z}+mxd\tilde{y}
\end{array}
$$
Here, the left-invariant one-forms on the Heisenberg group $G$ and on $\widetilde{G}=\R^3 $ are $P^m$ and $d\tilde{x}_m$ respectively. This case is a special example of the twisted torus case considered in section 2. Note that we can work in terms of the derivatives of the coordinates $dx^i$, instead of the frame forms $P^m$ and obtain the coordinate metric directly.

The algebra (\ref{nalgebra})
 is a Drinfel'd double and we can follow the general procedure outlined in section 2.6 to recover a conventional description of the background. The natural $G_R$- and $\widetilde{G}_L$-invariant forms in this polarisation are
$$
r^x=dx  \qquad  r^y=dy-mzdx \qquad  r^z=dz
$$
$$
{\ti \ell}_x=d{\ti x}   \qquad  {\ti \ell}_y=d{\ti y}   \qquad  {\ti \ell}_z=d{\ti z}
$$
(Note that $P^y\neq r^y$.) The $\cG_L$-invariant one-form ${\cal P}^{\hat{M}}$, in the nilfold polarisation, may be written as ${\cal P}^{\hat{M}}=\Phi^{\hat{N}}{\cal V}_{\hat{N}}{}^{\hat{M}}(x)$, where $\Phi^{\hat{M}}=(r^m,{\ti \ell}_m)$ and
$$
{\cal V}_{\hat{M}}{}^{\hat{N}}=\left(
           \begin{array}{cc}
             A_m{}^n & 0 \\
             0 & (A^{-1})^m{}_n \\
           \end{array}
         \right)
$$
The adjoint action of the Heisenberg group on itself, which appears in ${\cal V}$, is given by
$$
A_m{}^n=\left(
          \begin{array}{ccc}
            1 & mz & 0 \\
            0 & 1 & 0 \\
            0 & -mx & 1 \\
          \end{array}
        \right)
$$
The construction of  section 2.6 gives the $\widetilde{G}_L$-invariant metric
$$
ds^2=\delta_{mn}A^m{}_pA^n{}_q r^p\otimes r^q=g_{ij}dx^i\otimes dx^j
$$
where
$$
g_{ij}=\left(
\begin{array}{ccc}
1 & 0 & 0 \\
 0 & 1 & -mx \\
0 & -mx & 1+m^2x^2
                     \end{array}
\right)	
$$
and the  contribution from the generalised metric ${\cal H}_{MN}(x)$  to the $B$-field is zero. In fact, by substituting
$$
\frac{1}{2}d\left(r^m\wedge{\ti \ell}_m\right)=-\frac{1}{2}mdx\wedge d\tilde{y}\wedge dz    \qquad  {\cal K}=-mdx\wedge d\tilde{y}\wedge dz
$$
into (\ref{H conjecture}), we see that $H=0$ in this polarisation. The 3-dimensional nilfold geometry is thus recovered.

\noindent\textbf{$T^3$ with $H$-Flux}

\noindent We now consider acting with ${\cal O}\in O(2,2;\Z)$ to change the polarisation to that corresponding to the $T^3$ with constant
$H$-flux. The element ${\cal O}$ and polarisation tensor are given by
$$
{\cal O}=\Theta=\left(%
\begin{array}{cccccc}
  1 & 0 & 0 & 0 & 0 & 0 \\
  0 & 0 & 0 & 0 & 1 & 0 \\
  0 & 0 & 1 & 0 & 0 & 0 \\
  0 & 0 & 0 & 1 & 0 & 0 \\
  0 & 1 & 0 & 0 & 0 & 0 \\
  0 & 0 & 0 & 0 & 0 & 1 \\
\end{array}%
\right)
$$
As in the doubled torus bundle ${\cal T}$, the action of a T-duality along the $y$-direction has the effect of exchanging
$$
Z_y\leftrightarrow X^y  \qquad  y\leftrightarrow \tilde{y}  \qquad P^y\leftrightarrow Q_y
$$
relative to the nilfold polarisation. The gauge algebra of the resulting supergravity is, by construction, identical to that from the nilfold.
However, we now label the generators acting geometrically on the $T^3$ as $Z_m$ ($Z_x$ is the $U(1)$ acting on the $x$-circle etc) and the ones from $B$-field  transformations as $X^m$, so that the algebra is now
$$
[Z_x,Z_z]=mX^y  \qquad  [Z_x,Z_y]=mX^z  \qquad  [Z_z,Z_y]=-mX^x
$$
where all other commutators vanish.
This is of course   the same algebra as in (\ref{nalgebra}), after relabeling the generators.

This Lie algebra fixes the local structure of the doubled twisted torus and it is particularly useful to
consider the left-invariant one-forms on $\cX$, which may be written as
\begin{eqnarray}\label{H-flux}
\begin{array}{lll}
  P^x=dx  &\qquad  P^y=dy  &\qquad   P^z=dz\\
Q_x=d\tilde{x}-mzdy  &\qquad  Q_y=d\tilde{y}-mxdz &\qquad Q_z=d\tilde{z}+mxdy
\end{array}
\end{eqnarray}
 The one-forms $P^m$ tell us that the spacetime is locally $\R^3$. The action of $\G$ on the coordinates is
\begin{eqnarray}\label{h-flux gamma}
\begin{array}{lll}
 x\rightarrow  x+\alpha &\qquad  y\rightarrow y+\beta   &\qquad  z\rightarrow z+\gamma \\
\tilde{x}\rightarrow  \tilde{x}+m \gamma y+\tilde{\alpha} &\qquad  \tilde{y}\rightarrow  \tilde{y}+m\alpha z+\tilde{\beta} &\qquad\tilde{z}\rightarrow \tilde{z}-m\alpha
y+\tilde{\gamma}
\end{array}
\end{eqnarray}
The identification $g\sim h\cdot g$ imposes the following identifications on the coordinates
\begin{eqnarray}\label{identifications}
(x,y,z,\tilde{x},\tilde{y},\tilde{z})&\sim& (x+1,y,z,\tilde{x},\tilde{y}+mz,\tilde{z}-my) \nonumber\\
  (x,y,z,\tilde{x},\tilde{y},\tilde{z})&\sim&
(x,y+1,z,\tilde{x},\tilde{y},\tilde{z})\nonumber\\
(x,y,z,\tilde{x},\tilde{y},\tilde{z})&\sim& (x,y,z+1,\tilde{x}+my,\tilde{y},\tilde{z}) \nonumber\\
 (x,y,z,\tilde{x},\tilde{y},\tilde{z})&\sim&
(x,y,z,\tilde{x}+1,\tilde{y},\tilde{z})\nonumber\\
 (x,y,z,\tilde{x},\tilde{y},\tilde{z})&\sim&
(x,y,z,\tilde{x},\tilde{y}+1,\tilde{z})\nonumber\\
(x,y,z,\tilde{x},\tilde{y},\tilde{z})&\sim& (x,y,z,\tilde{x},\tilde{y},\tilde{z}+1)
\end{eqnarray}
We see from the identifications of the spacetime coordinates $(x,y,z)$ that the spacetime globally is a $T^3$.

The generators of the left action, $\cG_L$, are
\begin{eqnarray}
 \widetilde{Z}_x=\frac{\partial}{\partial x}+mz\frac{\partial}{\partial \tilde{y}}-my\frac{\partial}{\partial \tilde{z}}  \qquad  \widetilde{Z}_y&=&\frac{\partial}{\partial y} \qquad \widetilde{Z}_z=\frac{\partial}{\partial z}+my\frac{\partial}{\partial \tilde{x}}
\nonumber\\
\widetilde{X}^x=\frac{\partial}{\partial \tilde{x}} \qquad  \widetilde{X}^y&=&\frac{\partial}{\partial \tilde{y}} \qquad
\widetilde{X}^z=\frac{\partial}{\partial \tilde{z}}
\end{eqnarray}
These are not invariant under the action of $\G$ and transform as
\begin{eqnarray}
\begin{array}{ll}
 \widetilde{Z}_x\rightarrow \widetilde{Z}_x+m\gamma \widetilde{X}^y-m\beta \widetilde{X}^z  &\qquad \widetilde{X}^x\rightarrow \widetilde{X}^x\\
 \widetilde{Z}_y\rightarrow \widetilde{Z}_y+m\alpha \widetilde{X}^z-m\gamma \widetilde{X}^x  &\qquad \widetilde{X}^y\rightarrow \widetilde{X}^y\\
\widetilde{Z}_z\rightarrow \widetilde{Z}_z-m\alpha \widetilde{X}^y+m\beta \widetilde{X}^x   &\qquad \widetilde{X}^z\rightarrow
\widetilde{X}^z
\nonumber
\end{array}
\end{eqnarray}
The $\widetilde{X}^m$ generate an abelian group $\widetilde{G}_L\simeq\R^3\subset\cG_L$. We see that $\G$ preserves $\widetilde{G}_L$
and we can recover the spacetime as the quotient $T^3\simeq\cX/\widetilde{G}_L$.


 We now   recover the conventional   background from the doubled geometry following  section 2.6. As discussed in section 2.6, we use the  parameterisation in which an element of $\cG$ is written  as $h={\ti g}k$.
 Using the generators
$$
T_x=\left(\begin{array}{cc}
  t_x & 0 \\
  0 & t_x \\
\end{array}%
\right) \qquad  T_y=\left(%
\begin{array}{cc}
  0 & -t_z \\
  -t_x & 0 \\
\end{array}%
\right) \qquad  T_z=\left(%
\begin{array}{cc}
  t_z & 0 \\
  0 & t_z \\
\end{array}%
\right)
$$
$$
T^x=\left(%
\begin{array}{cc}
  0 & 0 \\
  t_y & 0 \\
\end{array}%
\right) \qquad  T^y=\left(%
\begin{array}{cc}
  t_y & 0 \\
  0 & t_y \\
\end{array}%
\right) \qquad  T^z=\left(%
\begin{array}{cc}
  0 & t_y \\
  0 & 0 \\
\end{array}%
\right)
$$
where the matrices $t_m$ are given in (\ref{t's}) and writing a general element of   $\cG$ as
$$h={\ti g}k=\exp({\ti x}_m\widetilde{T}^m)\exp(x^mT_m)$$
we find that
$$
h=\left(
    \begin{array}{cccccc}
      1 & mx & {\ti y} & 0 & 0 & {\ti z} \\
      0 & 1 & z & 0 & 0 & -y \\
      0 & 0 & 1 & 0 & 0 & 0 \\
      0 & -my & {\ti x} & 1 & mx & {\ti y}+\frac{1}{2}my^2 \\
      0 & 0 & 0 & 0 & 1 & z \\
      0 & 0 & 0 & 0 & 0 & 1 \\
    \end{array}
  \right)
$$
Then  the left-invariant forms $P^m$ and $Q_m$ in this parameterisation are
\begin{eqnarray}\label{me!}
\begin{array}{lll}
  P^x=dx  &\qquad  P^y=dy  &\qquad   P^z=dz\\
Q_x=d\tilde{x}+mydz  &\qquad  Q_y=d\tilde{y}-mxdz &\qquad Q_z=d\tilde{z}+mxdy
\end{array}
\end{eqnarray}
It is useful to have $k$ and ${\ti g}$ explicitly
$$
k=\left(
    \begin{array}{cccccc}
      1 & mx & 0 & 0 & 0 & 0 \\
      0 & 1 & z & 0 & 0 & -y \\
      0 & 0 & 1 & 0 & 0 & 0 \\
      0 & -my & 0 & 1 & mx & \frac{1}{2}my^2 \\
      0 & 0 & 0 & 0 & 1 & z \\
      0 & 0 & 0 & 0 & 0 & 1 \\
    \end{array}
  \right)   \qquad  {\ti g}=\left(
                              \begin{array}{cccccc}
                                1 & 0 & {\ti y} & 0 & 0 & {\ti z} \\
                                0 & 1 & 0 & 0 & 0 & 0 \\
                                0 & 0 & 1 & 0 & 0 & 0 \\
                                0 & 0 & {\ti x} & 1 & 0 & {\ti y} \\
                                0 & 0 & 0 & 0 & 1 & 0 \\
                                0 & 0 & 0 & 0 & 0 & 1 \\
                              \end{array}
                            \right)
$$
so that we can determine ${\ti g}^{-1}d{\ti g}={\ti \ell}={\ti \ell}_m\widetilde{T}^m$ and $dkk^{-1}=r=r^mT_m+r_m\widetilde{T}^m$ explicitly
\begin{eqnarray}
\begin{array}{lll}
  r^x=dx  &\qquad  r^y=dy  &\qquad   r^z=dz\\
r_x=mzdy  &\qquad  r_y=-mxdz &\qquad r_z=mxdy\\
{\ti \ell}_x=d{\ti x}  &\qquad  {\ti \ell}_y=d{\ti y} &\qquad {\ti \ell}_z=d{\ti z}\\
\end{array}
\end{eqnarray}
We note that
$$
Q_x={\ti q}_x+mydz-mzdy \qquad  Q_y={\ti q}_y+mzdx-mxdz \qquad  Q_z={\ti q}_z+mxdy-mydx
$$
where $Q_m$ is given in (\ref{me!}) and ${\ti q}_m={\ti \ell}_m+r_m$. We   write the left-invariant one-form on $\cG$ as ${\cal P}^{\hat{M}}=\Phi^{\hat{N}}{\cal V}_{\hat{N}}{}^{\hat{M}}$, where $\Phi^{\hat{M}}=(r^m,{\ti q}_m)$ and
$$
{\cal V}_{\hat{N}}{}^{\hat{M}}=\left(
              \begin{array}{cc}
                \delta_m{}^n & b_{mn} \\
                0 & \delta^m{}_n \\
              \end{array}
            \right)
$$
where
$$
b_{mn}=\left(\begin{array}{ccc}
                 0 & mz & -my \\
                  -mz & 0 & mx \\
                  my & -mx & 0
              \end{array}
            \right)
$$
The metric for this background is flat
$$
ds^2=\delta_{mn}dx^m\otimes dx^n
$$
 {and f}rom the identifications (\ref{identifications})   we  {see that globally the compact space is} $T^3$. We now consider the $H$-field. The antisymmetric matrix $b_{mn}$ in ${\cal V}_{\hat{N}}{}^{\hat{M}}$ defines a two-form
$$
b=mzdx\wedge dy+mydz\wedge dx+mxdy\wedge dz
$$
so that
$$
db=3mdx\wedge dy\wedge dz
$$
It is not hard to show that
$$
\frac{1}{2}d\left(r^m\wedge {\ti q}_m\right)=-\frac{3}{2}mdx\wedge dy\wedge dz    \qquad  {\cal K}=-mdx\wedge dy\wedge dz
$$
so that the physical $H$-field given by (\ref{H with K}) is
$$
H=mdx\wedge dy\wedge dz
$$
and we recover the $T^3$ with constant $H$-flux background as expected.

\noindent\textbf{T-Fold}

\noindent We now consider acting on the nilfold polarisation with a different ${\cal O}\in O(2,2;\Z)$ to change the polarisation to that of the
T-fold. The element ${\cal O}$ and polarisation tensor are given by
$$
{\cal O}=\Theta=\left(%
\begin{array}{cccccc}
  1 & 0 & 0 & 0 & 0 & 0 \\
  0 & 1 & 0 & 0 & 0 & 0 \\
  0 & 0 & 0 & 0 & 0 & 1 \\
  0 & 0 & 0 & 1 & 0 & 0 \\
  0 & 0 & 0 & 0 & 1 & 0 \\
  0 & 0 & 1 & 0 & 0 & 0 \\
\end{array}%
\right)
$$
Acting with this  ${\cal O}$, which corresponds to performing a T-duality along the $z$-direction, has the effect of exchanging
$$
Z_z\leftrightarrow X^z  \qquad  z\leftrightarrow \tilde{z}  \qquad P^z\leftrightarrow Q_z
$$
relative to the nilfold polarisation. The gauge algebra of the  {corresponding field theory }(\ref{O(d+1,d+1) Lagrangian}) is
$$
[Z_x,X^z]=mZ_y  \qquad  [Z_x,X^y]=mZ_z  \qquad  [X^z,X^y]=-mX^x
$$
where all other commutators vanish. The left-invariant one-forms corresponding to this algebra are
\begin{eqnarray}\label{T-fold}
\begin{array}{lll}
 P^x=dx  &\qquad  P^y=dy-mxd\tilde{z} &\qquad P^z=dz+mxd\tilde{y} \\
Q_x=d\tilde{x}-m\tilde{z}d\tilde{y} &\qquad  Q_y=d\tilde{y} &\qquad Q_z=d\tilde{z}
\end{array}
\end{eqnarray}
The global structure $\G$ of the doubled twisted torus $\cX$ is determined by the rigid left action on the coordinates
\begin{eqnarray}\label{T-fold gamma}
\begin{array}{lll}
 x\rightarrow  x+\alpha  &\qquad y\rightarrow  y+m\alpha\tilde{z}+\beta  &\qquad z\rightarrow  z-m\alpha\tilde{y}+\gamma \\
\tilde{x}\rightarrow  \tilde{x}+m\tilde{\gamma}\tilde{y}+\tilde{\alpha} &\qquad  \tilde{y}\rightarrow  \tilde{y}+\tilde{\beta} &\qquad \tilde{z}\rightarrow
\tilde{z}+\tilde{\gamma}
\end{array}
\end{eqnarray}
and we can identify $\cG$ under the action of $\G$ so that the coordinates on $\cX$ are subject to the identifications
\begin{eqnarray}\label{twistedtorusgamma}
(x,y,z,\tilde{x},\tilde{y},\tilde{z})&\sim& (x+1,y+m\tilde{z},z-m\tilde{y},\tilde{x},\tilde{y},\tilde{z}) \nonumber\\
  (x,y,z,\tilde{x},\tilde{y},\tilde{z})&\sim&
(x,y+1,z,\tilde{x},\tilde{y},\tilde{z})\nonumber\\
(x,y,z,\tilde{x},\tilde{y},\tilde{z})&\sim& (x,y,z+1,\tilde{x},\tilde{y},\tilde{z}) \nonumber\\
 (x,y,z,\tilde{x},\tilde{y},\tilde{z})&\sim&
(x,y,z,\tilde{x}+1,\tilde{y},\tilde{z})\nonumber\\
 (x,y,z,\tilde{x},\tilde{y},\tilde{z})&\sim&
(x,y,z,\tilde{x},\tilde{y}+1,\tilde{z})\nonumber\\
(x,y,z,\tilde{x},\tilde{y},\tilde{z})&\sim& (x,y,z,\tilde{x}+m\tilde{y},\tilde{y},\tilde{z}+1)
\end{eqnarray}
We see that, under the identification, the physical coordinates $(y,z)$ mix with the auxiliary coordinates $(\tilde{y},\tilde{z})$ so that,
whilst the polarisation of ${\cal P}^M$ into one-forms corresponding to differing maximally isotropic subgroups of $\cG$ is globally defined,
the polarisation of the coordinates $\mathbb{X}^I$ is not and so the distinction between spacetime coordinates $x^i=(x,y,z)$ and auxiliary
coordinates $\tilde{x}_i=(\tilde{x},\tilde{y},\tilde{z})$ can not be made globally. The generators of the left action $\cG_L$, in this polarisation, are
\begin{eqnarray}
 \widetilde{Z}_x=\frac{\partial}{\partial x}+m\tilde{z}\frac{\partial}{\partial y}-m\tilde{y}\frac{\partial}{\partial z}  \qquad  \widetilde{Z}_y&=&\frac{\partial}{\partial y} \qquad \widetilde{Z}_z=\frac{\partial}{\partial z}
\nonumber\\
\widetilde{X}^x=\frac{\partial}{\partial \tilde{x}} \qquad  \widetilde{X}^y&=&\frac{\partial}{\partial \tilde{y}} \qquad
\widetilde{X}^z=\frac{\partial}{\partial \tilde{z}}+m\tilde{y}\frac{\partial}{\partial \tilde{x}}
\end{eqnarray}
These are not invariant under $\G$ and transform as
\begin{eqnarray}
\begin{array}{ll}
 \widetilde{Z}_x\rightarrow \widetilde{Z}_x+m\tilde{\gamma} \widetilde{Z}_y-m\tilde{\beta} \widetilde{Z}_z &\qquad  \widetilde{X}^x\rightarrow \widetilde{X}^x\\
\widetilde{Z}_y\rightarrow \widetilde{Z}_y &\qquad \widetilde{X}^y\rightarrow \widetilde{X}^y+m\alpha \widetilde{Z}_z-m\tilde{\gamma}\widetilde{X}^x \\
\widetilde{Z}_z\rightarrow \widetilde{Z}_z &\qquad  \widetilde{X}^z\rightarrow \widetilde{X}^z-m\alpha \widetilde{Z}_y+m\tilde{\beta} \widetilde{X}^x
\end{array}\nonumber
\end{eqnarray}
The $\widetilde{X}^m$ generate the Heisenberg group $\widetilde{G}_L\subset\cG_{L}$.  We can identify the cover of the T-fold as the coset $C_T\simeq {\cal G}/\widetilde{G}_L$, but since $\G$ does not preserve $\widetilde{G}_L$,
the quotient $\cX/\widetilde{G}_L$ is not well-defined in a conventional sense. Instead, a patch of the
spacetime is recovered   as a patch of the coset $\cG/\widetilde{G}_L$. As remarked above, there is no global spacetime description of
the T-fold and we must glue these local spacetime descriptions together with the identifications (\ref{twistedtorusgamma}).

The action of $\widetilde{G}_L$ on the coordinates is
$$
\begin{array}{lll}
x\rightarrow x &\qquad y\rightarrow y &\qquad z\rightarrow z \\
\tilde{x}\rightarrow\tilde{x}+m\tilde{\gamma}\tilde{y}+\tilde{\alpha} &\qquad \tilde{y}\rightarrow\tilde{y}+\beta &\qquad \tilde{z}\rightarrow\tilde{z}+\tilde{\gamma}
\end{array}
$$
 {The natural left-invariant one-forms on $\widetilde{G}$ (the three-dimensional Heisenberg group) are}
$$
{\ti \ell}_x=Q_x  \qquad  {\ti \ell}_y=Q_y    \qquad  {\ti \ell}_z=Q_z
$$
Since the group $G=\R^3$ is abelian, the right- and left-invariant forms on $G$ coincide $r^m=\ell^m=dx^m$. The $\cG_L$-invariant one-forms may then be written, as in section 2.6, as ${\cal P}^{\hat{M}}=\Phi^{\hat{N}}{\cal V}_{\hat{N}}{}^{\hat{M}}$ where $\Phi^{\hat{M}}=(dx^m, {\ti \ell}_m)$ and
$$
{\cal V}_{\hat{M}}{}^{\hat{N}}=
               \left(
\begin{array}{cc}
\delta_m{}^n & 0 \\
\beta^{mn} & \delta^m{}_n
                     \end{array}
\right)
$$
where
$$
\beta^{mn}=\left(
\begin{array}{ccc}
0 & 0 & 0 \\
 0 & 0 & mx \\
0 & -mx & 0
                     \end{array}
\right)	
$$
The $\widetilde{G}_L$-invariant metric is then
$$
{\cal H}_{\hat{M}\hat{N}}=\left(
\begin{array}{cccccc}
1 & 0 & 0 & 0 & 0 & 0 \\
0 & 1 & 0 & 0 & 0 & -mx \\
0 & 0 & 1 & 0 & mx & 0 \\
0 & 0 & 0 & 1 & 0 & 0 \\
0 & 0 & mx & 0 & 1+(mx)^2 & 0 \\
0 & -mx & 0 & 0 & 0 & 1+(mx)^2
                     \end{array}
\right)
$$
from which we read off
\begin{equation}\label{bg}
g_{ij}=\frac{1}{1+(mx)^2}\left(
\begin{array}{ccc}
1+(mx)^2 & 0 & 0  \\
0 & 1 & 0  \\
0 & 0 & 1 \\
                     \end{array}
\right)	\qquad	b_{ij}=\frac{mx}{1+(mx)^2}\left(
\begin{array}{ccc}
0 & 0 & 0  \\
0 & 0 & -1  \\
0 & 1 & 0 \\
                     \end{array}
\right)
\end{equation}
 {Using
$$
\frac{1}{2}d\left(r^m\wedge {\ti \ell}_m\right)=-\frac{1}{2}mdx\wedge d\tilde{y}\wedge d\tilde{z}   \qquad  {\cal K}=-mdx\wedge d\tilde{y}\wedge d\tilde{z}
$$
in the expression (\ref{H conjecture}), we find that the $H$-field strength is simply given by $H=db$ with $b$ given in (\ref{bg}).}

\subsubsection{R-Flux}

The doubled twisted torus construction allows us to consider acting with an element  ${\cal O}$ of $O(3,3;\Z)$ to give a new polarisation $\Theta$
$$
{\cal O}=\Theta=\left(%
\begin{array}{cccccc}
  0 & 0 & 0 & 1 & 0 & 0 \\
  0 & 1 & 0 & 0 & 0 & 0 \\
  0 & 0 & 1 & 0 & 0 & 0 \\
  1 & 0 & 0 & 0 & 0 & 0 \\
  0 & 0 & 0 & 0 & 1 & 0 \\
  0 & 0 & 0 & 0 & 0 & 1 \\
\end{array}%
\right)
$$
This change of  polarisation exchanges
\begin{equation}\label{replacement}
Z_x\leftrightarrow X^x  \qquad  x\leftrightarrow \tilde{x}  \qquad P^x\leftrightarrow Q_x
\end{equation}
relative to the T-fold polarisation above. The left-invariant one-forms on $\cX$ are
\begin{eqnarray}\label{R-flux}
\begin{array}{lll}
P^x=dx-m\tilde{z}d\tilde{y}   &\qquad  P^y=dy-m\tilde{x}d\tilde{z}  &\qquad P^z=dz+m\tilde{x}d\tilde{y}  \\
 Q_x=d\tilde{x} &\qquad Q_y=d\tilde{y} &\qquad Q_z=d\tilde{z}
\end{array}
\end{eqnarray}
which describes the local structure of the doubled twisted torus. The global structure of $\cX$ is determined by the rigid left action of the
cocompact group $\G\subset\cG_L$, which acts on the coordinates as
\begin{eqnarray}\label{R-flux gamma}
\begin{array}{lll}
 x\rightarrow  x+m\tilde{\gamma}\tilde{y}+\alpha  &\qquad y\rightarrow  y+m\tilde{\alpha}\tilde{z}+\beta   &\qquad  z\rightarrow  z-m\tilde{\alpha}\tilde{y}+\gamma \\
\tilde{x}\rightarrow  \tilde{x}+\tilde{\alpha} &\qquad \tilde{y}\rightarrow  \tilde{y}+\tilde{\beta}  &\qquad \tilde{z}\rightarrow  \tilde{z}+\tilde{\gamma}
\end{array}
\end{eqnarray}
Identification of $\cG$ under $\G$ requires that the coordinates are subject to the identifications
\begin{eqnarray}
(x,y,z,\tilde{x},\tilde{y},\tilde{z})&\sim& (x+1,y,z,\tilde{x},\tilde{y},\tilde{z}) \nonumber\\
  (x,y,z,\tilde{x},\tilde{y},\tilde{z})&\sim&
(x,y+1,z,\tilde{x},\tilde{y},\tilde{z})\nonumber\\
(x,y,z,\tilde{x},\tilde{y},\tilde{z})&\sim& (x,y,z+1,\tilde{x},\tilde{y},\tilde{z}) \nonumber\\
 (x,y,z,\tilde{x},\tilde{y},\tilde{z})&\sim&
(x,y+m\tilde{z},z-m\tilde{y},\tilde{x}+1,\tilde{y},\tilde{z})\nonumber\\
 (x,y,z,\tilde{x},\tilde{y},\tilde{z})&\sim&
(x,y,z,\tilde{x},\tilde{y}+1,\tilde{z})\nonumber\\
(x,y,z,\tilde{x},\tilde{y},\tilde{z})&\sim& (x+m\tilde{y},y,z,\tilde{x},\tilde{y},\tilde{z}+1)
\end{eqnarray}
The left-acting group $\cG_L$ is generated by
\begin{eqnarray}
 \widetilde{Z}_x=\frac{\partial}{\partial x}  \qquad  \widetilde{Z}_y&=&\frac{\partial}{\partial y} \qquad \widetilde{Z}_z=\frac{\partial}{\partial z}
\nonumber\\
\widetilde{X}^x=\frac{\partial}{\partial \tilde{x}}+m\tilde{z}\frac{\partial}{\partial y}-m\tilde{y}\frac{\partial}{\partial z} \qquad
\widetilde{X}^y&=&\frac{\partial}{\partial \tilde{y}} \qquad \widetilde{X}^z=\frac{\partial}{\partial
\tilde{z}}+m\tilde{y}\frac{\partial}{\partial x}
\end{eqnarray}
which satisfy the commutation relation
$$
[\widetilde{X}^x,\widetilde{X}^z]=-m\widetilde{Z}_y  \qquad  [\widetilde{X}^x,\widetilde{X}^y]=-m\widetilde{Z}_z  \qquad
[\widetilde{X}^z,\widetilde{X}^y]=m\widetilde{Z}_x
$$
where all other commutators vanish. The $\widetilde{X}$'s do not close to form a sub-algebra and so there is no subgroup
$\widetilde{G}_L\subset\cG_L$, generated by $\widetilde{X}^m$, with which we can form a quotient $\cG/\widetilde{G}_L$ or $\cX/\widetilde{G}_L$.
In contrast to the previous polarisations, there is no way to recover a conventional description of spacetime, even locally.  {This may be seen in the action of $\widetilde{G}_L$ on the coordinates}
$$
\begin{array}{lll}
x\rightarrow x+m\tilde{\gamma}\tilde{y} &\qquad y\rightarrow y+m\tilde{\alpha}\tilde{z} &\qquad z\rightarrow z-m\tilde{\alpha}\tilde{y} \\
\tilde{x}\rightarrow\tilde{x}+\tilde{\alpha} &\qquad \tilde{y}\rightarrow\tilde{y}+\tilde{\beta} &\qquad \tilde{z}\rightarrow\tilde{z}+\tilde{\gamma}
\end{array}
$$
 {where we see that $\widetilde{G}_L$ acts on all of the coordinates, not just those we   identify as auxiliary coordinates. As described in section 2.6, it is still possible to write the left-invariant one-forms in the form ${\cal P}^{\hat{M}}=\Phi^{\hat{N}}{\cal V}_{\hat{N}}{}^{\hat{M}}$, where ${\cal V}_{\hat{N}}{}^{\hat{M}}$ is independent of the ${\ti x}_i$ and may be used to define a $\widetilde{G}_L$-invariant metric ${\cal H}_{\hat{M}\hat{N}}$; however, this does not give rise to a ${\ti x}_i$-independent metric and $H$-field strength.}

We  wish to interpret the coordinates $x,y,z$ as  spacetime coordinates and $\tilde x, \tilde y, \tilde z$ as dual coordinates conjugate to winding numbers.
The background fields, in this polarisation,
 depend explicitly on the dual coordinate ${\ti x}$ and so the background is not a conventional  geometry on the  three-dimensional space with coordinates $x,y,z$, and  can not be understood as a conventional spacetime, even locally. However, a dependence of the background fields on the auxiliary coordinates is quite natural in the doubled twisted torus description, and might be expected for general string solutions.

This example with $R$-flux arose from
 the replacement   (\ref{replacement}) of $x$ with $\ti{x}$ in the T-fold generalised metric (\ref{HT}), so that it  {may be viewed as}  a $T^2$ fibration over the dual coordinate $\tilde{x}\sim \mathbb{X}^1$ \cite{Dabholkar ``Generalised T-duality and non-geometric backgrounds''}.
 On doubling the fibres,  the T-fold is represented by a $T^4$  bundle ${\cal T}$ over the $x$-circle, while  the dual
 space is a
 5-dimensional space $\widetilde{\cal T}$ which is a  $T^4$ fibration over the dual circle $\widetilde{S}^1$
\begin{eqnarray}
T^4\hookrightarrow &\widetilde{\cal T}&\nonumber\\
&\downarrow&\nonumber\\
&\widetilde{S}^1&\nonumber
\end{eqnarray}
The coordinates $x$ and ${\ti x}$ appear together in the fully doubled six-dimensional  geometry   $\cX$ and   $\cX$ provides a universal description including both $\cal T$ and $\widetilde{\cal T}$.

 In general, the metric and flux, ${\cal H}$ and ${\cal K}$ on $\cX$, might depend on the dual coordinates $\tilde{x}_i$ as well as the spacetime coordinates $x^i$. When $R^{xab}=0$, and a description of the physics is locally possible in terms of a conventional spacetime, we do not expect the metric and $H$-field strength on the physical background to depend explicitly on the auxiliary coordinates. The fact that the physical background fields are invariant under the action of a group $\widetilde{G}_L$ when $R^{xab}=0$ allows for the $\tilde{x}_i$-dependence in ${\cal H}_{IJ}$ and ${\cal K}$ to be consistently removed and the ${\ti x}_i$-independent metric and $H$-field strength on the physical spacetime to be recovered for a given polarisation as discussed in section 2.6. In general, both the generalised metric ${\cal H}_{IJ}$ and the generalised flux ${\cal K}$ will contribute to the physical background and the fields $g_{ij}(x)$ and $H(x)$ emerge as a $\widetilde{G}_L$-invariant combination of the components of ${\cal H}_{IJ}(\mathbb{X})$ and ${\cal K}(\mathbb{X})$.

 For   $R$-flux examples such as that considered here, the generators $X^m$ do not form a closed subalgebra
 so that there is no  invariance of the physical background fields under a  subgroup
 that can be used to  recover a conventional metric and $H$-field strength, in keeping with the conclusion that such cases do not admit a conventional description as a three-dimensional Riemannian geometry, even locally.

\section{Sigma Model for the Doubled Torus Fibration}

Here, and in the following sections, we shall be interested in studying the doubled torus bundle ${\cal T}$ and the doubled twisted torus
$\cX$ as target spaces for two-dimensional sigma models.
Doubling some or all of the target space dimensions leads to extra degrees of freedom,
but a constraint reduces the degrees of freedom again to give the right content of the theory.
Of particular interest, will be the recovery of a conventional   world-sheet description of the sigma-model with target  the (undoubled) spacetime. We will examine in detail the recovery of the sigma-models for the
nilfold,
T-fold, and the $T^3$ with constant $H$-flux backgrounds from these doubled descriptions. In the next section, we shall see how these backgrounds are recovered from the doubled sigma model introduced in
\cite{Hull ``A geometry for non-geometric string backgrounds''} by an appropriate choice of polarisation of the five-dimensional doubled torus bundle $\mathcal{T}$. In section 6 this construction will be
generalised and we shall introduce a sigma model which describes the embedding of $\Sigma$ into the six-dimensional doubled twisted torus $\cX$.
This new sigma model allows for a description of compactifications with $R$-flux, for which no conventional spacetime description exists.

In this section, the sigma model for the doubled torus fibration is reviewed. We shall assume that the doubled torus fibres are $2d$-dimensional, for
general $d$, and only restrict to the case $d=2$ in the next section where the nilfold, T-fold and $T^3$ with constant $H$-flux examples are
considered explicitly.

\subsection{Non-Linear Sigma Model for a Conventional Torus Bundle}

Before proceeding to the doubled cases, we first review
  the conventional sigma model with   world-sheet, $\Sigma$ and a $(d+k)$-dimensional target space ${\cal N}$ which is a  $d$-dimensional torus fibration over a $k$-dimensional base manifold $M$.
\begin{eqnarray}
T^{d}\hookrightarrow &{\cal N}&\nonumber\\
&\downarrow&\nonumber\\
&M&\nonumber
\end{eqnarray}
We introduce  local coordinates $x^u$ on the base $M$ ($u,v=1,2,...k$ ) and
periodic coordinates $z^a$ on the   torus fibres ($a,b=k+1,k+2,...,k+d$). The metric $g_{ab}$ on the $T^d$ fibre is taken to be independent of the fibre coordinates $z^a$ but in general depends on the base coordinates $x^u$. It is convenient to write the metric on ${\cal N}$ as
\begin{eqnarray}\label{G}
G_{ij}=\left(\begin{array}{cc}
g_{uv}+g_{ab}A^a{}_{u}A^b{}_v  & g_{ac}A^c{}_v \\
g_{bc}A^c{}_u  & g_{ab} \\
\end{array} \right)
\end{eqnarray}
where the one-forms $A^a=A^a{}_udx^u$ are the $U(1)^d$ connections of the $T^d$ fibration   and $g_{uv}$ is a metric on $M$.

The sigma model  is given by the action
\begin{equation}
S_{\cal N}=\frac{1}{2}\oint_{\Sigma}g_{uv}dx^u\wedge *dx^v+\frac{1}{2}\oint_{\Sigma}g_{ab}(dz^a+A^a)\wedge *(dz^b+A^b)
\end{equation}
where $A^a=A^a{}_u\partial_{\alpha}x^ud\sigma^{\alpha}$ now denotes the pull-back of the connection one-form to the   world-sheet $\Sigma$ and $\sigma^{\alpha}=(\tau,\sigma)$ are coordinates on the   world-sheet. The exterior derivative is pulled back to the world-sheet, so that
$d=d\sigma^{\alpha}\partial_{\alpha}$, and we take the   world-sheet metric to be Lorentzian so that $*^2=1$. It is useful to write this action as
\begin{equation}\label{Ngensigma}
S_{\cal N}=\frac{1}{2}\oint_{\Sigma}G_{uv}dx^u\wedge *dx^v+\frac{1}{2}\oint_{\Sigma}g_{ab}dz^a\wedge *dz^b+\oint_{\Sigma}dz^a\wedge*J_a
\end{equation}
where $G_{uv}=g_{uv}+g_{ab}A^a{}_uA^b{}_v$ and $J_a=g_{ab}A^b{}_udx^u$.

Here, we shall be particularly interested in the case where the base $M$ is a circle $S^1_x$ with coordinate $x\sim x+1$ as considered in section 2. The monodromy of the fibration is $(e^f)^a{}_b$ where, in order for the geometry to be smooth, we require that this monodromy is an element of $SL(d;\Z)$ - the mapping class group of the $T^d$ fibres. The $(d+1)$-dimensional target space is then ${\cal N}=G/\G$, as described in section 2.1. The effect of this $SL(d;\Z)$-twist can be captured in the sigma model (\ref{Ngensigma}) by introducing the $G_L$-invariant   world-sheet one-forms
$$
P^a= (e^{fx})^a{}_bd\sigma^{\alpha}\partial_{\alpha}z^b  \qquad
 P^x=d\sigma^{\alpha}\partial_{\alpha}x
$$
The forms $P^m=(P^x,P^a)$ are pull-backs of the one-forms (\ref{forms}) and satisfy the Maurer-Cartan equations
$$
dP^x=0  \qquad  dP^a-f^a{}_bP^x\wedge P^b=0
$$
where $d=d\sigma^{\alpha}\partial_{\alpha}$ is a   world-sheet derivative.
The metric $g_{ab}$ is given in terms of a constant metric $h_{ab}$ on $T^d$
by $g_{ab}(x)=(e^{fx})_a{}^ch_{cd}(e^{fx})^d{}_b$.
The sigma model action is then
\begin{equation}\label{Nsigma}
S_{\cal N}=\frac{1}{2}\oint_{\Sigma}G P^x\wedge *P^x+\frac{1}{2}\oint_{\Sigma}h_{ab}P^a\wedge *P^b+\oint_{\Sigma}P^a\wedge *J_a
\end{equation}
where $G=1+h_{ab}A^a{}_xA^b{}_x$ and $J_a=h_{ab}A^a$. Note that $G$ and $A^a{}_x$ here are both independent of the base coordinate and the only explicit $x$-dependence is contained in $P^a$. The left action of $G$ is a rigid symmetry of the sigma model. If the background has an $H$-flux then we there is  also a Wess-Zumino term. We now consider how the physics of this sigma model, and those for more general monodromies, can be described in the doubled torus formalism.

\subsection{Sigma Model for the Doubled Torus Bundle}

As discussed in section 2, the $T^d$ fibration  with $B$-field over $M$ defines a $T^{2d}$ fibration over the base $M$. The metric $g_{ab}$ and $B$-field $B_{ab}$ of the $T^d$ fibres
specifies a generalised, $x^u$-dependent, metric ${\cal H}_{AB}$ on the $T^{2d}$ fibres of a doubled torus fibration over $M$. In addition, the connection one-form of the $T^d$ fibration $A^a=A^a{}_udx^u$ and the
$B$-field components $B_a=B_{au}dx^u$ determine  a connection ${\cal A}^A={\cal A}^A{}_udx^u$ for the
doubled torus fibration with field strength ${\cal F}^A=d{\cal A}^A$. The sigma model for the doubled torus fibration is the analogue of (\ref{Ngensigma}) for the conventional geometry where instead of the $(d+k)\times (d+k)$ metric $G_{ij}$ given in (\ref{g}) we have the $(2d+k)\times (2d+k)$ metric
$$
G=\left(\begin{array}{cc}
g_{uv}+\frac{1}{2}{\cal H}_{CD}{\cal A}^C{}_{u}{\cal A}^D{}_v  & \frac{1}{2}{\cal H}_{AC}{\cal A}^C{}_{v} \\
\frac{1}{2}{\cal H}_{BC}{\cal A}^C{}_u  & \frac{1}{2}{\cal H}_{AB} \\
\end{array} \right)
$$
The sigma model contains, in addition to a term $S_G$ given by integrating the   world-sheet metric induced by $G$ over $\Sigma$, a Wess-Zumino term
$$
S_{wz}=-\frac{1}{2}\int_VL_{AB}d\mathbb{X}^A\wedge{\cal F}^B=-\frac{1}{2}\oint_{\Sigma}L_{AB}d\mathbb{X}^A\wedge{\cal A}^B
$$
where $V$ is a three-dimensional extension of the   world-sheet such that $\partial V=\Sigma$ and $L_{AB}$ is the invariant of $O(d,d)$. It is also necessary to include a topological term \cite{Hull ``Doubled geometry and T-folds''}
$$
S_{\Omega}=\frac{1}{4}\oint_{\Sigma}\Omega_{AB}d\mathbb{X}^A\wedge d\mathbb{X}^B
$$
where
$$
\Omega=\left(%
\begin{array}{cc}
  0 & -\bid_d \\
  \bid_d & 0 \\
\end{array}%
\right)
$$
The topological $S_{\Omega}$ term does not contribute to the equations of motion but it does play an important role in the quantum
theory \cite{Hull ``Doubled geometry and T-folds''}. There may also be terms in the   world-sheet action corresponding to other  target space dimensions and the target space fields in general may depend on the corresponding coordinates. Such terms and dependencies will play no role in our
analysis and will be suppressed here, although these terms are important in constructing conformally-invariant backgrounds.

The action, $S=S_G+S_{wz}+S_{\Omega}$, for the sigma model on ${\cal T}$ is \cite{Hull ``Doubled geometry and T-folds''}
\begin{eqnarray}\label{Tgensigma}
S_{\cal T}&=&\frac{1}{2}\oint_{\Sigma}G_{uv}dx^u\wedge
*dx^v+\frac{1}{4}\oint_{\Sigma}{\cal H}_{AB}d\mathbb{X}^A\wedge*d\mathbb{X}^B-\frac{1}{2}\oint_{\Sigma}d\mathbb{X}^A\wedge *J_A\nonumber\\
&&+\frac{1}{4}\oint_{\Sigma}\Omega_{AB}d\mathbb{X}^A\wedge d\mathbb{X}^B
\end{eqnarray}
where
\begin{equation}\label{smiley}
J_A={\cal H}_{AB}{\cal A}^B-L_{AB}*{\cal A}^B
\end{equation}
The correct number of physical degrees of freedom is ensured by the imposition of the self-duality constraint \cite{Hull ``A geometry for non-geometric string backgrounds''}
\begin{equation}\label{constraint1}
d\mathbb{X}^A=L^{AB}\left({\cal H}_{BC}*d\mathbb{X}^C+*J_B\right)
\end{equation}
where ${\cal H}_{AB}$ and $J_A$ may depend on the base coordinate $x^u$. This constraint is consistent with the equations of motion of the sigma model (\ref{Tgensigma}).
The constraint can be thought of as imposing that half of the   $\mathbb{X}^A$ are right-moving and half left-moving, with the split into these two sectors varying over the base.

When the base $M$ is a circle $S^1_x$, with coordinate $x\sim x+1$, then the doubled torus bundle is the odd-dimensional twisted torus ${\cal T}\simeq\cG/\G$ discussed in section 2.3. On $\Sigma$ one can then define the $\cG_L$-invariant one-forms
\begin{equation}\label{peas}
P^x=d\sigma^{\alpha}\partial_{\alpha}x		\qquad	 {\cal P}^A=(e^{Nx})^A{}_Bd\sigma^{\alpha}\partial_{\alpha}\mathbb{X}^B
\end{equation}
which are the pull-backs of the $\cG_L$-invariant target space one-forms (\ref{doubled forms}) to the   world-sheet. These one-forms satisfy the pull-backs of the Maurer-Cartan equations
$$
dP^x=0  \qquad  d{\cal P}^A-N^A{}_BP^x\wedge {\cal P}^B=0
$$
where, in contrast to (\ref{doubled forms}), the exterior derivative here is that of the   world-sheet $d=d\sigma^{\alpha}\partial_{\alpha}$. The $x$-dependent doubled metric ${\cal H}_{AB}$ can be written in terms of the $x$-independent doubled metric ${\cal M}_{AB}$ appearing in the  {Lagrangian} (\ref{O(d,d) Lagrangian}) as
$$
{\cal H}_{AB}(x)=(e^{Nx})_A{}^C{\cal M}_{CD}(e^{Nx})^D{}_B
$$
The sigma model describing the embedding of $\Sigma$ into ${\cal T}$ may then be written as
\begin{eqnarray}\label{Ttwistedsigma}
S_{\cal T}&=&\frac{1}{2}\oint_{\Sigma}G P^x\wedge*P^x+\frac{1}{4}\oint_{\Sigma}{\cal M}_{AB}{\cal P}^A\wedge*{\cal P}^B+\frac{1}{2}\oint_{\Sigma}L_{AB}{\cal P}^A\wedge*J_A\nonumber\\
&&+\frac{1}{4}\oint_{\Sigma}\Omega_{AB}d\mathbb{X}^A\wedge d\mathbb{X}^B
\end{eqnarray}
where $G=1+\frac{1}{2}{\cal M}_{AB}{\cal A}^A{}_x{\cal A}^B{}_x$. Written in this way, the rigid invariance of the sigma model action under $\cG_L$ is manifest (note that the variation of the topological term under $\cG_L$ gives a total derivative). We shall see that the invariance of the action under subgroups $\widetilde{G}_L\subset\cG_L$ plays a crucial role in imposing the self-duality constraint (\ref{constraint1}) in the quantum theory. Indeed, the self-duality constraint (\ref{constraint1}) may also be written in a manifestly $\cG_L$-covariant form  \cite{Hull ``Doubled geometry and T-folds''}
\begin{equation}\label{constraint}
({\cal P}^A+{\cal A}^A)=L^{AB}{\cal M}_{BC}*({\cal P}^C+{\cal A}^C)
\end{equation}

\subsection{Polarisations and Constraints}

One may think of the sigma model (\ref{Tgensigma}) as a universal sigma model from which different dual sigma models on $(d+k)$-dimensional target spaces, all described as $T^d$ fibrations over $M$, can be recovered. A conventional sigma model is recovered by
specifying a choice of polarisation, $z^a=\Pi^a{}_A\mathbb{X}^A$, in the target space, as discussed in section 2.3. For locally geometric  backgrounds, only $d$ of the $2d$ $\mathbb{X}^A$ fields correspond to independent physical degrees of freedom and, once a polarisation is specified, those coordinates that are chosen to play the role of the auxiliary $\tilde{z}_a$ may be written in terms of the physical $(x,z^a)$,
provided the $\tilde{z}_a$ only appear through their derivative $d \tilde{z}_a$.
 The precise relationship between the (derivative of the) auxiliary $\tilde{z}_a$ and the physical coordinates $(x,z^a)$ is given by the self-duality constraint (\ref{constraint1}). It is this constraint (\ref{constraint1}) which ensures that, when
a global polarisation can be found, the physical sigma model can be described purely in terms of the local physical space-time coordinates
$x^i=(x,z^a)$ selected by this polarisation.

As explained in section 2.3, the polarisation $\Pi$ may not be globally defined and it may not always be
possible to globally choose which $d$ of the $2d$ fibre coordinates $\mathbb{X}^A$ will be identified as the physical coordinates $z^a$. Over a contractible patch of the base $M$, we can define a polarisation which selects coordinates $z^a$ from the doubled $\mathbb{X}^A$.
This then gives a patch of spacetime with coordinates $(x^u,z^a)$. As explained in section 2.3 and also in 2.4, these patches must be carefully glued together to obtain a global description of the target space.

Of particular interest is the case where $M=S^1_x$ and ${\cal T}=\cG/\G$, as described   in section 2. In this case it is possible to define a constant polarisation on the interval $I_x$, given by  $0\leq x < 1$, of the base. The metric and $B$-field on $I_x\times T^{2d}$ can be extended  to $\R_x\times T^{2d}$ by continuing in $x$, as was done for the five-dimensional doubled torus bundle in section 3.2.1. This gives a covering space ${\cal C}_{\cal T}$ of $\cal T$ in which the identification $(x,\mathbb{X}^A)\sim (x+1, (e^{-N})^A{}_B\mathbb{X}^B)$ is dropped and the cover may be thought of as the coset ${\cal C}_{\cal T}\simeq \cG/\G_{\cal C}$, where $\G_{\cal C}$ is the subgroup of $\G$ which gives the identifications of the $T^{2d}$ coordinates only.
The subgroup  $\widetilde{G}_L\simeq\R^d\subset\cG_L$ generated by $\widetilde{X}^a$
is   preserved by $\G_{\cal C}$,
so that the coset ${\cal C}_{\cal T}/ \widetilde{G}_L$ is well-defined. The self-duality constraint (\ref{constraint}) may be consistently imposed on the coordinates of this cover, eliminating all ${\ti z}_a$ dependence so that the sigma model is written solely in terms of the embedding into the target space directions with coordinates $x$ and $z^a$. The doubled sigma model (\ref{Ttwistedsigma}) then reduces to a sigma model with target space given by ${\cal C}_{\cal T}/ \widetilde{G}_L$ - a $T^d$ fibration over $\R_x$. As described in section 2.3, one may then replace $\R_x$ with $S^1_x$ by identifying the remaining target space coordinates under the full action of $\G$.

We now turn to consider in more detail how the constraint (\ref{constraint}) is imposed on the doubled torus sigma model (\ref{Ttwistedsigma}). In section 5 we shall consider several explicit applications of this formalism.

\subsection{The Classical Theory}

In this section and the next we   concentrate on the $(2d+1)$-dimensional doubled torus bundle ${\cal T}=\cG/\G$  for which {the base of the $T^d$ bundle is} $M=S^1_x$. We first consider the cover ${\cal C}_{\cal T}$ given by replacing $S^1_x$ with $\R_x$. A polarisation $\Pi$ can be globally defined on this cover. Once a polarisation is specified, the metric $g_{ab}$ and $B$-field $B_{ab}$ on the physical $T^d$ can be extracted from the
generalised metric ${\cal M}_{AB}$ (\ref{pol}). It is also useful to define the polarisation of the corresponding $\cG_L$-invariant one-forms
(\ref{peas}) $P^a=\Pi^a{}_A{\cal P}^A$ and $Q_a=\widetilde{\Pi}_{aA}{\cal P}^A$ where
\begin{equation}\label{PandQ}
P^a=(e^{Nx})^a{}_bdz^b+(e^{Nx})^{ab}d\tilde{z}_b \qquad  Q_a=(e^{Nx})_a{}^bd\tilde{z}_b+(e^{Nx})_{ab}dz^b
\end{equation}
The self-duality constraint (\ref{constraint}) may then be written as
\begin{equation}\label{constr}
Q_a=g_{ab}*(P^b+A^b)+B_{ab}P^b+B_{ax}P^x
\end{equation}
where $P^a$ and $Q_a$ are given by (\ref{PandQ}). Note that $\tilde{z}_a$ only appears as the derivative $d\tilde{z}_a$ in (\ref{PandQ}) and (\ref{constr}). In the classical theory, one considers the $\tilde{z}_a$ to be auxiliary fields and eliminates all $d\tilde{z}_a$-dependence in the equations of motion using
(\ref{constr}), leaving the equations of motion written in terms of $x$, $dx$ and $dz^a$ only. The requirement
that the polarisation selects a null space with respect to $L_{AB}$ ensures that the $\tilde{z}_a$ dependence may be completely eliminated from
the equations of motion using the self-duality constraint. If we now impose the identification $x\sim x+1$, in effect replacing $\R_x$ with $S^1_x$ again, we  must consider how the theory in the physical $T^d$ fibres is patched together. As explained in section 2.3, if the polarisation is not globally defined, i.e. if $\widetilde{G}_L$ does not preserve and is not preserved by $\G$, then this local description in terms of the coordinates $x$ and $z^a$ does not extend globally. Imposing the constraint (\ref{constraint}) in the quantum theory is more involved, as we shall now discuss.

\subsection{The Quantum Theory}

Let us consider first the $(2d+1)$-dimensional doubled target space ${\cal C}_{\cal T}$, which may be thought of as a cover of the doubled torus bundle ${\cal T}\simeq \cG/\G$. As seen in section 2, for a given polarisation, a cover of the $(d+1)$-dimensional physical target space is given by the coset ${\cal C}_{\cal T}/\widetilde{G}_L$, where $\widetilde{G}_L\subset\cG_L$ is the subgroup selected by the polarisation $\Pi$. More generally, a sigma model on the coset $H/K$ is obtained by gauging a   $K\subset H$ symmetry of the
sigma model on $H$.
 Applying this to the case here, a conventional sigma model description of the background is then recovered locally by gauging the left-acting abelian isometry group $\widetilde{G}_L\simeq \R^d\subset \cG_L$ (generated by
$\widetilde{X}^a=\Pi^{aA}\widetilde{T}_A$) of the  doubled formalism sigma model  (\ref{Ttwistedsigma}) with target space $\cG$ \cite{Hull ``Doubled geometry and T-folds''}.

Now let us consider the case where the doubled target space is the compact twisted torus ${\cal T}=\cG/\G$. As discussed in section 2, if $\G$ preserves and is preserved by $\widetilde{G}_L$, then the action of $\widetilde{G}_L$ is well-defined on ${\cal T}$ and the sigma model on the physical $T^d$ bundle is recovered globally as the sigma model on the quotient ${\cal T}/\widetilde{G}_L$. This sigma model is given by gauging the $\widetilde{G}_L\simeq\R^d$ subgroup of $\cG$ for the sigma model on ${\cal T}$. If $\widetilde{G}_L$ is not preserved by $\G$, then the physical background is not a $T^d$ bundle, but a T-fold. In this case the sigma model in the patch over $0\leq x<1$ may be given again by gauging the sigma model as described above. A global description of the target space is given by identifying $x\sim x+1$ as discussed in sub-section 2.3.

We now review   how the gauging imposes the constraint (\ref{constraint}) and selects a polarisation, following  \cite{Hull ``Doubled geometry and T-folds''}. We first consider a sigma model with target space ${\cal C}_{\cal T}$, the cover of ${\cal T}$, upon which a global polarisation may be defined. The polarisation selects a
subset of the coordinates $\tilde{z}_a$ which are the auxiliary ones to be eliminated, and a
subgroup $\widetilde{G}_L\subset\cG_L$. The generators
  $\widetilde{X}^a$ of $\widetilde{G}_L$
   act as shifts
  on the auxiliary coordinates $\tilde{z}_a$, $\delta\tilde{z}_a=\epsilon_a$, which are isometries of the doubled metric ${\cal H}$.
  The vector fields $\widetilde{X}^a$ will be well-defined
  on a cover ${\cal C}_{\cal T}$ in general.
 Gauging this isometry requires the introduction of the world-sheet gauge fields, which are one-forms  $C_a=C_{a\alpha}(\tau,\sigma) d\sigma^{\alpha}$, and allowing the parameters to depend
on the   world-sheet coordinates $\epsilon_a=\epsilon_a (\tau,\sigma)$. The one-forms transform under the local symmetry as $\delta C_a=-d\epsilon_a$ so that the derivatives
$$
{\cal D}\mathbb{X}^A= d\mathbb{X}^A+\Pi^{Aa}C_a
$$
are gauge-invariant. The minimal coupling $d\mathbb{X}^A\rightarrow{\cal D}\mathbb{X}^A$ is equivalent to the minimal coupling of the auxiliary fields $d\tilde{z}_a \rightarrow {\cal D}\tilde{z}_a=d\tilde{z}_a+C_a$, and   this allows $d\tilde{z}_a$ to be absorbed into a shift of $C_a$.

For the example considered here, where the bundle ${\cal T}$ is given by a duality-twist (\ref{mass}), it is useful to consider ${\cal C}^A$, related to the one-form $C_a$ by the action of the twist matrix on the projection with the polarisation tensor:
\begin{equation}\label{C}
{\cal C}^A=(e^{Nx})^A{}_B\Pi^{Ba}C_a
\end{equation}
If we then define ${\cal C}^a=\Pi^a{}_A{\cal C}^A$ and ${\cal C}_a=\widetilde{\Pi}_{aA}{\cal C}^A$ we may write
$$
{\cal C}^a=(e^{Nx})^{ab}C_b \qquad  {\cal C}_a=(e^{Nx})_a{}^bC_b
$$
For the choice of polarisation in which $Q^{ab}=\Pi^a{}_AN^A{}_B\Pi^{bB}$ is zero, $Q^{ab}=0$,
 we have ${\cal C}^a=0$ and $P^a$ is left unaltered by the minimal coupling. If $Q^{ab}\neq 0$, then ${\cal C}^a\neq 0$ and both $P^a$ and $Q_a$   receive  minimal coupling corrections.

 The gauged sigma
model is obtained by first introducing the minimal coupling
$$
{\cal P}^A\rightarrow {\cal P}^A+{\cal C}^A =(e^{Nx})^A{}_B{\cal D}\mathbb{X}^B
$$
of the one-forms in the kinetic term of (\ref{Ttwistedsigma}) and then adding the term
\begin{equation}\label{terms}
\frac{1}{2}\oint_{\Sigma}L_{AB}{\cal P}^A\wedge{\cal C}^B
\end{equation}
 {to}  (\ref{Ttwistedsigma}), as shown in \cite{Hull ``Doubled geometry and T-folds''}. The resulting gauged sigma model is
\begin{eqnarray}\label{Tgaugedtwistedsigma}
S_{{\cal C}_{\cal T}/\widetilde{G}_L}&=&\frac{1}{2}\oint_{\Sigma}GP^x\wedge* P^x+\frac{1}{4}\oint_{\Sigma}{\cal M}_{AB}{\cal P}^A\wedge*{\cal P}^B+\frac{1}{2}\oint_{\Sigma}{\cal
C}^A\wedge *{\cal J}_A+\frac{1}{2}\oint_{\Sigma}{\cal P}^A\wedge* J_A\nonumber\\
&&+\frac{1}{4}\oint_{\Sigma}\Omega_{AB}d\mathbb{X}^A\wedge d\mathbb{X}^B+\frac{1}{4}\oint_{\Sigma}{\cal M}_{AB}{\cal C}^A\wedge*{\cal C}^B
\end{eqnarray}
where
\begin{equation}\label{frowny}
{\cal J}_A={\cal M}_{AB}{\cal P}^B-L_{AB}*{\cal P}^B+J_A
\end{equation}
The self-duality constraint (\ref{constraint}) may be written as
$${\cal J}_A=0$$
 The gauging consists of adding the linear term
 $$
 \frac{1}{2}\oint_{\Sigma}{\cal
C}^A\wedge *{\cal J}_A
$$
which, because of the polarisation projector in the definition (\ref{C})  of ${\cal C}^A$, is a coupling to half of the components of  ${\cal J}_A$. Then further terms, including ones quadratic in the gauge field,  are added to obtain gauge invariance.
The   coupling ${\cal
C}^A\wedge *{\cal J}_A$ leads in the quantum theory to a BRST charge that imposes the constraint
${\cal J}_A=0$ on physical states, as
  will be shown in section 4.6.

The action (\ref{Tgaugedtwistedsigma}) can be expanded out by substituting in the expressions (\ref{pol}), (\ref{PandQ}) and (\ref{C}) for the chosen polarisation. By
completing the square in the one-forms $C_a$, one can show that the gauged action splits into two parts
$$
S_{{\cal C}_{\cal T}/\widetilde{G}_L}[x,\mathbb{X}^A,C_a]=S_1[x,z^a]+S_2[\lambda_a]
$$
where $\lambda_a=(e^{Nx})_a{}^b(C_b+d\tilde{z}_b)+...$.
If the polarisation selects a subgroup $\widetilde{G}_L$ that is null with respect to $L_{AB}$, then the Lagrangian for $S_2[\lambda_a]$ is quadratic in $\lambda_a$ and we may perform the
integration over the gauge fields $C_a=(e^{-Nx})_a{}^b\lambda_b+...$ to leave the action $S_1[x,z^a]$.
This action is that of the conventional sigma model embedding into the physical $(d+1)$-dimensional target space ${{\cal C}_{\cal T}/\widetilde{G}_L}$ with coordinates $x^i=(x,z^a)$.
Integrating out
 $C_a$ gives
  a determinant which   contributes to the dilaton term in the action so that the dilaton
of the conventional sigma model $\phi$ is related to that of the doubled sigma model $\Phi$ by\footnote{It is this, T-duality-invariant, dilaton $\Phi$ which plays the role of the string coupling in String Field Theory \cite{Alvarez  ``Is the string coupling constant invariant under T-duality?'',Kugo ``Target space duality as a symmetry of string field theory''}.}
$$
\Phi=\phi-\frac{1}{2}\ln(g(x))
$$
where $g(x)=\det(g_{ab}(x))$.

Here we have worked with the covering space, in which the vector fields $\widetilde{X}^a$  {which generate $\widetilde{G}_L$} are well-defined. If they are well-defined in the quotient, then we can make  the identification $x\sim x+1$ to obtain the theory on the quotient. If this is not the case, there is no global description, and one is led  to working with different polarisations  in different patches, as discussed in \cite{Hull ``A geometry for non-geometric string backgrounds''}.

Different choices of polarisation select different sets of generators $\widetilde{T}_A$ to be identified as the $\widetilde{X}^a$ and therefore a
different embedding $\widetilde{G}_L, \widetilde{G}_L',...$ of the abelian subgroup $\R^d\subset \cG_L$. In this way, different choices of
polarisation lead to different gaugings  and results in recovering sigma models for different backgrounds from the doubled sigma model (\ref{Ttwistedsigma}).

\subsubsection{Example: Recovering ${\cal N}$ From ${\cal T}$}

To see how this works in practice, we consider how the sigma model for a conventional $T^d$ bundle background given by the action (\ref{Nsigma}) is recovered from (\ref{Ttwistedsigma}) given a choice of polarisation. Consider the case discussed in section 2.1 in which the only non-trivial element of the twist matrix is $f^a{}_b=\Pi^a{}_AN^A{}_B\widetilde{\Pi}_b{}^B$,    and   the $B$-field is zero. In this example,  {general} elements $g\in\cG$ and $h\in\G$  {may be written as}
$$
g=\left(
    \begin{array}{ccc}
      (e^{fx})^a{}_b & 0 & z^a \\
      0 & (e^{-fx})_a{}^b & {\ti z}_a \\
      0 & 0 & 1 \\
    \end{array}
  \right)   \qquad  h=\left(
    \begin{array}{ccc}
      (e^{f\alpha})^a{}_b & 0 & \alpha^a \\
      0 & (e^{-f\alpha})_a{}^b & {\ti \alpha}_a \\
      0 & 0 & 1 \\
    \end{array}
  \right)
$$
 This polarisation selects a subgroup $\widetilde{G}_L\simeq\R^d\subset\cG_L$ generated by $\widetilde{X}^a$ where
$$
\widetilde{Z}_x=\frac{\partial}{\partial x}+f^a{}_bz^b\frac{\partial}{\partial z^a}-f_a{}^b{\ti z}_b\frac{\partial}{\partial {\ti z}_a}   \qquad  \widetilde{Z}_a=\frac{\partial}{\partial z^a}   \qquad  \widetilde{X}^a=\frac{\partial}{\partial {\ti z}_a}
$$
These vector fields are not invariant under $\G$, but transform as
$$
\widetilde{Z}_x\rightarrow\widetilde{Z}_x    \qquad  \widetilde{Z}_a\rightarrow (e^{-f\alpha})^b{}_a\widetilde{Z}_b \qquad  \widetilde{X}^a\rightarrow (e^{f\alpha})_b{}^a\widetilde{X}^b
$$
and we see that, as expected, $\widetilde{G}_L$ is preserved by $\G$. The quotient ${\cal N}\simeq {\cal T}/\widetilde{G}_L$ is therefore well-defined.

We introduce the one-forms $C_a$ and their $SL(d;\Z)$-twisted counterparts ${\cal C}^a=0$ and ${\cal C}_a=(e^{-fx})_a{}^bC_b$ and  the background fields
$$
{\cal M}_{\hat{A}\hat{B}}=\left(\begin{array}{cc}
h_{ab} & 0 \\
0 & h^{ab}
\end{array}\right)	\qquad	{\cal A}^{\hat{A}}{}_x=\left(\begin{array}{c}
A^a{}_x \\
0
\end{array}\right)
$$
 {In this polarisation }the currents (\ref{smiley}) and (\ref{frowny}) are  {given by}
$$
J^a=\Pi^{aA}J_A=-*A^a	\qquad	J_a=\widetilde{\Pi}_a{}^AJ_A=h_{ab}A^b\nonumber
$$
$$
{\cal J}^a=\Pi^{aA}{\cal J}_A=h^{ab}Q_b-*(P^a+A^a)	 \qquad	{\cal J}_a=\widetilde{\Pi}_a{}^A{\cal J}_A=h_{ab}(P^b+A^b)-*Q_a\nonumber
$$
where $P^a=(e^{fx})^a{}_bdz^b$ and $Q_a=(e^{-fx})_a{}^bd\tilde{z}_b$.

Using these expressions   for ${\cal M}_{\hat{A}\hat{B}}$, ${\cal A}^{\hat{A}}$, ${\cal P}^{\hat{A}}$, $J_{\hat{A}}$ and ${\cal J}_{\hat{A}}$ in  (\ref{Tgaugedtwistedsigma})   gives the Lagrangian
\begin{eqnarray}
{\cal L}_{{\cal T}/\widetilde{G}_L}&=&\frac{1}{2}\left(1+\frac{1}{2}h_{ab}A^a{}_xA^b{}_x\right)P^x \wedge* P^x +\frac{1}{4}h_{ab}P^a\wedge* P^b +\frac{1}{4}h^{ab}Q_a\wedge* Q_b\nonumber\\
&& +\frac{1}{2}\mathcal{C}_a\wedge* {\cal J}^a +\frac{1}{2}P^a\wedge* J_a+\frac{1}{2}Q_a\wedge* J^a +\frac{1}{2}P^a\wedge Q_a +\frac{1}{4}h^{ab}\mathcal{C}_a\wedge* \mathcal{C}_b\nonumber
\end{eqnarray}
where we have used the fact that $\Omega_{\hat{A}\hat{B}}d\mathbb{X}^{\hat{A}}\wedge d\mathbb{X}^{\hat{A}}=2dz^a\wedge d\tilde{z}_a=2P^a\wedge Q_a$. Completing the square in $C_a$ gives the action $S_{{\cal T}/\widetilde{G}_L}[x,\mathbb{X}^A,C_a]=S_1[x,z^a]+S_2[\lambda_a]$ where
$$
S_1[x,z^a]=\frac{1}{2}\oint_{\Sigma}(1+h_{ab}A^a{}_xA^b{}_x)P^x\wedge * P^x +\frac{1}{2}\oint_{\Sigma}h_{ab}P^a\wedge* P^b +\oint_{\Sigma}P^a\wedge* J_a
$$
is the standard sigma model (\ref{Nsigma}) on the torus bundle ${\cal N}$ given by a $T^d$ fibration over $S^1_x$ with metric in the torus fibres given by $g_{ab}(x)=(e^{fx})_a{}^ch_{cd}(e^{fx})^d{}_b$ and connections $(e^{-fx})^a{}_bA^b$. The action $S_2[\lambda_a]$ is
$$
S_2[\lambda_a]=\frac{1}{4}\oint_{\Sigma}h^{ab}\lambda_a\wedge* \lambda_b
$$
where $\lambda_a=Q_a+\mathcal{C}_a-h_{ab}*(P^b+A^b)$ appears quadratically. The fields $C_a$, not $\lambda_a$, appear in the measure of the path integral and integrating  out $C_a$ gives   a   shift to the dilaton
\begin{equation}\label{dilshift}
\phi\rightarrow\phi-\frac{1}{2}\ln(g(x))
\end{equation}
where $g(x)=\det(g_{ab}(x))$.

\subsection{Gauging, Quotient Spaces and the Self-Duality Constraint}

Gauging reduces the sigma model on $\cal T$ (or ${\cal C}_{\cal T}$) to that on  the quotient ${\cal T}/\widetilde{G}_L$ (or ${\cal C}_{\cal T}/\widetilde{G}_L$).
We now review how  this gauging imposes the constraint (\ref{constraint}), using
   BRST arguments. For simplicity,  consider the case of a trivial bundle where ${\cal A}^A=0$. The gauged action may be written as
\begin{eqnarray}
S_{{\cal C}_{\cal T}/\widetilde{G}_L}&=&\frac{1}{4}\oint_{\Sigma}{\cal H}_{AB}d\mathbb{X}^A\wedge*d\mathbb{X}^B+ \frac{1}{4}\oint_{\Sigma}\Omega_{AB}d\mathbb{X}^A\wedge*d\mathbb{X}^B+\frac{1}{2}\oint_{\Sigma}dx\wedge*dx\nonumber\\
&&+\frac{1}{2}\oint_{\Sigma}C_a\wedge *{\cal J}^a+\frac{1}{4}\oint_{\Sigma}g^{ab}C_a\wedge*C_b
\end{eqnarray}
where the current $\mathcal{J}^a=\Pi^{aA}(e^{-Nx})_A{}^B{\cal J}_B$ and we have written $g^{ab}(x)={\cal H}_{AB}(x)\Pi^{Aa}\Pi^{Bb}$.

The self-duality constraint (\ref{constraint}) is   ${\cal J}_A=0$ and it was shown in \cite{Hull ``Doubled geometry and T-folds''}
that this is implied by the apparently weaker constraint ${\cal J}^a=0$.
We now review how the gauging corresponding to a polarisation $\Pi$ constrains the current ${\cal J}^a$ to vanish in the quantum theory.

It is useful to write the action in   world-sheet light-cone coordinates
$\xi^{\pm}=\tau\pm \sigma$ where $ \partial_{\pm}=(\partial_{\tau}\pm \partial_{\sigma})/2$. The gauged action may then be written as
\begin{eqnarray}\label{lightcone}
S_{{\cal C}_{\cal T}/\widetilde{G}_L}&=&-\frac{1}{2}\oint_{\Sigma}d^2\xi\,\, {\cal H}_{AB}(x)\partial_{+}\mathbb{X}^A\partial_-\mathbb{X}^B -\frac{1}{2}\oint_{\Sigma}d^2\xi\,\,  \Omega_{AB}\partial_{+}\mathbb{X}^A\partial_-\mathbb{X}^B -\oint_{\Sigma}d^2\xi\,\,\partial_{+}x\partial_-x\nonumber\\
&&-\frac{1}{2}\oint_{\Sigma}d^2\xi\,\,\left(C_{-a}{\cal J}_+{}^a+C_{+a}{\cal J}_-{}^a\right)-\frac{1}{2}\oint_{\Sigma}d^2\xi\,\, g^{ab}(x)C_{-a}C_{+b}
\end{eqnarray}
For simplicity, we neglect global issues due to the action of $\G$ and consider the doubled target space of the ungauged sigma model to be the cover, ${\cal C}_{\cal T}$.

The action (\ref{lightcone}) is invariant under the infinitesimal $\widetilde{G}_L$ gauge transformations
$$
\delta_{\epsilon}\mathbb{X}^A=\Pi^{Aa}\epsilon_a	\qquad	 \delta_{\epsilon}C_{a\pm}=-\partial_{\pm}\epsilon	_a \qquad	 \delta_{\epsilon}x=0
$$
We will fix this with the gauge choice $C_{-a}=0$  (strictly speaking, in general one would need to set $C_{-a}$ to a constant modulus and integrate over that modulus).
We introduce a ghost field $c_a$ for these transformations, and an anti-ghost field $b_+{}^a$ and Lagrange multiplier field $\pi_+{}^a$.
The BRST transformations with Grassmann-odd constant parameter $\Lambda$ are then\begin{eqnarray}
\begin{array}{llll}
\delta_Q\mathbb{X}^A=\Lambda\Pi^{Aa}c_a	&\qquad	\delta_Q C_{-\pm a}=-\Lambda\partial_{\pm}c_a &\qquad \delta_Qc_a=0 &  \\
\delta_Q x=0 &\qquad	\delta_Q b_+{}^a=\Lambda\pi_{+}{}^a &\qquad \delta_Q{\pi}_+{}^a=0 &\qquad
\end{array}
\end{eqnarray}
We gauge fix by adding the BRST exact term $S_{Q}$ to the action given by
$$\Lambda S_{Q}=\delta_Q\oint_{\Sigma}d^2\xi\,\, b_+{}^aC_{-a}$$
so that
\begin{equation}
S_{Q}=\oint_{\Sigma}d^2\xi\,\, \pi_+{}^aC_{-a}+\oint_{\Sigma}d^2\xi\,\, b_+{}^a\partial_-c_a
\end{equation}

The Lagrange multiplier field $\pi_+{}^a$ imposes the gauge condition $C_{-a}=0$.
 The equation of motion for $C_{-a}$ gives the on-shell value of $\pi_+{}^a$ as
$$
\pi_+{}^a=\frac{1}{2}{\cal J}_+{}^a+\frac{1}{2}g^{ab}(x)C_{+b}
$$
Integrating out $\pi_+{}^a$, the action is now
\begin{eqnarray}\label{lightcone2}
S_{{\cal C}_{\cal T}/\widetilde{G}_L}&=&-\frac{1}{2}\oint_{\Sigma}d^2\xi\,\,{\cal H}_{AB}\partial_{+}\mathbb{X}^A\partial_-\mathbb{X}^B -\frac{1}{2}\oint_{\Sigma}d^2\xi\,\,\Omega_{AB}\partial_{+}\mathbb{X}^A\partial_-\mathbb{X}^B -\oint_{\Sigma}d^2\xi\,\,\partial_{+}x\partial_-x\nonumber\\
&&-\frac{1}{2}\oint_{\Sigma}d^2\xi\,\, C_{+a}{\cal J}_-{}^a +\oint_{\Sigma}d^2\xi\,\, b_+{}^a\partial_-c_a
\end{eqnarray}
which has the BRST symmetry
\begin{eqnarray}
\begin{array}{lll}
\delta_Q\mathbb{X}=\Lambda\Pi^{Aa}c_a	&\qquad	\delta_Qc_a=0 &\qquad \delta_QC_{+a}=0 \\
\delta_Q x=0 &\qquad	\delta_Q b_+{}^a=\frac{1}{2}\Lambda\left({\cal J}_+{}^a+g^{ab}C_{+b}\right) &\qquad
\end{array}
\end{eqnarray}
where $\pi_+{}^a$ has been replaced by its on-shell value. We see, from (\ref{lightcone2}), that $C_{+a}$ is a Lagrange multiplier field which enforces the constraint ${\cal J}_-{}^a=0$.
 Integrating out $C_{+a}$ in (\ref{lightcone2}) gives the action
\begin{eqnarray}
S_{{\cal C}_{\cal T}/\widetilde{G}_L}&=&-\frac{1}{2}\oint_{\Sigma}d^2\xi\,\,{\cal H}_{AB}\partial_{+}\mathbb{X}^A\partial_-\mathbb{X}^B -\frac{1}{2}\oint_{\Sigma}d^2\xi\,\,\Omega_{AB}\partial_{+}\mathbb{X}^A\partial_-\mathbb{X}^B \nonumber\\&& -\oint_{\Sigma}d^2\xi\,\,\partial_{+}x\partial_-x+\oint_{\Sigma}d^2\xi\,\,b_+{}^a\partial_-c_a
\end{eqnarray}
which has BRST symmetry
\begin{eqnarray}
\begin{array}{ll}
\delta_Q\mathbb{X}=\Lambda\Pi^{Aa}c_a	&\qquad	\delta_Qc_a=0 \\
\delta_Q x=0 &\qquad	\delta_Q b_+{}^a=\frac{1}{2}\Lambda{\cal J}_+{}^a\nonumber
\end{array}
\end{eqnarray}
generated by the BRST   charge
\begin{equation}\label{Q}
Q=\oint J_Q=\frac{1}{2}\oint d\xi^+c_a{\cal J}_+{}^a
\end{equation}
The physical states are the $Q$ cohomology classes
  with ghost number zero. A state of ghost number zero is annihilated by $b_+{}^a$, so that the physical state condition $Q|\Psi\rangle=0$ implies
$$
{\cal J}_+{}^a|\Psi\rangle =0
$$
We see then that the BRST constraints imply the ${\cal J}_+{}^a=0$ on  physical states, while ${\cal J}_-{}^a=0$ is imposed by a Lagrange multiplier.
 This completes the argument that the gauging imposes the constraint  ${\cal J}_\pm{}^a=0$, and this then implies (\ref{constraint}), as shown in \cite{Hull ``Doubled geometry and T-folds''}.

\section{Non-Linear Sigma Model Examples}

In this section we revisit the three locally geometric, three-dimensional, examples discussed in section 3 from the point of view of the
  world-sheet theory described in the previous section. Recall that each of the examples considered in section 3 could be thought of as a $T^2$ fibration over $S^1_x$. The base has coordinate $x\sim x+1$ and the coordinates on the $T^2$ fibres are $z^a=(y,z)$. These backgrounds can be equivalently written in terms of the five-dimensional $T^4$ bundle ${\cal T}$. The non-linear sigma model, describing the embedding of the   world-sheet $\Sigma$ into the
doubled torus bundle ${\cal T}$, is then given by the action (\ref{Ttwistedsigma}). In this section we shall only consider backgrounds in which $A^a{}_x$ and $B_{ax}$ are zero
 so that $J_A=0$. This is done for convenience and the generalisation to more general backgrounds is straightforward. We shall also choose the $x$-independent doubled metric to be ${\cal M}=\bid_{4}$. The $x$-dependent doubled metric and connection are then
$$
{\cal H}_{AB}=(e^{Nx})_A{}^C\delta_{CD}(e^{Nx})^D{}_B	\qquad	{\cal A}^A=0
$$
The gauged sigma model (\ref{Tgaugedtwistedsigma}) is
\begin{eqnarray}\label{six dim gauged sigma}
S&=&\frac{1}{2}\oint_{\Sigma}P^x\wedge* P^x+\frac{1}{4}\oint_{\Sigma}\delta_{AB}{\cal P}^A\wedge*{\cal P}^B+\frac{1}{2}\oint_{\Sigma}{\cal
C}^A\wedge *{\cal J}_A\nonumber\\
&&+\frac{1}{4}\oint_{\Sigma}\Omega_{AB}d\mathbb{X}^A\wedge d\mathbb{X}^B+\frac{1}{4}\oint_{\Sigma}\delta_{AB}{\cal C}^A\wedge*{\cal C}^B
\end{eqnarray}
where
$$
{\cal J}_A=\delta_{AB}{\cal P}^B-L_{AB}*{\cal P}^B	\qquad	{\cal C}^A=(e^{Nx})^A{}_B\Pi^{Ba}C_a
$$
where the two gauge fields $C_a$ are selected by the choice of polarisation. We choose coordinates $\mathbb{X}^A=(\mathbb{X}^1,\mathbb{X}^2,\mathbb{X}^3,\mathbb{X}^4)$ on the torus fibres so that the
twist matrix $N^A{}_B$ and monodromy matrix $e^N$
 can be written as
$$
N^A{}_B=\left(\begin{array}{cccc}
0 & 0 & 0 & 0 \\
-m & 0 & 0 & 0 \\
0 & 0 & 0 & m \\
0 & 0 & 0 & 0
\end{array}\right)	\qquad	(e^N)^A{}_B=\left(\begin{array}{cccc}
1 & 0 & 0 & 0 \\
-m & 1 & 0 & 0 \\
0 & 0 & 1 & m \\
0 & 0 & 0 & 1
\end{array}\right)
$$
where $m\in\Z$.
From (\ref{mass}), the twist matrix $N$ determines the structure constants of the algebra (\ref{O(d,d+16) Lie algebra}).
The left-invariant   world-sheet one-forms $P^x$ and ${\cal P}^A$ are
\begin{eqnarray}
\begin{array}{lll}
P^x=dx &\quad {\cal P}^1=d\mathbb{X}^1-mx\mathbb{X}^2 &\quad {\cal P}^2=d\mathbb{X}^2 \\
 &\quad {\cal P}^3=d\mathbb{X}^3 &\quad {\cal P}^4=d\mathbb{X}^4+mx\mathbb{X}^3
\end{array}
\end{eqnarray}
where  $\mathcal{P}^A=(\mathcal{P}^1,\mathcal{P}^2,\mathcal{P}^3,\mathcal{P}^4)$ are left-invariant one-forms on $\cG$.  {The generators of the left action $\cG_L$ also play an important role and, with this coordinate choice, may be written as}
$$
\widetilde{Z}_x=\frac{\partial}{\partial x}+m\mathbb{X}^2\frac{\partial}{\partial\mathbb{X}^1}-m\mathbb{X}^3\frac{\partial}{\partial\mathbb{X}^4}   \qquad  \widetilde{T}_A=\frac{\partial}{\partial\mathbb{X}^A}
$$

\subsection{Recovering the Nilfold from ${\cal T}$}

Given the coordinate choice on the $T^4$ fibres above, we recover the Nilfold by the choice of polarisation projector $\Pi$ and corresponding polarisation tensor $\Theta=(\Pi,\widetilde{\Pi})$ where
$$
\Pi^a{}_A=\left(\begin{array}{cc}
1 & 0 \\
0 & 1 \\
0 & 0 \\
0 & 0
\end{array} \right)	\qquad	\widetilde{\Pi}_{aA}=\left(\begin{array}{cc}
0 & 0 \\
0 & 0 \\
1 & 0 \\
0 & 1
\end{array} \right)
$$
This means that $ (\mathbb{X}^1,\mathbb{X}^2 )$
are selected as the physical coordinates $y,z$
and
$ ( \mathbb{X}^3,\mathbb{X}^4)$ are selected as the auxiliary coordinates $\ti y,\ti z$,
which we write as $\mathbb{X}^{\hat{A}}=(y,z,{\ti y},{\ti z})$.
From (\ref{O(d,d+16) Lie algebra}) and (\ref{mass}), this   polarisation leads to  the only non-vanishing structure constants of the gauge algebra (\ref{O(d,d+16) Lie algebra}) being $f_{xz}{}^y=\widetilde{\Pi}_{zA}N^A{}_B\Pi^{yB}=m\in\Z$.
 The left-invariant one-forms are  {then}
 \begin{equation}\label{Big P}
\mathcal{P}^A=(P^y,P^z,Q_y,Q_z)
\end{equation}
where
\begin{eqnarray}\label{P}
\begin{array}{lll}
P^x=dx &\quad P^y=dy-mxdz &\quad Q_y=d\tilde{y} \\
 &\quad P^z=dz &\quad Q_z=d\tilde{z}+mxd\tilde{y}
\end{array}
\end{eqnarray}
The polarisation projector which acts as $\widetilde{X}^a=\Pi^{aA}\widetilde{T}_A$, where $\Pi^{aA}=\Pi^a{}_BL^{AB}$ can be written in this basis as
$$
\Pi^{aA}=\left(\begin{array}{cccc}
0 & 0 & 1 & 0 \\
0 & 0 & 0 & 1
\end{array} \right)	\qquad	\widetilde{\Pi}_a{}^{A}=\left(\begin{array}{cccc}
1 & 0 & 0 & 0 \\
0 & 1 & 0 & 0
\end{array} \right)
$$
where $\widetilde{\Pi}_a{}^A=\widetilde{\Pi}_{aB}L^{AB}$ has been included for completeness. The projector $\Pi^{aA}$ selects out the abelian subgroup $\widetilde{G}_L\simeq\R^2\subset\cG_L$ generated by the vector fields $\widetilde{X}^y$ and $\widetilde{X}^z${, so that t}he generators of the left action $\widetilde{T}_A=(\widetilde{T}_1,\widetilde{T}_2,\widetilde{T}_3,\widetilde{T}_4)$ are, in this polarisation, given by $\widetilde{T}_{\hat{A}}=(\widetilde{Z}_y,\widetilde{Z}_z,\widetilde{X}^y,\widetilde{X}^z)$. The vector fields generating $\widetilde{G}_L$ are not globally defined on ${\cal T}$ and under the shift $x\rightarrow x+\alpha$ they transform as
$$
\widetilde{X}^y\rightarrow \widetilde{X}^y+m\alpha\widetilde{X}^z \qquad \widetilde{X}^z\rightarrow \widetilde{X}^z
$$
however, we see that $\G$ preserves the subgroup $\widetilde{G}_L\simeq \R^2\subset\cG_L$ generated by $(\widetilde{X}^y,\widetilde{X}^z)$. The quotient ${\cal T}/\widetilde{G}_L$ is therefore a well-defined submanifold of ${\cal T}$.

As described in the previous section, the sigma model on ${\cal T}/\widetilde{G}_L$ is given by gauging the $\widetilde{G}_L\subset\cG_L$ rigid symmetry of the sigma model (\ref{six dim gauged sigma}). We introduce the
  world-sheet one-forms $C_y$ and $C_z$ and, as described in section 4.5, the duality-twisted gauge fields ${\cal C}^A=(e^{Nx})^A{}_B\Pi^{Ba}C_a$. Using the polarisation projectors and the expression for the monodromy matrix in (\ref{f}) and (\ref{nilodromy}), it is not hard to show that the twisted gauge fields ${\cal C}^A$ and the constraint current ${\cal J}_A$ are written in this polarisation as
\begin{eqnarray}\label{C and J}
{\cal C}^{\hat{A}}=\left(\,0\,,\, 0\,, C_y \,, \,C_z+mxC_y\,\right)	\qquad	{\cal J}_{\hat{A}}=\left(\begin{array}{c}
P^y-*Q_y \\
P^z-*Q_z \\
Q_y-*P^y \\
Q_z-*P^z
\end{array} \right)
\end{eqnarray}
from which it is clear that the vanishing of the current ${\cal J}_{\hat{A}}$ implies $P^y=*Q_y$ and $P^z=*Q_z$.  {As argued in the previous section}, it is enough to show that if ${\cal J}^a=\Pi^{aA}{\cal J}_A$ is constrained to vanish then ${\cal J}_A$ must also vanish. The minimal coupling ${\cal P}^A\rightarrow {\cal P}^A+{\cal C}^A=(e^{Nx})^A{}_B{\cal D}\mathbb{X}^B$ introduces gauge-invariant derivatives
for the dual fibre coordinates;
$$
d\tilde{y}\rightarrow d\tilde{y}+C_y	\qquad	 d\tilde{z}\rightarrow d\tilde{z}+C_z
$$
Substituting (\ref{Big P}) and (\ref{C and J}) into the gauged doubled torus sigma model (\ref{Tgaugedtwistedsigma}) and noting that the topological term may be written as
$$
\frac{1}{4}\Omega_{\hat{A}\hat{B}}d\mathbb{X}^{\hat{A}}\wedge d\mathbb{X}^{\hat{B}}=\frac{1}{2}P^y\wedge Q_y+\frac{1}{2}P^z\wedge Q_z
$$
the Lagrangian of the gauged action can be expanded out to give
\begin{eqnarray}
{\cal L}_{{\cal T}/\widetilde{G}_f}&=&\frac{1}{2}P^x\wedge *P^x+\frac{1}{4}P^y\wedge *P^y+\frac{1}{4}P^z\wedge *P^z
\nonumber\\
&&+\frac{1}{4}Q_y\wedge *Q_y+\frac{1}{4}Q_z\wedge *Q_z +\frac{1}{2}{\cal C}_y\wedge* Q_y+\frac{1}{2}{\cal C}_z\wedge* Q_z +\frac{1}{4}{\cal C}_y\wedge *{\cal C}_y+\frac{1}{4}{\cal C}_z\wedge *{\cal C}_z\nonumber\\
&&-\frac{1}{2}{\cal C}_y\wedge P^y-\frac{1}{2}{\cal C}_z\wedge P^z+\frac{1}{2}P^y\wedge Q_y+\frac{1}{2}P^z\wedge Q_z
\nonumber
\end{eqnarray}
After a little rearrangement, this can be written as
\begin{eqnarray}
{\cal L}_{{\cal T}/\widetilde{G}_f}&=&\frac{1}{2}P^x\wedge *P^x+\frac{1}{4}P^y\wedge *P^y+\frac{1}{4}P^z\wedge *P^z
\nonumber\\
&&+\frac{1}{4}(Q_y+{\cal C}_y)\wedge *(Q_y+{\cal C}_y)+\frac{1}{4}(Q_z+{\cal C}_z)\wedge *(Q_z+{\cal C}_z)\nonumber\\
&&\frac{1}{2} P^y\wedge(Q_y+{\cal C}_y)+\frac{1}{2} P^z\wedge(Q_z+{\cal C}_z)
\nonumber
\end{eqnarray}
Completing the square in $Q_y+{\cal C}_y$ and $Q_z+{\cal C}_z$  gives the action
$$S_{{\cal T}/\widetilde{G}_f}=S_1[x,z^a]+S_2[\lambda_a]$$
 where
$$
S_1[x,z^a]=\frac{1}{2}\oint_{\Sigma}P^x\wedge *P^x+\frac{1}{2}\oint_{\Sigma}P^y\wedge *P^y+\frac{1}{2}\oint_{\Sigma}P^z\wedge *P^z
$$
  is the action for the sigma model with the nilfold as  target space and
$$
S_2[\lambda_a]=\frac{1}{4}\oint_{\Sigma}\lambda_y\wedge *\lambda_y+\frac{1}{4}\oint_{\Sigma}\lambda_z\wedge *\lambda_z
$$
where
$$
\lambda_y=C_y+d\tilde{y}-*dy	\qquad	\lambda_z=C_z+d\tilde{z}+mxC_y+mxd\tilde{y}-*dz
$$
The topological term $S_{\Omega}$ does not contribute to the equations of motion but plays an important role in the quantum theory as it
allowed us to complete the square   and separate the action into two distinct parts without dropping surface terms.
Eliminating the auxiliary fields $C_a$ leaves the action $S_1$ for the nilfold sigma model, as required.

\subsection{Recovering the $T^3$ with $H$-flux background from ${\cal T}$}

Given the coordinate choice on the $T^4$ fibres above, we recover the $T^3$ background with constant $H$-flux by the choice of polarisation projector $\Pi$ and corresponding polarisation tensor $\Theta=(\Pi,\widetilde{\Pi})$, where
\begin{eqnarray}\label{pi2}
\Pi^a{}_A=\left(\begin{array}{cc}
0 & 0 \\
0 & 1 \\
1 & 0 \\
0 & 0
\end{array} \right)	\qquad	\widetilde{\Pi}_{aA}=\left(\begin{array}{cc}
1 & 0 \\
0 & 0 \\
0 & 0 \\
0 & 1
\end{array} \right)
\end{eqnarray}
This means that $ (\mathbb{X}^3,\mathbb{X}^2 )$
are selected as the physical coordinates $(y,z)$
and
$ ( \mathbb{X}^1,\mathbb{X}^4)$ are selected as the auxiliary coordinates $(\ti y,\ti z)$, so that
 $\mathbb{X}^{ {A}}=({\ti y},z,y,{\ti z})$ and the corresponding  one-forms are $\mathcal{P}^{ {A}}=(Q_y,P^z,P^y,Q_z)$ respectively.
From (\ref{mass}) and (\ref{pi2}), this   polarisation leads to  the only non-vanishing structure constants of the gauge algebra (\ref{O(d,d+16) Lie algebra}) being $K_{xyz}=\widetilde{\Pi}_{yA}N^A{}_B\widetilde{\Pi}_z{}^B=m\in\Z$, and   the left-invariant one-forms may be written
\begin{eqnarray}\label{P2}
\begin{array}{lll}
P^x=dx &\quad P^y=dy &\quad P^z=dz \\
 &\quad Q_y=d\tilde{y}-mxdz &\quad Q_z=d\tilde{z}+mxdy
\end{array}
\end{eqnarray}
The generators of the left action $\widetilde{T}_A=(\widetilde{T}_1,\widetilde{T}_2,\widetilde{T}_3,\widetilde{T}_4)$ are, in this polarisation, given by $\widetilde{T}_{\hat{A}}=(\widetilde{X}^y,\widetilde{Z}_z,\widetilde{Z}_y,\widetilde{X}^z)$. The polarisation projectors $\Pi^{aA}$ and $\widetilde{\Pi}_a{}^A$ can be written in this basis as
$$
\Pi^{aA}=\left(\begin{array}{cccc}
1 & 0 & 0 & 0 \\
0 & 0 & 0 & 1
\end{array} \right)	\qquad	\widetilde{\Pi}_a{}^{A}=\left(\begin{array}{cccc}
0 & 0 & 1 & 0 \\
0 & 1 & 0 & 0
\end{array} \right)
$$
The projector $\Pi^{aA}$ selects out the abelian subgroup $\widetilde{G}_L\simeq\R^2\subset\cG$ generated by the vector fields $\widetilde{X}^y$ and $\widetilde{X}^z$. These vector fields are globally defined on ${\cal T}$ and so the quotient ${\cal T}/\widetilde{G}_L$ is therefore a well-defined sub-manifold of ${\cal T}$.

The sigma model on ${\cal T}/\widetilde{G}_L$ is given by gauging the $\widetilde{G}_L\subset\cG_L$ rigid symmetry of the sigma model (\ref{six dim gauged sigma}). We introduce the
  world-sheet one-forms $C_y$ and $C_z$ and, as described in section 4.5, the duality-twisted gauge fields ${\cal C}^A=(e^{Nx})^A{}_B\Pi^{Ba}C_a$. The minimal coupling introduces gauge-invariant derivatives
for the dual fibre coordinates;
$$
d\tilde{y}\rightarrow d\tilde{y}+C_y	\qquad	 d\tilde{z}\rightarrow d\tilde{z}+C_z
$$
The twisted gauge fields are written in this polarisation as
\begin{eqnarray}\label{C2}
{\cal C}^{\hat{A}}=\left(\,
0\, ,\, 0\, , C_y , C_z \right)
\end{eqnarray}
Substituting (\ref{P2}) and (\ref{C2}) into the gauged doubled torus sigma model (\ref{Tgaugedtwistedsigma}) and noting that the topological term may be written as
$$
\frac{1}{4}\Omega_{\hat{A}\hat{B}}d\mathbb{X}^{\hat{A}}\wedge d\mathbb{X}^{\hat{B}}=\frac{1}{2}P^y\wedge Q_y+\frac{1}{2}P^z\wedge Q_z+mxdy \wedge dz
$$
the Lagrangian of the gauged action can be expanded out to give
\begin{eqnarray}
{\cal L}_{{\cal T}/\widetilde{G}_K}&=&\frac{1}{2}P^x\wedge *P^x+\frac{1}{4}P^y\wedge *P^y+\frac{1}{4}P^z\wedge *P^z+\frac{1}{2} P^y\wedge(Q_y+{\cal C}_y)+\frac{1}{2} P^z\wedge(Q_z+{\cal C}_z)
\nonumber\\
&&+\frac{1}{4}(Q_y+{\cal C}_y)\wedge *(Q_y+{\cal C}_y)+\frac{1}{4}(Q_z+{\cal C}_z)\wedge *(Q_z+{\cal C}_z)+mx dy \wedge dz
\nonumber
\end{eqnarray}
If we now complete the square in $Q_y+{\cal C}_y$ and $Q_z+{\cal C}_z$ the action splits in two $S_{{\cal T}/\widetilde{G}_K}=S_1[x,z^a]+S_2[\lambda_a]$ where
$$
S_1[x,z^a]=\frac{1}{2}\oint_{\Sigma}dx\wedge *dx+\frac{1}{2}\oint_{\Sigma}dy\wedge *dy+\frac{1}{2}\oint_{\Sigma}dz\wedge *dz+\int_Vmdx\wedge dy\wedge dz
$$
The term involving the  $B$-field $B=mxdy\wedge dz$ has been written as a three-dimensional integral of the $H$-field strength $H=mdx\wedge dy\wedge dz$, pulled back to a three-dimensional extension of the   world-sheet $V$ where $\partial V=\Sigma$. $S_2[\lambda_a]$ is given by
$$
S_2[\lambda_a]=\frac{1}{4}\oint_{\Sigma}\lambda_y\wedge
*\lambda_y+\frac{1}{4}\oint_{\Sigma}\lambda_z\wedge
*\lambda_z
$$
where
$$
\lambda_y=Q_y+C_y-*P^y  \qquad  \lambda_z=Q_z+C_z-*P^z
$$
Eliminating the auxiliary fields $C$ leaves the action $S_1$ for the   sigma model whose target space is a $T^3$ with constant flux
$H=mdx\wedge dy\wedge dz$, as required.

\subsection{Recovering the T-Fold from ${\cal T}$}

We recover the T-fold background by the choice of polarisation projector $\Pi$ and corresponding polarisation tensor $\Theta=(\Pi,\widetilde{\Pi})$
\begin{eqnarray}\label{pi3}
\Pi^a{}_A=\left(\begin{array}{cc}
1 & 0 \\
0 & 0 \\
0 & 0 \\
0 & 1
\end{array} \right)	\qquad	\widetilde{\Pi}_{aA}=\left(\begin{array}{cc}
0 & 0 \\
0 & 1 \\
1 & 0 \\
0 & 0
\end{array} \right)
\end{eqnarray}
We will work with the covering space of the T-fold, so that $x$ is for the moment regarded as non-compact.
This means that $ (\mathbb{X}^1,\mathbb{X}^4 )$
are selected as the physical coordinates $(y,z)$, so
 $\mathbb{X}^{ {A}}=(y,{\ti z},{\ti y},z)$ while $\mathcal{P}^{ {A}}=(P^y,Q_z,Q_y,P^z)$.
From (\ref{mass}) and (\ref{pi3}), this   polarisation leads to  the only non-vanishing structure constants of the gauge algebra (\ref{O(d,d+16) Lie algebra}) being   $Q_x{}^{yz}=\Pi^y{}_AN^A{}_B\Pi^{zB}=m\in\Z$.
The left-invariant one-forms may be written as
\begin{eqnarray}\label{P}
\begin{array}{lll}
P^x=dx &\quad P^y=dy-mxd\tilde{z} &\quad P^z=dz+mxd\tilde{y} \\
 &\quad Q_y=d\tilde{y} &\quad Q_z=d\tilde{z}
\end{array}
\end{eqnarray}
The generators of the left action $\widetilde{T}_A=(\widetilde{T}_1,\widetilde{T}_4,\widetilde{T}_3,\widetilde{T}_4)$ are, in this polarisation, given by $\widetilde{T}_{ {A}}=(\widetilde{Z}_y,\widetilde{X}^z,\widetilde{X}^y,\widetilde{Z}_z)$. The polarisation projectors can be written in this basis as
$$
\Pi^{aA}=\left(\begin{array}{cccc}
1 & 0 & 0 & 0 \\
0 & 0 & 0 & 1
\end{array} \right)	\qquad	\widetilde{\Pi}^{aA}=\left(\begin{array}{cccc}
0 & 0 & 1 & 0 \\
0 & 1 & 0 & 0
\end{array} \right)
$$
The projector $\Pi^{aA}$ selects out the abelian subgroup $\widetilde{G}_L\simeq\R^2\subset\cG$ generated by the vector fields $\widetilde{X}^y$ and $\widetilde{X}^z$.
These vector fields are not well-defined under the shift $x\rightarrow x+\alpha$ and transform as
$$
\widetilde{X}^y\rightarrow \widetilde{X}^y+m\alpha\widetilde{Z}^z \qquad \widetilde{X}^z\rightarrow \widetilde{X}^z-m\alpha\widetilde{Z}^y
$$
and $\G$ does not preserve the subgroup $\widetilde{G}_L$ generated by $(\widetilde{X}^y,\widetilde{X}^z)$. These generators are well-defined on the cover ${\cal C}_{\cal T}\simeq\cG/\G_{\cal C}$, where $\G_{{\cal C}}$ is the subgroup of $\G$ which leaves $x$ invariant, and so we consider the sigma model with target space ${\cal C}_{\cal T}$ initially here. As before, it is useful to introduce the left-invariant gauge one-forms ${\cal
C}^A=(e^{Nx})^A{}_B\Pi^{Ba}C_a$, where here we have
$$
{\cal C}^{\hat{A}}=\left(\,-mxC_z \,,\, mxC_y \,,\, C_y \,,\, C_z\,\right)
$$
Introducing the gauge-invariant derivatives ${\cal D}y=dy+C_y$ and ${\cal D}z=dz+C_z$, the minimal coupling ${\cal P}^A\rightarrow {\cal P}^A+{\cal C}^A$ may then be written as
\begin{eqnarray}
\begin{array}{ll}
  P^y\rightarrow dy-mx\mathcal{D}\tilde{z} &\qquad Q_y\rightarrow \mathcal{D}\tilde{y} \\
  P^z\rightarrow dz+mx\mathcal{D}\tilde{y} &\qquad Q_z\rightarrow \mathcal{D}\tilde{z}
\end{array}
\end{eqnarray}
The Lagrangian for the $\widetilde{G}_L$-gauging of the sigma model on the cover ${\cal C}_{\cal T}$ is given by the Lagrangian
\begin{eqnarray}
{\cal L}_{{\cal C}_{\cal T}/\widetilde{G}_Q}&=&\frac{1}{2}dx\wedge *dx+\frac{1}{4}dy\wedge *dy +\frac{1}{4}dz\wedge*dz\nonumber\\
&&+\frac{1}{4}\left(1+(mx)^2\right){\cal D}\tilde{y}\wedge*{\cal D}\tilde{y}+\frac{1}{4}\left(1+(mx)^2\right){\cal D}\tilde{z}\wedge*{\cal D}\tilde{z}\nonumber\\
&&-\frac{1}{2}{\cal D}\tilde{y}\wedge (dy-mx*dz) -\frac{1}{2}{\cal D}\tilde{z}\wedge(dz+mx*dy)
\end{eqnarray}
Completing the square in $C_y$ and $C_z$ as before, the gauged theory may be written as
\begin{eqnarray}
{\cal L}_{{\cal C}_{\cal T}/\widetilde{G}_Q}&=&\frac{1}{2}dx\wedge *dx +\frac{1}{2(1+(mx)^2)}\left(dy\wedge *dy+dz\wedge
*dz\right)\nonumber\\
&&+\frac{mx}{1+(mx)^2}dy\wedge dz +\frac{1}{4}(1+(mx)^2)\left(\lambda_y\wedge *\lambda_y+\lambda_z\wedge *\lambda_z\right)
\end{eqnarray}
where
$$
\lambda_y=C_y+Q_y-\frac{1}{1+(mx)^2}\left(*dy-mxdz\right)\qquad
\lambda_z=C_z+Q_z-\frac{1}{1+(mx)^2}\left(*dz+mxdy\right)
$$
so we see that the action splits into two parts $S_{{\cal C}_{\cal T}}=S_1[x,z^a]+S_2[\lambda_a]$ where
$$
S_1[x,z^a]=\frac{1}{2}\oint_{\Sigma}dx\wedge *dx+\frac{1}{2}\oint_{\Sigma}g_{ab}dz^a\wedge *dz^b +\frac{1}{2}\oint_{\Sigma}\hat{B}_{ab}dz^a\wedge dz^b
$$
The metric and $B$-field on the $T^2$ fibres are
$$
g_{ab}=\frac{1}{1+(mx)^2}\left(\begin{array}{cc}
 1& 0 \\
  0 &  1
\end{array}\right)	\qquad	 B_{ab}=\frac{mx}{1+(mx)^2}\left(\begin{array}{cc}
 0 & 1 \\
 -1 & 0
\end{array}\right)
$$
which is the background (\ref{gb}), and
$$
S_2[\lambda_a]=\frac{1}{4}\oint_{\Sigma}(1+(mx)^2)\delta^{ab}\lambda_a\wedge *\lambda_b
$$
The $C_a$ are again auxiliary fields that can be eliminated in the classical theory.
In the quantum theory, the Jacobean between the $\lambda_a$ and the $C_a$ is trivial but the integration over the $\lambda_a$ gives a non-trivial $x$-dependent
shift of the dilaton due to the factor of $(1+(mx)^2)/4$ in front of the $\lambda_a$ terms. The correction to the dilaton is
$$
\phi\rightarrow\Phi=\phi-\ln(1+(mx)^2)
$$
The result is the sigma model with action $S_1$ plus a dilaton term, so that the  target space   is a $T^2$ fibration over the line $\R_x$ - a cover of the T-fold. The conventional T-fold background is
recovered by the identification $x\sim x+1$ so that $\R_x\rightarrow S^1_x$ as described in section 3.

\section{Worldsheet Theory for the Doubled Twisted Torus}

As discussed in sections 2.4, 3.3 and  \cite{Hull ``Gauge Symmetry T-Duality and Doubled Geometry''},
 we propose to extend the doubled torus
construction  for the models of sections 2.4 and 3.3 by introducing an additional direction with coordinate $\tilde{x}$ that is conjugate to the winding number on the $x$-circle.
This then gives a full geometric interpretation to the gauge algebra $(\ref{O(d,d+16) Lie algebra})$
as the generators then all act geometrically on the enlarged space.
From the group-theoretic point of view, this extension is natural and we shall see that the models of section 3.1 and section 5 are recovered. However, this extended formalism also suggests a formulation of models that have non-trivial R-flux that might arise from the action of a generalised T-duality of the kind proposed in \cite{Dabholkar ``Generalised T-duality and non-geometric backgrounds''}.
In this section, we discuss the   world-sheet theory for  the sigma-model whose target is this doubled space, and the constraint that halves the doubled degrees of freedom and  allows the conventional formulation to be recovered, at least for the locally geometric backgrounds. However, it also leads to a formulation on backgrounds that are not even locally geometric.

  We represent the
Lie algebra (\ref{O(d,d+16) Lie algebra})  as acting on the $2(d+1)$ coordinates $(x, \ti x, \mathbb{X}^A)$ of $\cX$, where $\mathbb{X}^A$ are the
coordinates on the doubled torus fibre $T^{2d}$, as
\begin{eqnarray}\label{left generators}
Z_x=\frac{\partial}{\partial x}+N^A{}_B\mathbb{X}^B\frac{\partial}{\partial\mathbb{X}^A}   \qquad X^x=\frac{\partial}{\partial{\ti x}} \qquad T_A=\frac{\partial}{\partial\mathbb{X}^A}-\frac{1}{2}N_{AB}\mathbb{X}^B\frac{\partial}{\partial{\ti x}}
\end{eqnarray}
The one-forms dual to these  {left-invariant} vector fields   satisfy the Maurer-Cartan equations
\begin{equation}
d{\cal P}^A-N^A{}_BP^x\wedge {\cal P}^B=0   \qquad \qquad   dQ_x-\frac{1}{2}N_{AB}{\cal P}^A\wedge{\cal P}^B=0  \qquad\qquad    dP^x=0
\end{equation}
which are solved by
\begin{equation}
{\cal P}^A=\left(e^{Nx}\right)^A{}_Bd\mathbb{X}^B   \qquad \qquad   Q_x=d\tilde{x}+\frac{1}{2}N_{AB}\mathbb{X}^Ad\mathbb{X}^B  \qquad\qquad
P^x=dx
\end{equation}
It is useful to define ${\cal P}^M={\cal P}^M{}_Id\mathbb{X}^I$ as the one-forms on $\cX$ satisfying  the  Maurer-Cartan equations
\begin{eqnarray}\label{bianchi}
d{\cal P}^M+\frac{1}{2}t_{NP}{}^M{\cal P}^N\wedge{\cal P}^P=0
\end{eqnarray}
where $t _{xB}{}^A=-N^A{}_B$ and $t_{x[AB]}=-N_{AB}$.

The global identifications of the $z^a$, $\tilde{z}_a$ and $x$ coordinates are fixed by identification with the doubled torus formalism. The
global identification of the $\tilde{x}$ coordinate remains to be determined. From a comparison with the case $t_{MN}{}^P=0$, where
$\cX=T^{2(d+1)}$, in which we know that the radius of $\tilde{x}$ is the inverse to that of $x$ (in appropriate units), we expect the entire space
$\cX$ to be compact.

More generally we   consider a general $2D$-dimensional twisted torus $\cX=\cG/\G$ for a group with Lie algebra
\begin{equation}\label{Bigalg}
[T_M,T_N]=t_{MN}{}^PT_P
\end{equation}
which are not necessarily of the form $t _{xB}{}^A=-N^A{}_B$ and $t_{x[AB]}=-N_{AB}$. For example, a conventional compactification on the
$D$-dimensional twisted torus ${\cal N}=G/\G'$, where $\G'$ is a cocompact subgroup of $G$ and $G$ has Lie algebra
$$
[Z_m,Z_n]=-f_{mn}{}^pZ_p
$$
gives rise to a doubled group $\cG=G\ltimes \R^d$ where $t _{mn}{}^p=-f_{mn}{}^p$. Unless the twisted torus ${\cal N}$ is a torus
bundle the algebra of the doubled group will not be of the form (\ref{nilfold algebra}). Furthermore, if a left-invariant $H$-flux, $K_{mnp}$, is also included in
the reduction then the algebra is deformed further and the structure constants for the algebra are $t _{mn}{}^p=-f_{mn}{}^p$ and
$t_{mnp}=K_{mnp}$ so that \cite{Hull ``Flux compactifications of string theory on twisted tori''}
\begin{equation}\label{Kflux}
[Z_m,Z_n]=-f_{mn}{}^pZ_p+K_{mnp}X^p	\qquad	[Z_m,X^n]=-f_{mp}{}^nX^p	\qquad	[X^m,X^n]=0
\end{equation}
Backgrounds that are not torus bundles  cannot be described by the doubled torus bundle ${\cal T}$.  However,
they can be incorporated into  a doubled twisted torus $\cX$.

\subsection{Non-Linear Sigma-Model for the Doubled Twisted Torus}

\noindent The Action describing the embedding of a closed string   world-sheet $\Sigma$ into the target space $\cX$ is
\begin{eqnarray}\label{fibre action}
S_{\cX}&=&\frac{1}{4}\oint_{\Sigma}d^2\sigma\sqrt{h}h^{\alpha\beta}{\cal H}_{IJ}\partial_{\alpha}\mathbb{X}^I\partial_{\beta}\mathbb{X}^J +\frac{1}{12}\int_Vd^3\sigma'\varepsilon^{\alpha'\beta'\gamma'}{\cal K}_{IJK}\partial_{\alpha'}\mathbb{X}^I\partial_{\beta'}\mathbb{X}^J\partial_{\gamma'}\mathbb{X}^K\nonumber\\
&&+\frac{1}{2\pi}\oint_{\Sigma}d^2\sigma\sqrt{h}\phi R(h)
\end{eqnarray}
where $V$ is an extension of the   world-sheet, with coordinates $\sigma^{\alpha'}$, such that $\partial V=\Sigma$. We shall choose a gauge in which the   world-sheet metric $h_{\alpha\beta}$ is flat and Lorentzian and so the  {  world-sheet} Ricci scalar
$R(h)$ is zero and the   world-sheet Hodge star is an almost product structure $*^2=+1$. The metric ${\cal H}_{IJ}={\cal H}_{IJ}(\mathbb{X})$ and Wess-Zumino
field strength are given by
$$
{\cal H}_{IJ}={\cal M}_{MN}{\cal P}^M{}_I{\cal P}^N{}_J    \qquad  {\cal K}_{IJK}=t_{MNP}{\cal P}^M{}_I{\cal P}^N{}_J{\cal P}^P{}_K
$$
so that the line element and three-form on the twisted torus $\cX$ may be written as
$$
ds_{\cX}^2={\cal M}_{MN}{\cal P}^M\otimes {\cal P}^N    \qquad  {\cal K}=\frac{1}{6}t_{MNP}{\cal P}^M\wedge {\cal P}^N\wedge{\cal
P}^P
$$
where ${\cal P}={\cal G}^{-1}d{\cal G}$ are the left-invariant one-forms, $\mathcal{M}_{MN}$ takes values in the coset $O(D,D)/O(D)\times
O(D)$ and is taken to be independent of $\mathbb{X}^I$ and $t_{MNP}=L_{MQ}t_{NP}{}^Q$ are the structure constants for the Lie algebra (\ref{Bigalg}). We
can write the Wess-Zumino field strength as ${\cal K}_{IJK}=t_{MNP}{\cal P}^M{}_I{\cal P}^N{}_J{\cal P}^P{}_K=t_{MNP}\widetilde{{\cal
P}}^M{}_I\widetilde{{\cal P}}^N{}_J\widetilde{{\cal P}}^P{}_K$, where $\widetilde{\cal P}=d\cG\cG^{-1}$ is the right-invariant one-form for the
group $\cG$. We see then that the sigma model has a manifest, left-acting $\cG_L$ symmetry.  {The Wess-Zumino term is invariant under
$\cG_L\times \cG_R$, but the kinetic term which includes the metric ${\cal H}_{IJ}(\mathbb{X})$ is only invariant under $\cG_L$.} We recall that, on the twisted torus $\cX=\cG/\G$, only that
subgroup of $\cG_L$ which is preserved by $\G$ will have a well-defined action. Note also that the Wess-Zumino three-form ${\cal K}$ satisfies $d{\cal K}=0$ by
virtue of the Jacobi identity $t_{[MN}{}^Qt_{P]Q}{}^T=0$. An open string version of this theory is considered in \cite{Albertsson:2008gq}.

\subsubsection{The Constraint}

\noindent  The model has double the required degrees of freedom, so we seek a generalisation of the constraint (\ref{constraint}) to halve these degrees of freedom to leave the correct number.
Under infinitesimal variations in $\mathbb{X}^I$, the  {left-invariant }one-forms change as
\begin{equation}
\delta {\cal P}^M={\cal P}^M{}_Id(\delta\mathbb{X}^I)+(\partial_J{\cal P}^M{}_I)\delta\mathbb{X}^J d\mathbb{X}^I
\end{equation}
The equations of motion  {of the action (\ref{fibre action}) }are then given by
\begin{eqnarray}\label{eom}
d*{\cal M}_{MN}{\cal P}^N+\mathcal{M}_{NP}t_{MQ}{}^P{\cal P}^Q\wedge *{\cal P}^N+L_{MN}d{\cal P}^N=0
\end{eqnarray}
The equations of motion (\ref{eom}) and Maurer-Cartan equations
 (\ref{bianchi}) are consistent with $d({\cal P}^M-L^{MN}{\cal M}_{NP}*{\cal P}^P)=0$.
We shall then impose the   constraint
\begin{equation}\label{constraintA}
{\cal P}^M=L^{MN}{\cal M}_{NP}*{\cal P}^P
\end{equation}
generalising (\ref{constraint}).

\subsubsection{The Constraint from Gauging}

\noindent From section 2.4, the conventional spacetime is recovered locally from the doubled twisted torus as a patch of the coset $\cG/\widetilde{G}_L$ where
$\widetilde{G}\subset\cG_L$ is a left acting subgroup that is also
maximally isotropic (i.e.   the Lie-subalgebra is a maximally null subspace of the Lie algebra of $\widetilde{G}$
with respect to the metric $L_{MN}$ of signature $(D,D)$).
A non-linear sigma model with target space $\cG/\widetilde{G}_L$ is obtained by gauging the left-acting $\widetilde{G}_L\subset\cG_L$ isometry subgroup of a non-linear sigma
model for the target space $\cG$. The sigma model
\begin{eqnarray}\label{action}
S_{\cG}&=&\frac{1}{4}\oint_{\Sigma}{\cal H}_{IJ}d\mathbb{X}^I\wedge*d\mathbb{X}^J +\frac{1}{12}\int_V{\cal K}_{IJK}d\mathbb{X}^I\wedge d\mathbb{X}^J\wedge
d\mathbb{X}^K
\end{eqnarray}
  has rigid $\cG_L$ symmetry, generated by the vector field
$$
\widetilde{T}_M=(\widetilde{{\cal P}}^{-1})_M{}^I\frac{\partial}{\partial \mathbb{X}^I}
$$
We shall be interested in gauging the null subgroup $\widetilde{G}$, which acts as $\cG\rightarrow \tilde{g}\cG$ for ${\ti g}\in\widetilde{G}$.
$\widetilde{G}_L$ is generated by the vector field $\widetilde{X}^m=\Pi^{mM}\widetilde{T}_M$ so that
$$
\widetilde{X}^m=\Pi^{mM}(\widetilde{{\cal P}}^{-1})_M{}^I\frac{\partial}{\partial \mathbb{X}^I}
$$
Suppose for now that $R^{mnp}=\Pi^{mM}\Pi^{nN}\Pi^{pP}t_{MNP}=0$ so that the $\widetilde{X}^m$ generate a group $\widetilde{G}_L$ with Lie algebra
$$
[\widetilde{X}^m,\widetilde{X}^n]=-f^{mn}{}_p\widetilde{X}^p
$$
We will return to the case when $R\ne 0$ later.
Under the action of the isometry the embedding fields transform infinitesimally as
$$
\delta\mathbb{X}^I=\epsilon_m\widetilde{X}^{m}\mathbb{X}^I=\Pi^{mM}(\widetilde{{\cal P}}^{-1})_M{}^I\epsilon_m
$$
where the parameter is now local, $\epsilon\rightarrow\epsilon(\tau,\sigma)$. We introduce Lie algebra valued one-forms $C_m$ which transform as connections
under the gauge symmetry
\begin{equation}\label{C trans}
\delta C_m=d\epsilon_m-f^{np}{}_m\epsilon_pC_n
\end{equation}
and the covariant derivatives
$$
\mathcal{D}\mathbb{X}^I=d\mathbb{X}^I+\widetilde{X}^mC_m\mathbb{X}^I=d\mathbb{X}^I+(\widetilde{\cal P}^{-1})_M{}^I\Pi^{Mm}C_m
$$
The kinetic term in (\ref{action}) can be made gauge invariant simply by minimal coupling giving the gauge-invariant kinetic term
$$
S_{Kin}=\frac{1}{4}\oint_{\Sigma}{\cal H}_{IJ}\mathcal{D}\mathbb{X}^I\wedge*\mathcal{D}\mathbb{X}^J
$$

The gauging of the Wess-Zumino term
is not simply a minimal coupling as the $B$ field is only invariant under the isometry action up to a gauge transformation. The gauging is achieved following the general prescription of
 \cite{Hull:1989jk}.
Under an infinitesimal gauge transformation, the Wess-Zumino term changes by
$$
\delta_{\epsilon}S_{wz}=\frac{1}{2}\int_V\delta_{\epsilon}{\cal K}=\frac{1}{2}\oint_{\Sigma}i_{\epsilon}{\cal K}
$$
where
$i_{\epsilon}$ is the contraction with the vector field
$\epsilon=\epsilon_m\widetilde{X}^m$ and can be written as $i_{\epsilon}=\epsilon_m\Pi^{mM}(\widetilde{{\cal P}}^{-1})_M{}^Ii_I$.
We have used the fact that $d{\cal K}=0$ so that
$\delta_{\epsilon}{\cal
K}=(i_{\epsilon}d+di_{\epsilon}){\cal K}=di_{\epsilon}{\cal K}$.
It is useful to define a one-form $v^m=v^m{}_Id\mathbb{X}^I$ on ${\cal G}$ by
$$
v^m=\Pi^{mM}L_{MN}\widetilde{\cal P}^N
$$
which satisfies
$$
\epsilon_mdv^m=i_{\epsilon}{\cal K}
$$
Then the variation of the Wess-Zumino term can be written
$$
\delta_{\epsilon}S_{wz}=-\frac{1}{2}\oint_{\Sigma}d\epsilon_m\wedge v^m
$$
This variation can be canceled
 by adding the term
\begin{equation}\label{SC}
S_{c}=\frac{1}{2}\oint_{\Sigma}C_m\wedge v^m
\end{equation}
where $C_m$ is the gauge field transforming as (\ref{C trans}). It is not difficult to show that
$$
\delta_{\epsilon}v^m={\cal L}_{\epsilon}v^m=-\epsilon_nf^{mn}{}_pv^p+L_{MN}\Pi^{mM}\Pi^{nN}d\epsilon_n
$$
so that
$$
\delta_{\epsilon}S_{c}=\frac{1}{2}\oint_{\Sigma}\left(d\epsilon_m\wedge v^m+L_{MN}\Pi^{mM}\Pi^{nN}C_m\wedge d\epsilon_n\right)
$$
The first term cancels the variation of the Wess-Zumino term so that
$$
\delta_{\epsilon}(S_{wz}+S_{c})=\frac{1}{2}c^{mn}\oint_{\Sigma}C_m\wedge d\epsilon_n  \qquad  c^{mn}=L_{MN}\Pi^{mM}\Pi^{nN}
$$
Since we require that the polarisation $\Pi^{mM}$ is null with respect to $L_{MN}$, the coefficient $c^{mn}$ vanishes and $S_{wz}+S_{c}$ is
gauge invariant. The full gauged non-linear sigma model on $\cG$ is then
$$
S_{\cG/\widetilde{G}}=\frac{1}{4}\oint_{\Sigma}{\cal H}_{IJ}\mathcal{D}\mathbb{X}^I\wedge*\mathcal{D}\mathbb{X}^J+\frac{1}{2}\oint_{\Sigma}C_m\wedge
v^m+\frac{1}{12}\int_V{\cal K}_{IJK}d\mathbb{X}^I\wedge d\mathbb{X}^J\wedge d\mathbb{X}^K
$$
We stress the fact that the gauging requires that $v^m$ is globally defined and the gauge group $\widetilde{G}_L\subset\cG_L$ is maximally
isotropic, i.e. the polarisation is null.

We   define the one-forms ${\cal C}=\cG^{-1}C\cG$ so that $C_m\wedge v^m=L_{MN}{\cal C}^M\wedge {\cal P}^N$ and the gauged theory
can be written as
$$
S_{\cG/\widetilde{G}}=\frac{1}{4}\oint_{\Sigma}{\cal M}_{MN}{\cal P}^M\wedge *{\cal P}^N+\frac{1}{2}\oint_{\Sigma}\mathcal{C}^M\wedge *{\cal
J}_M+\frac{1}{4}\oint_{\Sigma}{\cal M}_{MN}{\cal C}^M\wedge *{\cal C}^N+\frac{1}{12}\int_Vt_{MNP}{\cal P}^M\wedge {\cal P}^N\wedge{\cal P}^P
$$
where
$$
{\cal J}_M={\cal M}_{MN}{\cal P}^N-L_{MN}*{\cal P}^N
$$
We note that the constraint (\ref{constraintA}) may be written as ${\cal J}_M=0$.

As in the doubled torus construction, the conventional undoubled
theory is recovered by eliminating the gauge fields $\mathcal{C}_m$, which again appear quadratically as auxiliary fields. In the quantum theory, integrating out $\mathcal{C}_m$
 generates a shift in the dilaton.

As an example, let us consider a general twisted torus with $H$-flux as discussed  {at the beginning of this section and in \cite{Kaloper ``The O(dd) story of massive supergravity'',Hull ``Flux compactifications of string theory on twisted tori''}.} The doubled group $\cG$ in this case is generated by the Lie algebra (\ref{Kflux}) where the non-zero structure constants of the algebra (\ref{Bigalg}) are
$$
\widetilde{\Pi}_m{}^M\widetilde{\Pi}_n{}^N\Pi^p{}_Pt_{MN}{}^P=f_{mn}{}^p   \qquad
\widetilde{\Pi}_m{}^M\widetilde{\Pi}_n{}^N\widetilde{\Pi}_p{}^Pt_{MNP}=K_{mnp}
$$
The Maurer-Cartan equations
 for the left-invariant one-forms on $\cG$ are
\begin{equation}\label{example}
dP^m+\frac{1}{2}f_{np}{}^mP^n\wedge P^p=0   \qquad  dQ_m- f_{mn}{}^pQ_p\wedge P^n-\frac{1}{2}K_{mnp}P^n\wedge P^p=0
\end{equation}
These one-forms are dual to the vector fields generating the right-acting gauge algebra (\ref{Kflux}). The right-invariant left action, $\cG_L$ is generated by $\widetilde{T}_{\hat{A}}=(\widetilde{Z}_m,\widetilde{X}^m)$ which satisfy the Lie algebra
$$
[\widetilde{Z}_m,\widetilde{Z}_n]=f_{mn}{}^p\widetilde{Z}_p-K_{mnp}\widetilde{X}^p	\qquad	[\widetilde{Z}_m,\widetilde{X}^n]=f_{mp}{}^n\widetilde{X}^p	 \qquad	[\widetilde{X}^m,\widetilde{X}^n]=0
$$
Gauging the left acting subgroup $\widetilde{G}_L$ generated by $\widetilde{X}^m$ requires the introduction of the one-form fields $C_m$ by minimal coupling
\begin{equation}
{\cal P}=\cG^{-1} d\cG\rightarrow \cG^{-1} {\cal D}\cG=\cG^{-1} \left(d+C\right)\cG
\end{equation}
It is useful to define ${\cal C}=\cG^{-1} C\cG$, so that
\begin{equation}
{\cal P}^M\rightarrow{\cal P}^M+\mathcal{C}^M \qquad  {\cal C}^M=\left(\cG^{-1}C\cG\right)^M
\end{equation}
In the current example (\ref{example}) one can show that $\Pi_M{}^m\mathcal{C}^M=0$ and we may write
\begin{equation}
{\cal P}^M\rightarrow \widehat{{\cal P}}^M={\cal P}^M+\Pi^{Mm}\mathcal{C}_m    \qquad  \Leftrightarrow \qquad Q_m\rightarrow Q_m+\mathcal{C}_m
\end{equation}
where ${\cal C}_m=\widetilde{\Pi}_{mM}{\cal C}^M$.

The gauged action is  {then}
\begin{eqnarray}
S=\frac{1}{4}\oint_{\Sigma}{\cal M}_{MN}\widehat{{\cal P}}^M\wedge *\widehat{{\cal P}}^N+\frac{1}{2}\oint_{\Sigma}L_{MN}{\cal
P}^M\wedge\Pi^{Nm}\mathcal{C}_m+\frac{1}{12}\int_Vt_{MNP}{\cal P}^M\wedge {\cal P}^N\wedge{\cal P}^P\nonumber
\end{eqnarray}
 {where, f}or this example, the Wess-Zumino term is
\begin{eqnarray}\label{WZW}
S_{wz}&=&\frac{1}{4}\int_Vf_{np}{}^mQ_m\wedge P^n\wedge P^p-\frac{1}{12}\int_VK_{mnp}P^m\wedge P^n\wedge P^p\nonumber\\
&=&\frac{1}{2}\oint_{\Sigma}P^m\wedge Q_m+\frac{1}{6}\int_VK_{mnp}P^m\wedge P^n\wedge P^p
\end{eqnarray}
Note that we have used the fact that $P^m\wedge Q_m$ is globally defined to write the two-dimensional term. Expanding the gauged action using (\ref{WZW}) and then completing the square in $\mathcal{C}_m$, the doubled action may be written
\begin{eqnarray}
S&=&\oint_{\Sigma}\left(\frac{1}{2}g_{mn}P^m\wedge *P^n+\frac{1}{2}B_{mn}P^m\wedge P^n+\frac{1}{4}g^{mn}\lambda_m\wedge
*\lambda_n\right)+\int_V\frac{1}{6}K_{mnp}P^m\wedge P^n\wedge P^p\nonumber
\end{eqnarray}
where
\begin{eqnarray}
\lambda_m&=&Q_m+\mathcal{C}_m-g_{mn}*P^n-B_{mn}P^n
\end{eqnarray}
{The $\lambda_m$ can then be integrated out  to give a theory whose target space is a twisted torus with $H$-flux. As in the doubled torus construction of section 5, the change in variables from $C_m$ to $\lambda_m$ introduces a determinant in the path integral which gives a shift to the dilaton.}

\noindent\textbf{Recovering the Doubled Torus}

\noindent Upon gauging the left action $\tilde{x}\rightarrow\tilde{x}+\epsilon$  {generated by $\widetilde{X}^x$} and integrating out the corresponding gauge field $C_x$, the
doubled twisted torus formalism for the duality twist construction reduces to the doubled torus formalism of section 2 as we now show. The $x$-independent
tensor ${\cal M}_{MN}$ for an $O(d,d)$-twisted reduction is given by \cite{Hull ``Gauge Symmetry T-Duality and Doubled Geometry''}
\begin{equation} {\cal M}_{MN}= \left(\begin{array}{ccc}
1+{\cal M}_{AB}{\cal A}^A{}_x{\cal A}^B{}_x+b^2 & b & bL_{AC}{\cal A}^C{}_x+{\cal M}_{AC}{\cal A}^C{}_x \\
b & 1 & L_{AC}{\cal A}^C{}_x  \\
 bL_{BC}{\cal A}^C{}_x+{\cal M}_{BC}{\cal A}^C{}_x   &  L_{AC}{\cal A}^C{}_x & {\cal
 M}_{AB}+L_{AC}L_{BD}{\cal A}^C{}_x{\cal A}^D{}_x
\end{array}\right)\nonumber
\end{equation}
where $b=\frac{1}{2}L_{AB}{\cal A}^A{}_x{\cal A}^B{}_x$.   The WZW term for the doubled group sigma model (\ref{action}) with $t_{xA}{}^B=-N^B{}_A$ and $t_{xAB}=-N_{AB}$ is
\begin{eqnarray}
S_{wz}&=&-\frac{1}{4}\int_VN_{AB}P^x\wedge{\cal P}^A\wedge{\cal P}^B=\frac{1}{2}\oint_{\Sigma}P^x\wedge Q_x
\end{eqnarray}
 {and t}he action for the doubled group sigma model may then be written as the integral over the two-dimensional Lagrangian
\begin{eqnarray}
{\cal L}&=&\frac{1}{4}{\cal M}_{MN}({\cal P}^M+{\cal C}^M)\wedge *({\cal P}^N+{\cal C}^N)+\frac{1}{2}P^x\wedge C_x+\frac{1}{2}P^x\wedge Q_x
\end{eqnarray}
where
$$
{\cal P}^{\hat{M}}+{\cal C}^{\hat{M}}=\left( {\cal P}^A , P^x , Q_x+C_x\right)
$$
Expanding this out and completing the square in $C_x$, the expression may be simplified considerably
\begin{eqnarray}
{\cal L}&=&\frac{1}{4}{\cal M}_{AB}{\cal P}^A\wedge*{\cal P}^B-\frac{1}{2}{\cal P}^A \wedge*J_A+\frac{1}{2}gP^x\wedge*P^x+\frac{1}{4}\lambda_x\wedge*\lambda_x
\end{eqnarray}
where $J_A={\cal M}_{AB}{\cal A}^B-L_{AB}*{\cal A}^B$, we have defined ${\cal A}^A={\cal A}^A{}_xdx$, and
\begin{eqnarray}
\lambda_x&=&Q_x+C_x-*P^x+b P^x+L_{AB}{\cal A}^A{}_x{\cal P}^B\nonumber\\
g&=&1 +\frac{1}{2}{\cal M}_{AB}{\cal A}^A{}_x{\cal A}^B{}_x
\end{eqnarray}
Integrating out $\lambda_x$ gives the doubled torus sigma model (\ref{Ttwistedsigma}) of \cite{Hull ``A geometry for non-geometric string
backgrounds''}.

\subsection{Recovering the conventional background}

In this section we derive the prescription for constructing the physical metric and $H$-field strength from the doubled geometry  that was presented in section 2.6. In particular, the strange expression for the $H$-field strength (\ref{H conjecture}) arises quite naturally from the   world-sheet point of view. We shall assume here that the generators $X^m$   close to generate a subgroup $\widetilde{G}$ and so a conventional description does exist, i.e. we take $R^{mnp}=0$. We shall generalise to  the $R^{mnp}\neq 0$ case in the following section.

 {We recall from (\ref{decomposition})  that  the left-invariant one-forms on $\cX$ may be written as
\begin{equation}\label{P again}
{\cal P}^{\hat{M}}=\Phi^{\hat{N}}{\cal V}_{\hat{N}}{}^{\hat{M}}(x)
\end{equation}
where
$$
\Phi^{\hat{M}}=(r^m,{\ti q}_m)
$$
where $r^m$ and $\tilde{q}_m$ are defined in section 2.6. The ${\cal V}_{\hat{N}}{}^{\hat{M}}(x)$ may then be used to define the metric
${\cal H}_{\hat{M}\hat{N}}(x)={\cal M}_{\hat{P}\hat{Q}}{\cal V}^{\hat{M}}{}_{\hat{P}}{\cal V}^{\hat{Q}}{}_{\hat{N}}$
whose components define a metric $g_{mn}$ and $B$-field $B_{mn}$ by
\begin{equation}\label{H again}
{\cal H}_{\hat{M}\hat{N}}(x)=
\left(\begin{array}{cc}
g_{mn}+B_{mp}g^{pq}B_{qn} & B_{mp}g^{pn} \\
g^{mp}B_{np} & g^{mn}
                          \end{array}\right)
\end{equation}
The gauging of the $\widetilde{G}_L$ subgroup may then be achieved by the minimal coupling $\tilde{q}_m\rightarrow \tilde{q}_m+{\cal C}_m$ and the addition of the term $S_c$ given in (\ref{SC}). Expanding the gauged sigma model action using (\ref{P again}) and (\ref{H again}) gives
\begin{eqnarray}
S&=&\frac{1}{4}\oint_{\Sigma}\left(g_{mn}+B_{mp}g^{pq}B_{qn}\right)r^m\wedge *r^n+\frac{1}{2}\oint_{\Sigma}B_{mp}g^{pn}r^m\wedge *({\ti q}+{\cal C})_n\nonumber\\
 &&+\frac{1}{4}\oint_{\Sigma}g^{mn}({\ti q}+{\cal C})_m\wedge *({\ti q}+{\cal C})_n +\frac{1}{2}\oint_{\Sigma}r^m\wedge {\cal C}_m +\frac{1}{2}\int_V {\cal K}\nonumber
\end{eqnarray}
Completing the square in ${\cal C}_m$, this action may be written as
\begin{eqnarray}
S&=&\frac{1}{2}\oint_{\Sigma}g_{mn}r^m\wedge *r^n+\frac{1}{2}\oint_{\Sigma}B_{mn}r^m\wedge *r^n -\frac{1}{2}\oint_{\Sigma}r^m\wedge {\ti q}_m \nonumber\\
&&+\frac{1}{4}\oint_{\Sigma}g^{mn}\lambda_m\wedge *\lambda_n +\frac{1}{2}\int_V {\cal K}\nonumber
\end{eqnarray}
where
$$
\lambda_m={\ti q}_m+{\cal C}_m-g_{mn}*r^n-B_{mn}r^n
$$
Integrating over $\lambda_m$ gives a shift in the dilaton as in (\ref{dilshift}) and we recover a conventional   world-sheet description of the theory
$$
S=\frac{1}{2}\oint_{\Sigma}g_{ij}dx^i\wedge *dx^j +\int_V H
$$
where
$$
g_{ij}=g_{mn}r^m{}_ir^n{}_j \qquad  H=dB-\frac{1}{2}d\left(r^m\wedge {\ti q}_m\right)+\frac{1}{2}{\cal K}
$$
as claimed in section 2.6.}

\subsection{Compactifications with $R$-Flux}

 {In section 2 we discussed the doubled torus description of target spaces that could be constructed as $T^d$ fibrations over $S^1_x$ with monodromy in $O(d,d;\Z)$. The natural action of $O(d,d;\Z)$ on the theory in the fibres related different polarisations by T-duality. There is some evidence that there should still be a T-duality on the base circle   \cite{ Dabholkar ``Generalised T-duality and
non-geometric backgrounds''}
 that
 exchanges $Z_x$ with $X^x$ and would act on the structure constants in the gauge algebra (\ref{algebra fHQ}) as}
\begin{equation}
K_{xab}\rightarrow f_{ab}{}^x    \qquad    f_{xa}{}^b\rightarrow Q_a{}^{xb}    \qquad  Q_x{}^{ab}\rightarrow R^{xab}
\end{equation}
to give the algebra
\begin{eqnarray}
[X^x,Z_a]=Q_{a}{}^{xb}Z_b+f_{ab}{}^xX^b   \qquad  [X^x,X^a]=-Q^{xa}{}_{b}X^b+R^{xab}Z_b\nonumber
\end{eqnarray}
\begin{eqnarray}\label{hey}
[Z_a,Z_b ]=f_{ab}{}^xZ_x  \qquad [X^a,Z_b]=-Q^{xa}{}_{b}Z_x \qquad  [X^a,X^b]=R^{xab}Z_x
\end{eqnarray}
As discussed in section 3.3.2, it was conjectured in \cite{Dabholkar ``Generalised T-duality and non-geometric backgrounds''} that the structure constant $R^{xab}$
(`$R$-flux') corresponds to a background constructed with a   twist over a dual circle $\widetilde{S}^1$ (with coordinate $\tilde x$ conjugate
to the winding number). An example of such a background is that which arises from the conjectured T-duality of the T-fold in section 3 along the
$x$ direction \cite{Dabholkar ``Generalised T-duality and non-geometric backgrounds''}. The algebra (\ref{hey}) in this case is
\begin{eqnarray}\label{R algebra}
\left[X^x,X^y\right]=mZ_z   \qquad \left[X^y,X^z\right]=mZ_x   \qquad \left[X^z,X^x\right]=mZ_y
\end{eqnarray}
All other commutators vanish. The generators $\widetilde{X}^m$ (the right-invariant counterparts to $X^m$ above) do not close to form a subalgebra. As such, we cannot
integrate out the $\tilde{x}_m$ completely to get a target space described solely in terms of the $x^m$. In the classical theory however, we can
use the self-duality constraint (\ref{constraintA}) to remove the $d\tilde{x}_m$ dependence and write the doubled theory in terms of the Lagrangian ${\cal
L}(\tilde{x},dx)$.

We choose to write the one-forms on $\cX$ in a way that makes manifest the cyclic symmetry of the coordinates
\begin{eqnarray}
\begin{array}{ll}
P^x=dx-n\tilde{y}d\tilde{z}+n\tilde{z}d\tilde{y}   &\qquad Q_x=d\tilde{x} \\
P^y=dy-n\tilde{z}d\tilde{x}+n\tilde{x}d\tilde{z}   &\qquad Q_y=d\tilde{y} \\
P^z=dz-n\tilde{x}d\tilde{y}+n\tilde{y}d\tilde{x}   &\qquad Q_z=d\tilde{z}
\end{array}
\end{eqnarray}
where
$$
R^{xyz}=m=2n\in\Z
$$
With a little work, the self-duality constraints ($Q_m=*\delta_{mn}P^n$), given by (\ref{constraintA}) with ${\cal M}_{MN}=\delta_{MN}$, can be written in terms of $\tilde{x}_m$ and $dx^m$ only
\begin{eqnarray}\label{hard work}
Q_x&=&\frac{1}{T}\left(\zeta_x*dx+n^2\tilde{x}\tilde{y}*dy+n^2\tilde{z}\tilde{x}*dz+n\tilde{z}dy-n\tilde{y}dz\right)\nonumber\\
Q_y&=&\frac{1}{T}\left(\zeta_y*dy+n^2\tilde{y}\tilde{z}*dz+n^2\tilde{x}\tilde{y}*dx+n\tilde{x}dz-n\tilde{z}dx\right)\nonumber\\
Q_z&=&\frac{1}{T}\left(\zeta_z*dz+n^2\tilde{z}\tilde{x}*dx+n^2\tilde{y}\tilde{z}*dy+n\tilde{y}dx-n\tilde{x}dy\right)
\end{eqnarray}
where
\begin{equation}
T\equiv1+n^2(\tilde{x}^2+\tilde{y}^2+\tilde{z}^2)
\end{equation}
and
\begin{equation}
\zeta_x=1+(n\tilde{x})^2\qquad  \zeta_y=1+(n\tilde{y})^2\qquad  \zeta_z=1+(n\tilde{z})^2
\end{equation}
Note that using the constraints $Q_m=*\delta_{mn}P^n$, one can show
\begin{equation}
\tilde{x}dx+\tilde{y}dy+\tilde{z}dz=*\left(\tilde{x}Q_x+\tilde{y}Q_y+\tilde{z}Q_z\right)
\end{equation}
This result is useful in determining the expressions (\ref{hard work}). The classical equation of motion for the $x$-coordinate is then given by
the  Maurer-Cartan equation
 $dQ_x=0$ and (\ref{hard work})  {so that, for example, the $x$ equation of motion is}
\begin{eqnarray}
d\left(\frac{1}{T}\left(\zeta_x*dx+n^2\tilde{x}\tilde{y}*dy+n^2\tilde{z}\tilde{x}*dz+n\tilde{z}dy-n\tilde{y}dz\right)\right)=0
\end{eqnarray}
The $y$ and $z$ equations of motion are given by cyclic permutations of this. These equations of motion may be recovered from the action
\begin{eqnarray}
S&=&\frac{1}{2}\oint_{\Sigma}g_{mn}dx^m\wedge *dx^n+\frac{1}{2}\oint_{\Sigma}B_{mn}dx^m\wedge dx^n
\end{eqnarray}
where the metric and $B$-field are
\begin{eqnarray}
g=\frac{1}{T}\left(%
\begin{array}{ccc}
  \zeta_x & n^2\tilde{x}\tilde{y} &  n^2\tilde{z}\tilde{x} \\
   n^2\tilde{x}\tilde{y} & \zeta_y &  n^2\tilde{y}\tilde{z} \\
  n^2\tilde{z}\tilde{x} &  n^2\tilde{y}\tilde{z} & \zeta_z \\
\end{array}%
\right) \qquad  B=\frac{n}{T}\left(%
\begin{array}{ccc}
  0 & \tilde{z} & -\tilde{y} \\
  -\tilde{z} & 0 & \tilde{x} \\

  \tilde{y} & -\tilde{x} & 0 \\
\end{array}%
\right)
\end{eqnarray}
We see that  it is possible to remove the $d\tilde{x}^m$ dependence and give a Lagrangian which depends explicitly on the
`winding' coordinates $\tilde{x}_m$ and $dx^m$. This is reminiscent of the results found in \cite{Harvey:2005ab,Gregory:1997te,Tong:2002rq}.


\begin{center}
\textbf{Acknowledgements}
\end{center}
The work of RR was supported by the Deutsche Forschungsgemeinschaft (DFG) in the SFB 676 ``Particles, Strings and the Early Universe". RR thanks the string theory group at Queen Mary, University of London for their kind hospitality during the final stages of this project.


\begin{thebibliography}{03}





\bibitem{Hellerman:2002ax}
  S.~Hellerman, J.~McGreevy and B.~Williams,
  ``Geometric constructions of non-geometric string theories,''
  JHEP {\bf 0401} (2004) 024
  [arXiv:hep-th/0208174].


\bibitem{Dabholkar ``Duality twists orbifolds and fluxes''}
  A.~Dabholkar and C.~Hull,
  ``Duality twists, orbifolds, and fluxes,''
  JHEP {\bf 0309}, 054 (2003)
  [arXiv:hep-th/0210209].



\bibitem{Kachru:2002sk}
  S.~Kachru, M.~B.~Schulz, P.~K.~Tripathy and S.~P.~Trivedi,
  ``New supersymmetric string compactifications,''
  JHEP {\bf 0303} (2003) 061
  [arXiv:hep-th/0211182].


\bibitem{Flournoy:2004vn}
  A.~Flournoy, B.~Wecht and B.~Williams,
  ``Constructing non-geometric vacua in string theory,''
  Nucl.\ Phys.\  B {\bf 706} (2005) 127
  [arXiv:hep-th/0404217].

\bibitem{Narain:1990mw}
  K.~S.~Narain, M.~H.~Sarmadi and C.~Vafa,
  ``Asymmetric orbifolds: Path integral and operator formulations,''
  Nucl.\ Phys.\  B {\bf 356} (1991) 163.

\bibitem{Kumar:1996zx}
  A.~Kumar and C.~Vafa,
  ``U-manifolds,''
  Phys.\ Lett.\  B {\bf 396} (1997) 85
  [arXiv:hep-th/9611007].

\bibitem{Hull ``A geometry for non-geometric string backgrounds''}
  C.~M.~Hull,
  ``A geometry for non-geometric string backgrounds,''
  JHEP {\bf 0510} (2005) 065
  [arXiv:hep-th/0406102].

\bibitem{Hull ``Doubled geometry and T-folds''}
  C.~M.~Hull,
  ``Doubled geometry and T-folds,''
  JHEP {\bf 0707} (2007) 080
  [arXiv:hep-th/0605149].

\bibitem{Buscher ``A Symmetry of the String Background Field Equations''}
  T.~H.~Buscher,
  ``A Symmetry of the String Background Field Equations,''
  Phys.\ Lett.\  B {\bf 194}, 59 (1987).

\bibitem{Dabholkar ``Generalised T-duality and non-geometric backgrounds''}
  A.~Dabholkar and C.~Hull,
  ``Generalised T-duality and non-geometric backgrounds,''
  JHEP {\bf 0605}, 009 (2006)
  [arXiv:hep-th/0512005].


\bibitem{Shelton ``Nongeometric flux compactifications''}
  J.~Shelton, W.~Taylor and B.~Wecht,
  ``Nongeometric flux compactifications,''
  JHEP {\bf 0510}, 085 (2005)
  [arXiv:hep-th/0508133].

\bibitem{Shelton:2006fd}
  J.~Shelton, W.~Taylor and B.~Wecht,
  ``Generalized flux vacua,''
  JHEP {\bf 0702} (2007) 095
  [arXiv:hep-th/0607015].

\bibitem{Courant}
T.~Courant,
``Dirac manifolds,"
Trans. Amer. Math. Soc., 319:631-661, (1990).

\bibitem{Gualtieri}
M.~Gualtieri,
``Generalized complex geometry,"
PhD Thesis (2004).
arXiv:math/0401221v1 [math.DG]

\bibitem{Gregory:1997te}
  R.~Gregory, J.~A.~Harvey and G.~W.~Moore,
  ``Unwinding strings and T-duality of Kaluza-Klein and H-monopoles,''
  Adv.\ Theor.\ Math.\ Phys.\  {\bf 1} (1997) 283
  [arXiv:hep-th/9708086].


\bibitem{Tong:2002rq}
  D.~Tong,
  ``NS5-branes, T-duality and   world-sheet instantons,''
  JHEP {\bf 0207}, 013 (2002)
  [arXiv:hep-th/0204186].


\bibitem{Harvey:2005ab}
  J.~A.~Harvey and S.~Jensen,
  ``Worldsheet instanton corrections to the Kaluza-Klein monopole,''
  JHEP {\bf 0510} (2005) 028
  [arXiv:hep-th/0507204].

\bibitem{Okuyama:2005gx}
  K.~Okuyama,
  ``Linear sigma models of H and KK monopoles,''
  JHEP {\bf 0508} (2005) 089
  [arXiv:hep-th/0508097].



\bibitem{Scherk ``How To Get Masses From Extra Dimensions''}
  J.~Scherk and J.~H.~Schwarz,
  ``How To Get Masses From Extra Dimensions,''
  Nucl.\ Phys.\  B {\bf 153}, 61 (1979).

\bibitem{Hull ``Massive string theories from M-theory and F-theory''}
  C.~M.~Hull,
  ``Massive string theories from M-theory and F-theory,''
  JHEP {\bf 9811} (1998) 027
  [arXiv:hep-th/9811021].

\bibitem{Hull ``Flux compactifications of string theory on twisted tori''}
  C.~M.~Hull and R.~A.~Reid-Edwards,
  ``Flux compactifications of string theory on twisted tori,''
  arXiv:hep-th/0503114.

\bibitem{Hull ``Gauge Symmetry T-Duality and Doubled Geometry''}
  C.~M.~Hull and R.~A.~Reid-Edwards,
  ``Gauge Symmetry, T-Duality and Doubled Geometry,''
  JHEP {\bf 0808} (2008) 043
  [arXiv:0711.4818 [hep-th]].

\bibitem{Dall'Agata:2007sr}
  G.~Dall'Agata, N.~Prezas, H.~Samtleben and M.~Trigiante,
  ``Gauged Supergravities from Twisted Doubled Tori and Non-Geometric String
  Backgrounds,''
  arXiv:0712.1026 [hep-th].

\bibitem{Giveon
  ``On nonAbelian duality''}
  A.~Giveon and M.~Rocek,
  ``On nonAbelian duality,''
  Nucl.\ Phys.\  B {\bf 421}, 173 (1994)
  [arXiv:hep-th/9308154].

\bibitem{Klimcik  ``Dual Nonabelian Duality And The Drinfel'd Double''}
  C.~Klimcik and P.~Severa,
  ``Dual Nonabelian Duality And The Drinfel'd Double,''
  Phys.\ Lett.\  B {\bf 351}, 455 (1995)
  [arXiv:hep-th/9502122].

\bibitem{Klimcik:1995ux}
  C.~Klimcik and P.~Severa,
  ``Dual Nonabelian Duality And The Drinfel'd Double,''
  Phys.\ Lett.\  B {\bf 351} (1995) 455
 [arXiv:hep-th/9502122].

\bibitem{Drinfel'd1}
V.~G.~Drinfel'd,
``On Poisson homogeneous spaces of Poisson-Lie groups,''
Teoreticheskaya i Mathematicheskaya Fizika, {\bf 95}, 2 (1993) 226-227,
translated in Theoretical and Mathematical Physics, Springer Verlag New York, {\bf 95}, 2 (1993) 524-525
%


\bibitem{Drinfel'd:1986in}
  V.~G.~Drinfel'd,
  ``Quantum groups,''
  J.\ Sov.\ Math.\  {\bf 41} (1988) 898





\bibitem{SemenovTianShansky:1993ws}
  M.~A.~Semenov-Tian-Shansky,
  ``Poisson Lie groups, quantum duality principle, and the quantum double,''
Theor.\ Math.\ Phys.\  {\bf 93} (1992) 1292
  [arXiv:hep-th/9304042].


\bibitem{Weinstein}
  J. Lu and A. Weinstein,
  ``Poisson-Lie Groups, Dressing Transformations and Bruhat Decompositions,''
J. Differential Geometry {\bf 31} (1990) 526
  [arXiv:hep-th/9502122].

\bibitem{Maharana ``Noncompact symmetries in string theory''}
  J.~Maharana and J.~H.~Schwarz,
  ``Noncompact symmetries in string theory,''
  Nucl.\ Phys.\  B {\bf 390}, 3 (1993)
  [arXiv:hep-th/9207016].


\bibitem{Hull ``Unity of superstring dualities''}
  C.~M.~Hull and P.~K.~Townsend,
  ``Unity of superstring dualities,''
  Nucl.\ Phys.\  B {\bf 438} (1995) 109
  [arXiv:hep-th/9410167].


\bibitem{Kaloper ``The O(dd) story of massive supergravity''}
  N.~Kaloper and R.~C.~Myers,
  ``The O(dd) story of massive supergravity,''
  JHEP {\bf 9905}, 010 (1999)
  [arXiv:hep-th/9901045].

\bibitem{Hull ``Flux compactifications of M-theory on twisted tori''}
  C.~M.~Hull and R.~A.~Reid-Edwards,
  ``Flux compactifications of M-theory on twisted tori,''
  JHEP {\bf 0610} (2006) 086
  [arXiv:hep-th/0603094].




\bibitem{ReidEdwards ``Geometric and non-geometric compactifications of IIB supergravity''}
  R.~A.~Reid-Edwards,
  ``Geometric and non-geometric compactifications of IIB supergravity,''
  arXiv:hep-th/0610263.

\bibitem{ReidEdwards:2008rd}
  R.~A.~Reid-Edwards and B.~Spanjaard,
  ``N=4 Gauged Supergravity from Duality-Twist Compactifications of String
  Theory,''
  arXiv:0810.4699 [hep-th].



\bibitem{Quevedo
  ``Duality symmetries from nonAbelian isometries in string theory''}
  X.~C.~de la Ossa and F.~Quevedo,
  ``Duality symmetries from nonAbelian isometries in string theory,''
  Nucl.\ Phys.\  B {\bf 403}, 377 (1993)
  [arXiv:hep-th/9210021].

\bibitem{Ron}
R A Reid-Edwards, in preparation.




\bibitem{Hull ``Global Aspects of T-Duality Gauged Sigma Models and T-Folds''}
  C.~M.~Hull,
  ``Global Aspects of T-Duality, Gauged Sigma Models and T-Folds,''
  arXiv:hep-th/0604178.


\bibitem{Alvarez  ``Is the string coupling constant invariant under T-duality?''}
  E.~Alvarez and Y.~Kubyshin,
  ``Is the string coupling constant invariant under T-duality?,''
  Nucl.\ Phys.\ Proc.\ Suppl.\  {\bf 57} (1997) 44
  [arXiv:hep-th/9610032].

\bibitem{Kugo ``Target space duality as a symmetry of string field theory''}
  T.~Kugo and B.~Zwiebach,
  ``Target space duality as a symmetry of string field theory,''
  Prog.\ Theor.\ Phys.\  {\bf 87} (1992) 801
  [arXiv:hep-th/9201040].

\bibitem{Albertsson:2008gq}
  C.~Albertsson, T.~Kimura and R.~A.~Reid-Edwards,
  ``D-branes and doubled geometry,''
  arXiv:0806.1783 [hep-th].



\bibitem{Hull:1989jk}
  C.~M.~Hull and B.~J.~Spence,
  ``The gauged nonlinear sigma model with Wess-Zumino term,''
  Phys.\ Lett.\  B {\bf 232} (1989) 204.





\bibitem{Hull ``Gauged D = 9 supergravities and Scherk-Schwarz reduction''}
  C.~M.~Hull,
  ``Gauged D = 9 supergravities and Scherk-Schwarz reduction,''
  Class.\ Quant.\ Grav.\  {\bf 21}, 509 (2004)
  [arXiv:hep-th/0203146].

\bibitem{d=4}
  B.~de Wit, H.~Samtleben and M.~Trigiante,
  ``The maximal D = 4 supergravities,''
  JHEP {\bf 0706} (2007) 049
  [arXiv:0705.2101 [hep-th]].




\bibitem{Tseytlin:1990va}
  A.~A.~Tseytlin,
  ``Duality Symmetric Closed String Theory And Interacting Chiral Scalars,''
  Nucl.\ Phys.\  B {\bf 350}, 395 (1991).


\bibitem{Samtleben Review}
  H.~Samtleben,
  ``Lectures on Gauged Supergravity and Flux Compactifications,''
  arXiv:0808.4076 [hep-th].

\bibitem{Giveon:1994fu}
  A.~Giveon, M.~Porrati and E.~Rabinovici,
  ``Target space duality in string theory,''
  Phys.\ Rept.\  {\bf 244} (1994) 77
  [arXiv:hep-th/9401139].

\bibitem{Dall'Agata:2008qz}
  G.~Dall'Agata and N.~Prezas,
  ``Worldsheet theories for non-geometric string backgrounds,''
  arXiv:0806.2003 [hep-th].



\end{thebibliography}
\end{document}